\documentclass[twocolumn]{aastex631} 
\usepackage{CLASSY_Preamble}

\usepackage{amssymb}	
\usepackage{longtable}
\usepackage{scalerel}

\mathchardef\mhyphen="2D

\newcommand{\feii}{Fe\,{\sc ii}}

\newcommand{\siv}{S\,{\sc iv}}
\newcommand{\siiv}{Si\,{\sc iv}}

\newcommand{\cii}{[C\,{\sc ii}]}
\newcommand{\ciii}{C\,{\sc iii}}
\newcommand{\civ}{C\,{\sc iv}}
\newcommand{\mgii}{Mg\,{\sc ii}}

\newcommand{\angstrom}{\text{ \normalfont\AA}}

\newcommand{\overbar}[1]{\mkern 1.5mu\overline{\mkern-1.5mu#1\mkern-1.5mu}\mkern 1.5mu}

\mathchardef\mhyphen="2D

\def\lya{Ly$\alpha$}
\def\ly{$\lambda$}

\def\hi{H\,{\sc i}}

\def\cii{C\,{\sc ii}}
\def\ciii{C\,{\sc iii}}
\def\civ{C\,{\sc iv}}
\def\nii{N\,{\sc ii}}

\def\nv{N\,{\sc v}}
\def\oi{O\,{\sc i}}

\def\nai{Na\,{\sc i}}

\def\mgii{Mg\,{\sc ii}}
\def\alii{Al\,{\sc ii}}

\def\Sii{Si\,{\sc i}}
\def\siiv{Si\,{\sc iv}}
\def\Siii{Si\,{\sc ii}}

\def\siiii{Si\,{\sc iii}}

\def\sixiii{Si\,{\sc xiii}}

\def\sii{S\,{\sc ii}}
\def\siv{S\,{\sc iv}}

\def\feii{Fe\,{\sc ii}}

\def\Q0059{Q0059--2735}

\def\S2S3{S2S3}

\definecolor{blk}{rgb}{0.0,0.0,0.0}
\definecolor{red}{rgb}{0.75,0.0,0.0}
\definecolor{yel}{rgb}{0.65,0.65,0.0}
\definecolor{grn}{rgb}{0.0,0.75,0.0}
\definecolor{blu}{rgb}{0.0,0.0,0.75}
\definecolor{gry}{rgb}{0.75,0.75,0.75}

\def\nh{\ifmmode n_\mathrm{\scriptscriptstyle H} \else $n_\mathrm{\scriptscriptstyle H}$\fi}
\def\ne{\ifmmode n_\mathrm{\scriptstyle e} \else $n_\mathrm{\scriptstyle e}$\fi}
\def\Te{\ifmmode T_\mathrm{\scriptstyle e} \else $T_\mathrm{\scriptstyle e}$\fi}
\def\Qh{\ifmmode Q_\mathrm{\scriptstyle H} \else $Q_\mathrm{\scriptstyle H}$\fi}
\def\Uh{\ifmmode U_\mathrm{\scriptstyle H} \else $U_\mathrm{\scriptstyle H}$\fi}
\def\Nh{\ifmmode N_\mathrm{\scriptstyle H} \else $N_\mathrm{\scriptstyle H}$\fi}
\def\NSi{\ifmmode N_\mathrm{\scriptstyle si} \else $N_\mathrm{\scriptstyle Si}$\fi}
\def\Uhhp{\ifmmode U_\mathrm{\scriptstyle H,HP} \else $U_\mathrm{\scriptstyle H,HP}$\fi}
\def\Nhhp{\ifmmode N_\mathrm{\scriptstyle H,HP} \else $N_\mathrm{\scriptstyle H,HP}$\fi}
\def\Uhvhp{\ifmmode U_\mathrm{\scriptstyle H,VHP} \else $U_\mathrm{\scriptstyle H,VHP}$\fi}
\def\Nhvhp{\ifmmode N_\mathrm{\scriptstyle H,VHP} \else $N_\mathrm{\scriptstyle H,VHP}$\fi}
\def\Nion{\ifmmode N_\mathrm{\scriptstyle ion} \else $N_\mathrm{\scriptstyle ion}$\fi}

\def\Zsun{\ifmmode {\rm Z}_{\odot} \else Z$_{\odot}$\fi}
\def\Msun{\ifmmode {\rm M}_{\odot} \else M$_{\odot}$\fi}
\def\kms{\ifmmode {\rm km~s}^{-1} \else km~s$^{-1}$\fi}
\def\Lya{\ifmmode {\rm Ly}\alpha \else Ly$\alpha$\fi}
\def\Lyb{\ifmmode {\rm Ly}\beta \else Ly$\beta$\fi}
\def\Lyg{\ifmmode {\rm Ly}\gamma \else Ly$\gamma$\fi}
\def\Lyd{\ifmmode {\rm Ly}\delta \else Ly$\delta$\fi}
\def\neaod{\ifmmode n_\mathrm{\scriptscriptstyle AOD} \else $n_\mathrm{\scriptscriptstyle AOD}$\fi}
\def\necrit{\ifmmode n_\mathrm{\scriptstyle cr} \else $n_\mathrm{\scriptstyle cr}$\fi}
\def\ncr{\ifmmode n_\mathrm{\scriptstyle cr} \else $n_\mathrm{\scriptstyle cr}$\fi}
\def\nepi{\ifmmode n_\mathrm{\scriptscriptstyle PI} \else $n_\mathrm{\scriptscriptstyle PI}$\fi}
\def\gtorder{\mathrel{\raise.3ex\hbox{$>$}\mkern-14mu\lower0.6ex\hbox{$\sim$}}}
\def\ltorder{\mathrel{\raise.3ex\hbox{$<$}\mkern-14mu\lower0.6ex\hbox{$\sim$}}}

\def\vro{\ifmmode v_\mathrm{\scriptscriptstyle 1, \scriptstyle r} \else $v_\mathrm{\scriptscriptstyle 1, \scriptstyle r}$\fi}
\def\vrc{\ifmmode v_\mathrm{\scriptscriptstyle 2, \scriptstyle r} \else $v_\mathrm{\scriptscriptstyle 2, \scriptstyle r}$\fi}
\def\vzo{\ifmmode v_\mathrm{\scriptscriptstyle 1, \scriptstyle z} \else $v_\mathrm{\scriptscriptstyle 1, \scriptstyle z}$\fi}
\def\vzc{\ifmmode v_\mathrm{\scriptscriptstyle 2, \scriptstyle z} \else $v_\mathrm{\scriptscriptstyle 2, \scriptstyle z}$\fi}

\newcommand{\Gone}{G$_{1}^{*}$}
\newcommand{\Gtwo}{G$_{2}^{*}$}
\newcommand{\Gthree}{G$_{3}^{*}$}
\newcommand{\Vout}{$V_\text{out}$}
\newcommand{\FWHMout}{FWHM$_\text{out}$}
\newcommand{\CF}{C$_\text{f}$}
\newcommand{\CFv}{C$_\text{f}(v)$}
\newcommand{\barCFv}{$\overline{C_\text{f}}$}
\newcommand{\Vcir}{$V_\text{cir}$}
\newcommand{\VMax}{$V_\text{max}$}
\newcommand{\Mstar}{M$_{\star}$}
\newcommand{\MdotSi}{$\dot{M}_\text{Si,out}$}
\newcommand{\MdotSiStar}{$\dot{M}_\text{Si,*}$}
\newcommand{\Mdot}{$\dot{M}_\text{out}$}
\newcommand{\Pdot}{$\dot{p}_\text{out}$}
\newcommand{\PdotStar}{$\dot{p}_{\star}$}
\newcommand{\PdotCrit}{$\dot{p}_\text{crit}$}
\newcommand{\Edot}{$\dot{E}_\text{out}$}
\newcommand{\EdotStar}{$\dot{E}_{\star}$}

\newcommand{\Nbar}{$\overbar{\text{N}}$}

\newcommand{\PI}{Paper~I}
\newcommand{\PII}{Paper~II}
\newcommand{\PIII}{Paper~III}
\newcommand{\PaperV}{Arellano-C\'ordova et al. submitted}
\defcitealias{Berg22}{\PI}
\defcitealias{James22}{\PII}
\defcitealias{Xu2022}{\PIII}

\begin{document}

\submitjournal{AASJournal ApJ}
\shortauthors{Xu et al.}
\shorttitle{CLASSY~III: Properties of Galactic Outflows}

\title{CLASSY III: \footnote{
Based on observations made with the NASA/ESA Hubble Space Telescope,
obtained from the Data Archive at the Space Telescope Science Institute, which
is operated by the Association of Universities for Research in Astronomy, Inc.,
under NASA contract NAS 5-26555.} The Properties of Starburst-Driven Warm Ionized Outflows}

\author[0000-0002-9217-7051]{Xinfeng Xu}
\affiliation{Center for Astrophysical Sciences, Department of Physics \& Astronomy, Johns Hopkins University, Baltimore, MD 21218, USA}

\author[0000-0003-1127-7497]{Timothy Heckman}
\affiliation{Center for Astrophysical Sciences, Department of Physics \& Astronomy, Johns Hopkins University, Baltimore, MD 21218, USA}

\author[0000-0002-6586-4446]{Alaina Henry}
\affiliation{Center for Astrophysical Sciences, Department of Physics \& Astronomy, Johns Hopkins University, Baltimore, MD 21218, USA}
\affiliation{Space Telescope Science Institute, 3700 San Martin Drive, Baltimore, MD 21218, USA}

\author[0000-0002-4153-053X]{Danielle A. Berg}
\affiliation{Department of Astronomy, The University of Texas at Austin, 2515 Speedway, Stop C1400, Austin, TX 78712, USA}

\author[0000-0002-0302-2577]{John Chisholm}
\affiliation{Department of Astronomy, The University of Texas at Austin, 2515 Speedway, Stop C1400, Austin, TX 78712, USA}

\author[0000-0003-4372-2006]{Bethan L. James}
\affiliation{AURA for ESA, Space Telescope Science Institute, 3700 San Martin Drive, Baltimore, MD 21218, USA}

\author[0000-0001-9189-7818]{Crystal L. Martin}
\affiliation{Department of Physics, University of California, Santa Barbara, Santa Barbara, CA 93106, USA}

\author[0000-0001-6106-5172]{Daniel P. Stark}
\affiliation{Steward Observatory, The University of Arizona, 933 N Cherry Ave, Tucson, AZ, 85721, USA}


\author[0000-0003-4137-882X]{Alessandra Aloisi}
\affiliation{Space Telescope Science Institute, 3700 San Martin Drive, Baltimore, MD 21218, USA}

\author[0000-0001-5758-1000]{Ricardo O. Amor\'{i}n}
\affiliation{Instituto de Investigaci\'{o}n Multidisciplinar en Ciencia y Tecnolog\'{i}a, Universidad de La Serena, Raul Bitr\'{a}n 1305, La Serena 2204000, Chile}
\affiliation{Departamento de Astronom\'{i}a, Universidad de La Serena, Av. Juan Cisternas 1200 Norte, La Serena 1720236, Chile}

\author[0000-0002-2644-3518]{Karla Z. Arellano-C\'{o}rdova}
\affiliation{Department of Astronomy, The University of Texas at Austin, 2515 Speedway, Stop C1400, Austin, TX 78712, USA}

\author[0000-0002-3120-7173]{Rongmon Bordoloi}
\affiliation{Department of Physics, North Carolina State University, 421 Riddick Hall, Raleigh, NC 27695-8202, USA}

\author[0000-0003-3458-2275]{St\'{e}phane Charlot}
\affiliation{Sorbonne Universit\'{e}, CNRS, UMR7095, Institut d'Astrophysique de Paris, F-75014, Paris, France}

\author[0000-0002-2178-5471]{Zuyi Chen}
\affiliation{Steward Observatory, The University of Arizona, 933 N Cherry Ave, Tucson, AZ, 85721, USA}

\author[0000-0001-8587-218X]{Matthew Hayes}
\affiliation{Stockholm University, Department of Astronomy and Oskar Klein Centre for Cosmoparticle Physics, AlbaNova University Centre, SE-10691, Stockholm, Sweden}

\author[0000-0003-2589-762X]{Matilde Mingozzi}
\affiliation{Space Telescope Science Institute, 3700 San Martin Drive, Baltimore, MD 21218, USA}

\author[0000-0001-6958-7856]{Yuma Sugahara}
\affiliation{Institute for Cosmic Ray Research, The University of Tokyo, Kashiwa-no-ha, Kashiwa 277-8582, Japan}
\affiliation{National Astronomical Observatory of Japan, 2-21-1 Osawa, Mitaka, Tokyo 181-8588, Japan}
\affiliation{Waseda Research Institute for Science and Engineering, Faculty of Science and Engineering, Waseda University, 3-4-1, Okubo, Shinjuku, Tokyo 169-8555, Japan}


\author[0000-0001-8152-3943]{Lisa J. Kewley}
\affiliation{Research School of Astronomy and Astrophysics, Australian National University, Cotter Road, Weston Creek, ACT 2611, Australia; ARC Centre of Excellence for All Sky Astrophysics in 3 Dimensions (ASTRO 3D), Canberra, ACT 2611, Australia}

\author[0000-0002-1049-6658]{Masami Ouchi}
\affiliation{National Astronomical Observatory of Japan, 2-21-1 Osawa, Mitaka, Tokyo 181-8588, Japan}
\affiliation{Institute for Cosmic Ray Research, The University of Tokyo, Kashiwa-no-ha, Kashiwa 277-8582, Japan}
\affiliation{Kavli Institute for the Physics and Mathematics of the Universe (WPI), University of Tokyo, Kashiwa, Chiba 277-8583, Japan}

\author[0000-0002-9136-8876]{Claudia Scarlata}
\affiliation{Minnesota Institute for Astrophysics, University of Minnesota, 116 Church Street SE, Minneapolis, MN 55455, USA}

\author[0000-0002-4834-7260]{Charles C. Steidel}
\affiliation{Cahill Center for Astronomy and Astrophysics, California Institute of Technology, MC249-17, Pasadena, CA 91125, USA}

\correspondingauthor{Xinfeng Xu} 
\email{xinfeng@jhu.edu}


\begin{abstract}
We report the results of analyses of galactic outflows in a sample of 45 low-redshift starburst galaxies in the COS Legacy Archive Spectroscopic SurveY (CLASSY), augmented by five additional similar starbursts with COS data. The outflows are traced by blueshifted absorption-lines of metals spanning a wide range of ionization potential. The high quality and broad spectral coverage of CLASSY data enable us to disentangle the absorption due to the static ISM from that due to outflows. We further use different line multiplets and doublets to determine the covering fraction, column density, and ionization state as a function of velocity for each outflow. We measure the outflow's mean velocity and velocity width, and find that both correlate in a highly significant way with the star-formation rate, galaxy mass, and circular velocity over ranges of four orders-of-magnitude for the first two properties. We also estimate outflow rates of metals, mass, momentum, and kinetic energy. We find that, at most, only about 20\% of silicon created and ejected by supernovae in the starburst is carried in the warm phase we observe. The outflows' mass-loading factor increases steeply and inversely with both circular and outflow velocity (log-log slope $\sim$ --1.6), and reaches $\sim 10$ for dwarf galaxies. We find that the outflows typically carry about 10 to 100\% of the momentum injected by massive stars and about 1 to 20 \% of the kinetic energy. We show that these results place interesting constraints on, and new insights into, models and simulations of galactic winds. 
\end{abstract} 



\section{Introduction} 
\label{sec:intro}

\color{black}

We live in a time of challenges and opportunities in the quest to understand the evolution of galaxies. We have a very successful theory for the development of the large-scale structure of the dark matter scaffolding in which the galaxies form and grow \citep[e.g.,][]{Wechsler18}. We also know the overall cosmic history of the rate of galaxy build-up through measurements of the star-formation rate (SFR) per unit co-moving volume element \citep[e.g.,][]{Madau14} and through measurements of the evolution of the cosmic inventory of baryons \citep[e.g.,][]{Peroux20}.

In the current paradigm of galaxy evolution \citep[e.g.,][]{Somerville15,Naab17}, baryons flow with the dark matter on large scales, and some are incorporated into the halos. Unlike the dark matter, the baryons can lose energy by radiation, and sink deeper into the potential well defined by the dark matter. In the simplest picture, this inflow is halted by centrifugal forces reflecting conservation of angular momentum. Stars form within the central regions of these disks.  Galaxies continue to grow over billions of years, primarily through continuing accretion of gas from the cosmic web, and secondarily through mergers with other dark matter halos and their baryonic contents. 

While this picture is simple and compelling, it does not account quantitatively for even the most basic properties of the baryonic content of galaxies. Some of the key unsolved problems are \citep[e.g.,][]{Somerville15,Behroozi19,Maiolino19,McGaugh00}:

\begin{enumerate}
\item
Why are only about 10\% of the baryons accreted (as gas) into the dark matter halos incorporated into stars, and why is this efficiency largely independent of redshift ($z$)?
\item Why does this efficiency reach its peak value over a relatively narrow range in dark matter halo masses of $\sim 10^{12}$ M$_{\odot}$, and why is this value roughly constant with $z$?
\item Why is there such a tight correlation between a galaxy’s stellar mass (\Mstar) and its SFR at a given epoch (the “star-forming main sequence”)? 
\item
Why is there such small scatter in the correlation between \Mstar\ and the galaxy’s chemical composition (metallicity) at a given epoch?
\item
Why is the scatter so small in the relationships between \Mstar, internal velocity dispersion and/or rotation speed, and radius in galaxies at a given z, and why do they evolve with z?
\item
How does the intergalactic medium (IGM) get enriched with metals?
\end{enumerate}

In all current theoretical models and numerical simulations of galaxies, these questions are dealt with through the rubric of ``feedback'': the effects of the return of energy, momentum, and heavy elements from massive stars and black holes on the surrounding gas \citep[e.g.,][]{Somerville15,Naab17}. The tightness of the scaling relations for galaxies (questions 3, 4, and 5 above), requires true two-way feedback that leads to self-regulating processes.

As noted above, feedback can be provided by either stars or supermassive black holes. In this paper, we focus on the former. Stellar feedback is dominated by massive stars, and is supplied in the form of radiation and stellar ejecta. For a young stellar population, the kinetic energy and momentum are primarily supplied by a combination of stellar winds from hot, massive stars and through core-collapse supernovae \citep[e.g.,][]{Leitherer99}. In order to affect the structure and content of a galaxy, either the momentum or kinetic energy provided by these stars must couple to the gas supply of the galaxy.

Undoubtedly, the most spectacular manifestations of feedback from populations of massive stars are global-scale galactic winds \citep[e.g.,][]{Heckman17a,Veilleux20}. These winds play a crucial role in the evolution of galaxies and the IGM \citep[e.g.,][]{Somerville15,Naab17}. In particular, the selective loss of gas and metals from the shallower potential wells of low-mass dark matter halos is believed to be responsible for both shaping the low-mass end of the galaxy stellar mass function and for establishing the mass-metallicity relation (questions 2, and 4 above). By carrying away low-angular-momentum gas, they also shaped the mass-radius relation (question 5). These outflows heated and polluted the circum-galactic medium (CGM) and IGM with metals and may have suppressed the accretion of gas passing from the CGM into the star-forming disk (questions 1 and 6). 

Simply put, the evolution of galaxies cannot be understood without first understanding galactic winds. Given their importance, and given both the large amounts of data that have been collected and the increasing quality of numerical simulations, it is perhaps surprising that we still have a very incomplete understanding of the processes that create outflowing gas, and of the impact the outflow has on the galaxy that launches it.  

It is crucial to emphasize that feedback processes associated with galactic winds cannot be spatially-resolved in numerical cosmological simulations. This problem has been long-recognized \citep[e.g.,][]{White91,katz96,Springel03,Hopkins14}. Instead the feedback processes are implemented numerically using ``sub-grid physics'' \citep[recipes, see e.g.,][]{Somerville15,Naab17}. These recipes often depend on things like the SFR and mass of the galaxy. It is clear that observations of feedback in-action are essential in guiding these choices and revealing the actual dependences on galaxy properties. Such data are also required to test models which attempt to simulate galactic winds with high enough spatial resolution to allow more {\it ab initio} calculation of the relevant physics \citep[e.g.,][]{Schneider18,Schneider20}. 


To date, the bulk of the data on winds across cosmic time have come from analysis of interstellar absorption-lines that trace outflowing cool or warm gas through the blue-shifted absorption-lines it produces \citep{Heckman00,Shapley03,Rupke05,Martin05,Grimes09,Sato09,Weiner09,Rubin10,Steidel10,Chen10,Erb12,Kornei12,Martin12,Bordoloi14,Rubin14,Heckman15,Zhu15,Chisholm15,Heckman16,Chisholm16a,Chisholm17,Sugahara17,Steidel18,Chisholm18,Sugahara19}. 

Clearly, there have been many prior investigations. In this paper, we seek to significantly improve the usefulness of such data in two respects. First, we use the COS Legacy Archive Spectroscopy SurveY (CLASSY) atlas \citep[][hereafter, \protect\citetalias{Berg22}]{Berg22}, which is a data set specifically designed to span a vast range in the most fundamental galaxy properties: stellar mass (and hence galaxy circular velocity), star-formation rate, and metallicity (see details in Section \ref{sec:obs}). Reaching low mass is especially important since feedback effects should be stronger in these shallow potential wells. This makes CLASSY ideal for testing how the fundamental outflow properties (outflow velocities, column densities, ionization state, and metal, mass, momentum, and kinetic energy outflow rates) depend on the galaxy properties, thereby providing a crucial test of the theoretical models and simulations. Second, the CLASSY data, by design, are high signal-to-noise ratio (SNR) spectra and cover a wide wavelength range in the UV that encompasses many interstellar lines that span wide ranges in ionization state and optical depth. Both points enable us to analyze the data more rigorously than before, leading to more robust measurements of outflow properties.

The structure of the paper is as follows. In Section \ref{sec:obs}, we introduce the CLASSY project and briefly describe the observations. In Section \ref{sec:reduction}, we go through various data reduction processes. In Section \ref{sec:analyses}, we present analyses to isolate the blue-shifted absorption lines for galactic outflows, and we also discuss the ancillary parameters. In Section \ref{sec:results}, we present the major results for the observed outflows, including covering fraction, column density, and ionization state as a function of velocity, and we also derive outflow rates of metals, mass, momentum, and kinetic energy. Furthermore, we compare the derived outflow properties to various host galaxy characteristics. Finally, in Section \ref{sec:discussion}, we discuss and compare our results to common models for galactic winds, as well as semi-analytic models and numerical simulations. We summarize the paper in Section \ref{sec:conclusion}.

We adopt a cosmology with H$_{0}$ = 69.6 km s$^{-1}$ Mpc$^{-1}$, $\Omega_m$ = 0.286, and $\Omega_{\Lambda}$ = 0.714, and we use Ned Wright's Javascript Cosmology Calculator website \citep{Wright06}. 

\section{Observations}
\label{sec:obs}
CLASSY is a Hubble Space Telescope (HST) treasury program (GO: 15840, PI: Berg), which provides the first high-resolution high-SNR restframe far-ultraviolet (FUV) spectral catalog of 45 local star-forming galaxies (0.002 $<$ z $<$ 0.182). These galaxies were selected to span a wide range of important galaxy properties, including  stellar mass (log \Mstar\ $\sim6-10\ M_\odot$), SFR ($\sim 0.01 - 100$ M$_\odot$ yr$^{−1}$), metallicity (12+log(O/H)$\sim7-9$), and electron density ($n_e\sim10^1-10^3$ cm$^{−3}$). For each galaxy, CLASSY completes its FUV wavelength coverage (1200 -- 2000\angstrom\ observed frame) by utilizing the G130M+G160M+G185M/G225M gratings of HST/COS. Overall, CLASSY combines 135 orbits of new HST data with 177 orbits of archival HST data to complete the first atlas of high-quality restframe FUV spectra of the proposed 45 galaxies. We define the CLASSY data hereafter as this combined dataset. We refer readers to \citetalias{Berg22} for the detailed sample selection, observations, and basic properties of these galaxies.

\section{Data Reduction}
\label{sec:reduction}

After the observations, all data were reduced locally using the COS data-reduction package CalCOS v.3.3.10\footnote{\url{https://github.com/spacetelescope/calcos/releases}}. These include both new data from CLASSY itself (GO: 15840) and all archival data that were included in CLASSY. Therefore, the whole CLASSY dataset was reduced and processed in a self-consistent way. The details of the data reduction have been presented in \citetalias{Berg22} and \citetalias{James22}, including spectra extraction, wavelength calibration, and vignetting.

Given the final reduced and co-added spectra for each galaxy \citepalias{Berg22}, we analyze the galactic outflow properties from various absorption and associated emission lines in this paper. Several additional steps in the reduction of the data are necessary, and are discussed in this section. In Section \ref{sec:stellarContinuum}, we discuss the fits of the stellar continuum for each galaxy, which are used to remove the starlight contamination of the spectra. In Section \ref{sec:system}, we discuss other systematic effects in the analyses of outflows.


To enlarge the sample size at the highest star-formation rates, we have added five Lyman Break Analog (LBA) galaxies from \cite{Heckman15} that were {\it not} already in the CLASSY sample but also had HST/COS observations. We have checked that these LBAs satisfy all selection criteria of the CLASSY sample. We processed and analyzed these data in exactly the same way as the CLASSY data. This gives us a total sample size of 50 galaxies. HST/COS G130M and G160M gratings have the original spectral resolutions $R$ $\sim$ 20,000, which is more than necessary for our outflow analyses. Therefore, for all spectra, we re-sample them into bins of 0.18\angstrom\ ($R$ $\sim$ 6000 -- 10000 from the blue to red end) to gain higher S/N.








\subsection{Starlight Normalization}
\label{sec:stellarContinuum}
By assuming that the observed spectra are combinations of multiple bursts of single-age, single-metallicity stellar populations, one can fit the stellar continuum of galaxies by linear combinations of stellar models, e.g., from Starburst99 \citep{Leitherer99}. We do so by following the same methodology laid out in \cite{Chisholm19}. Then, for each galaxy, we normalize the spectra by the best-fit stellar continuum.

\begin{figure}[h]
\center
	\includegraphics[angle=0,trim={0.5cm 1.4cm 0.0cm 4.5cm},clip=true,width=1\linewidth,keepaspectratio]{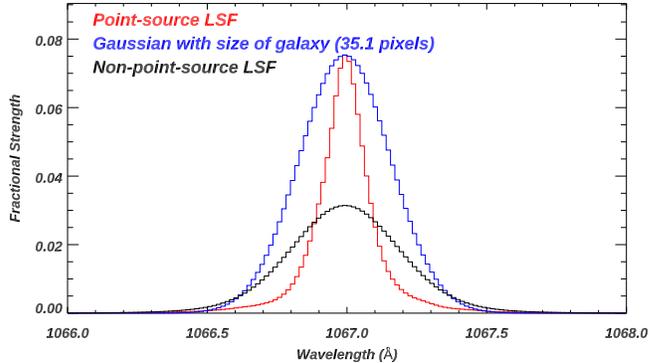}

\caption{\normalfont{Comparisons of the line spread functions (LSFs) for galaxy J1148+2546. The point-source LSF is shown in red, which is for the HST/COS G130M grating given the center wavelength of 1222\angstrom\ and at Lifetime Position 4 (LP4). The Gaussian representing the NUV light profile in the dispersion direction of this galaxy is shown in blue. The approximate LSF used in our fits is shown in black, which is the convolution between the blue and red curves. This galaxy has a relatively large NUV size in our sample, so the differences between the two LSFs are noticeable. See discussion in Section \ref{sec:2GFits}.} }
\label{fig:LSF}
\end{figure}

\begin{figure*}
\center
	\includegraphics[angle=0,trim={0.3cm 0.0cm 0.0cm 0.5cm},clip=true,width=0.45\linewidth,keepaspectratio]{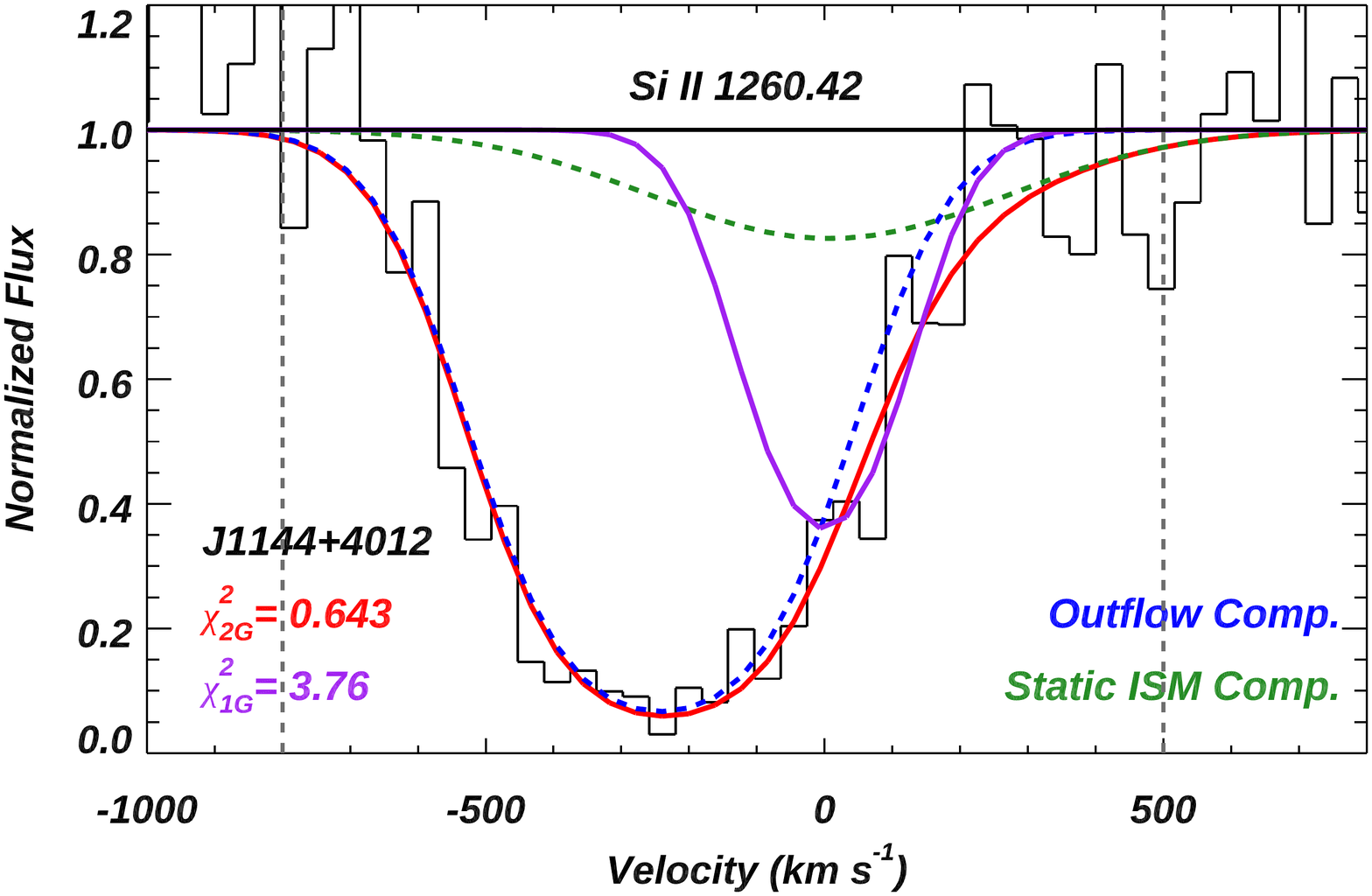}
	\includegraphics[angle=0,trim={0.3cm 0.0cm 0.0cm 0.5cm},clip=true,width=0.45\linewidth,keepaspectratio]{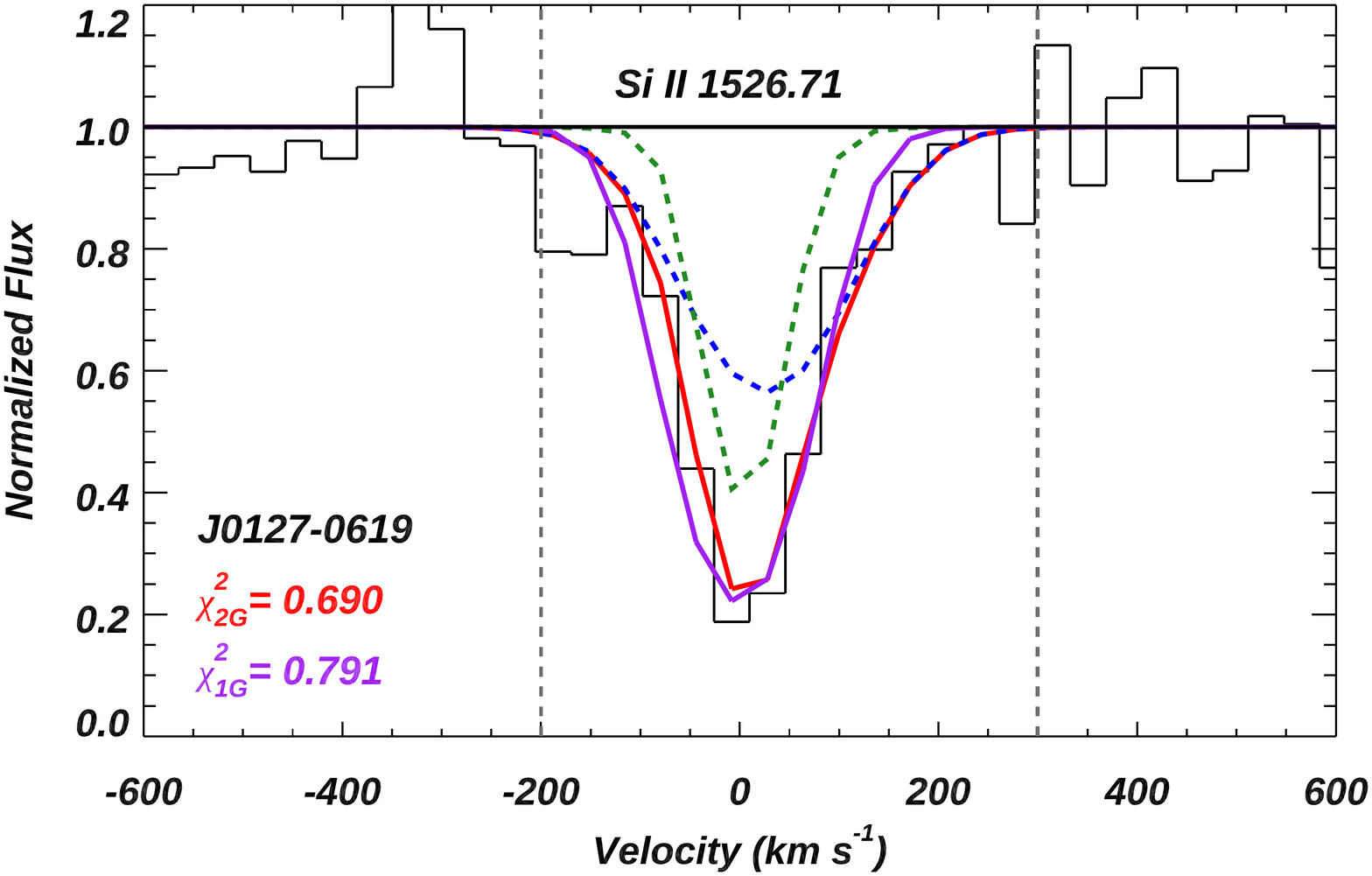}
	
\caption{\normalfont{Examples of double-Gaussian fits and F-tests to the absorption troughs. The normalized fluxes are shown in black histograms. For double-Gaussian fits, the profiles for the outflow and static ISM components are shown in blue and green dashed lines, respectively, and their summation is shown in red. For single-Gaussian fits, the models are shown in purple solid lines. All models have already considered the LSF effects discussed in Section \ref{sec:2GFits}. The $\chi^2$ values for single- and double-Gaussian fits are listed at the bottom left corner of each panel in purple and red, respectively. The fitting ranges are within the two vertical gray dashed lines. \textbf{Left:} A case when the absorption trough passed the F-test. This \Siii\ \ly 1260 trough has a significant blue shift. Thus, the second Gaussian (in blue) is necessary to fit the trough and represent the outflow. \textbf{Right:} A case when the trough failed the F-test. In this case, we don't gain significant improvement by introducing the second Gaussian, so we label this trough as ``no outflow". See details in Section \ref{sec:2GFits}. } }
\label{fig:2GFits}
\end{figure*}

\subsection{Other Systematic Effects}
\label{sec:system}
To get robust measurements of the outflow properties from the absorption lines, we need to take account of multiple systematic effects or contamination: 1) the static ISM component that is centered at $v$ = 0 km s$^{-1}$. This component represents the ISM gas that is not accelerated by the galactic winds, but it can blend with the lower velocity portions of the outflows; 2) the HST/COS line-spread-functions (LSFs), which describe the light distribution at the focal plane as a function of wavelength in response to the light source. This can slightly broaden and reshape the observed absorption troughs; and 3) the infilling effects from corresponding resonantly-scattered emission lines. We quantify each of these points in our analyses in Section \ref{sec:analyses} below.


\section{Basic Analyses}
\label{sec:analyses}
We begin with presenting a brief justification of the major methods adopted in Section \ref{sec:method}. Then in Section \ref{sec:2GFits}, we show the double-Gaussian fits to the observed outflow absorption troughs. In Section \ref{sec:infilling}, we discuss the infilling of absorption troughs by corresponding emission lines. Finally, in Section \ref{sec:anc}, we discuss various ancillary parameters that are adopted in the rest of the paper.

\subsection{Justifications of Methodology}
\label{sec:method}

For each galaxy in our sample, the final reduced, co-added, and starlight subtracted spectra cover $\sim$ 1200\angstrom\ -- 2000\angstrom\ in the observed frame. In this region, various lines from galactic outflows are detected as absorption troughs, including from low-ionization transitions, e.g., \oi\ \ly 1302, \cii\ \ly 1334, and \Siii\ multiplet (\ly1190, 1193, 1260, 1304, and 1526), and from higher-ionization transitions, e.g., \siiii\ \ly 1206, and \siiv\ \ly\ly 1393, 1402. We focus on these lines in our fitting method described in this section. In Table \ref{tab:atomic}, we list important atomic information for these lines.

To determine the basic properties of the outflows, we first fit the observed absorption troughs (Section \ref{sec:2GFits}). In the meantime, we need to take account of the spectral line-spread function (LSF) which is convolved with the intrinsic absorption-line profile to produce the observed profile. The LSF has contributions from both the COS optics and the spatial distribution of the UV continuum in the COS aperture. Our approach is to fit the observed profiles by using a simple analytic form for the intrinsic line profile, which has been convolved with our calculated LSF for each galaxy. We will describe this in more detail below. Here we note that we have taken a Gaussian to describe the intrinsic absorption-line profiles for both a component associated with the static ISM and one associated with the outflow. The choice of a Gaussian is motivated by several considerations. First, it is a simple analytic function, unlike the Voigt profile \citep{Draine11}. Second, it provides an excellent fit to the data (as we will show). 



\begin{table}
	\centering
	\caption{Atomic Data for Ions Measured in Outflows$^{(*)}$}
	\label{tab:atomic}
	\begin{tabular}{lcccc} 
		\hline
		\hline
		Ions & Vac. Wave. & f$_{lk}$ & A$_{kl}$   &E$_{l}$ -- E$_{k}$ \\
		 (1) & (2) & (3) & (4) & (5) \\
		\hline
		\hline
		\oi\  & 1302.17     & 5.20 $\times$ 10$^{-2}$& 3.41 $\times$ 10$^{8}$   & 0.0 - 9.52\\
		\cii\  & 1334.53     & 1.29 $\times$ 10$^{-1}$& 2.41 $\times$ 10$^{8}$   & 0.0 - 9.29\\	
		\Siii\  & 1190.42   & 2.77 $\times$ 10$^{-1}$& 6.53 $\times$ 10$^{8}$   & 0.0 - 10.41\\
		\Siii\  & 1193.29   & 5.75 $\times$ 10$^{-1}$& 2.69 $\times$ 10$^{9}$   & 0.0 - 10.39\\
		\Siii\  & 1260.42   & 1.22 $\times$ 10$^{-1}$& 2.57 $\times$ 10$^{9}$   & 0.0 - 9.84\\
		\Siii\  & 1304.37   & 9.28 $\times$ 10$^{-2}$& 3.64 $\times$ 10$^{8}$   & 0.0 - 9.50\\
		\Siii\  & 1526.71   & 1.33 $\times$ 10$^{-1}$& 3.81 $\times$ 10$^{8}$   & 0.0 - 8.12\\
		\siiii\  & 1206.51   & 1.67                 & 2.55 $\times$ 10$^{9}$   & 0.0 - 10.27\\
		\siiv\  & 1393.76   & 5.13$\times$ 10$^{-1}$  & 8.80 $\times$ 10$^{8}$   & 0.0 - 8.90\\
		\siiv\  & 1402.77   & 2.55$\times$ 10$^{-1}$   & 8.63 $\times$ 10$^{8}$   & 0.0 - 8.84\\

		\hline
	\multicolumn{5}{l}{%
  	\begin{minipage}{8cm}%
	Note. --\\
	    \textbf{(*).}\ \ Data obtained from National Institute of Standards and Technology (NIST) atomic database \citep{Kramida18}. \\
    	\textbf{(2).}\ \ Vacuum wavelengths in units of \angstrom.\\
        \textbf{(3).}\ \ Oscillator strengths.\\
        \textbf{(4).}\ \ Einstein A coefficients in units of s$^{-1}$.\\
        \textbf{(5).}\ \ Energies from lower to higher levels in units of eV.\\
  	\end{minipage}%
	}\\
	\end{tabular}
	\\ [0mm]
	
\end{table}

\subsection{Double-Gaussian Fits of the Absorption Troughs}
\label{sec:2GFits}

The theoretical LSF provided in HST/COS website\footnote{\url{https://www.stsci.edu/hst/instrumentation/cos/performance/spectral-resolution}} is good for a point source, but CLASSY galaxies are largely resolved. Therefore, we need to construct non-point source LSF for each galaxy, separately. The steps along with the double-Gaussian fitting process are as follows.
\begin{enumerate}
    \item First of all, we would like to generate a LSF for each galaxy based on the COS LSF for a point source (LSF$_{0}$) and the spatial distribution of the FUV continuum in the COS aperture in the dispersion direction. For the latter, we consider the galaxy size that is measured from HST/COS NUV acquisition images (FWHM$_{uv}$, see Table 2 in \citetalias{Berg22}). An example is shown in Figure \ref{fig:LSF}). We assume the galaxy has Gaussian profile with FWHM$_{uv}$ (G$_{uv}$, in blue), which is sufficient for our analyses. Then we convolve LSF$_{0}$ (in red) with G$_{uv}$, which results in an approximate non-point source LSF (LSF$_{uv}$, in black) for a given galaxy. This LSF$_{uv}$ is then used in the double-Gaussian fittings below to properly account for the LSF of each galaxy.
    
    
    \item We then use a double-Gaussians model to fit the observed absorption troughs adopting the fitting routine \textit{mpfit} \citep{Markward09}. Two examples are shown in Figure \ref{fig:2GFits}. Instead of using the standard Gaussian profile, we convolve it with LSF$_{uv}$ measured in step 1 to take into account the effects of LSFs (hereafter, we use \Gone\ and \Gtwo\ for the two convolved Gaussian profiles). \Gone\ has a fixed velocity center at $v$ = 0 km s$^{-1}$, which accounts for the static ISM component (green lines in Figure \ref{fig:2GFits}). \Gtwo\ has a velocity center $<$ 0 km s$^{-1}$ that represents the outflow component (blue lines in Figure \ref{fig:2GFits}). 
 
    \item For each fitted absorption trough, to check if \Gtwo\ is necessary (i.e., if there exists an outflow component), we conduct an F-test:
    \begin{equation}\label{eq:Ftest}
        F = \frac{(\chi^{2}_{1} - \chi^{2}_{2})/(p_{2} - p_{1})}{ \chi^{2}_{2}/(n - p_{2})}
    \end{equation}
    where $\chi^{2}_{1}$ and $\chi^{2}_{2}$ are the chi-squares from a single-Gaussian model (fits of the trough by only \Gone) and a double-Gaussian model (fits of the trough by \Gone\ and \Gtwo), respectively. $p_{1}$ and $p_{2}$ are the number of free parameters in single- and double-Gaussian models, respectively, and $n$ is the total number of bins of the fitted trough. We then compare this calculated $F$ value with the theoretical one from F-distribution table given significance level $\alpha$ = 0.05 and degree of freedom of ($p_{2}$ -- $p_{1}$, $n$ -- $p_{2}$). If the fitted $F$ value is greater than the theoretical one, we reject the null hypothesis (i.e., model 2 does not provide a significant better fit than model 1). This indicates that the inclusion of \Gtwo\ is necessary to fit the observed trough. Therefore, we treat this trough hosting blueshifted outflow(s). In Figure \ref{fig:2GFits}, we show examples of passing/failing the F-test in the left/right panels, respectively.
    

\end{enumerate}

\begin{figure*}
\center
	\includegraphics[angle=0,trim={2.8cm 0.8cm 0.5cm 0.8cm},clip=true,width=0.45\linewidth,keepaspectratio]{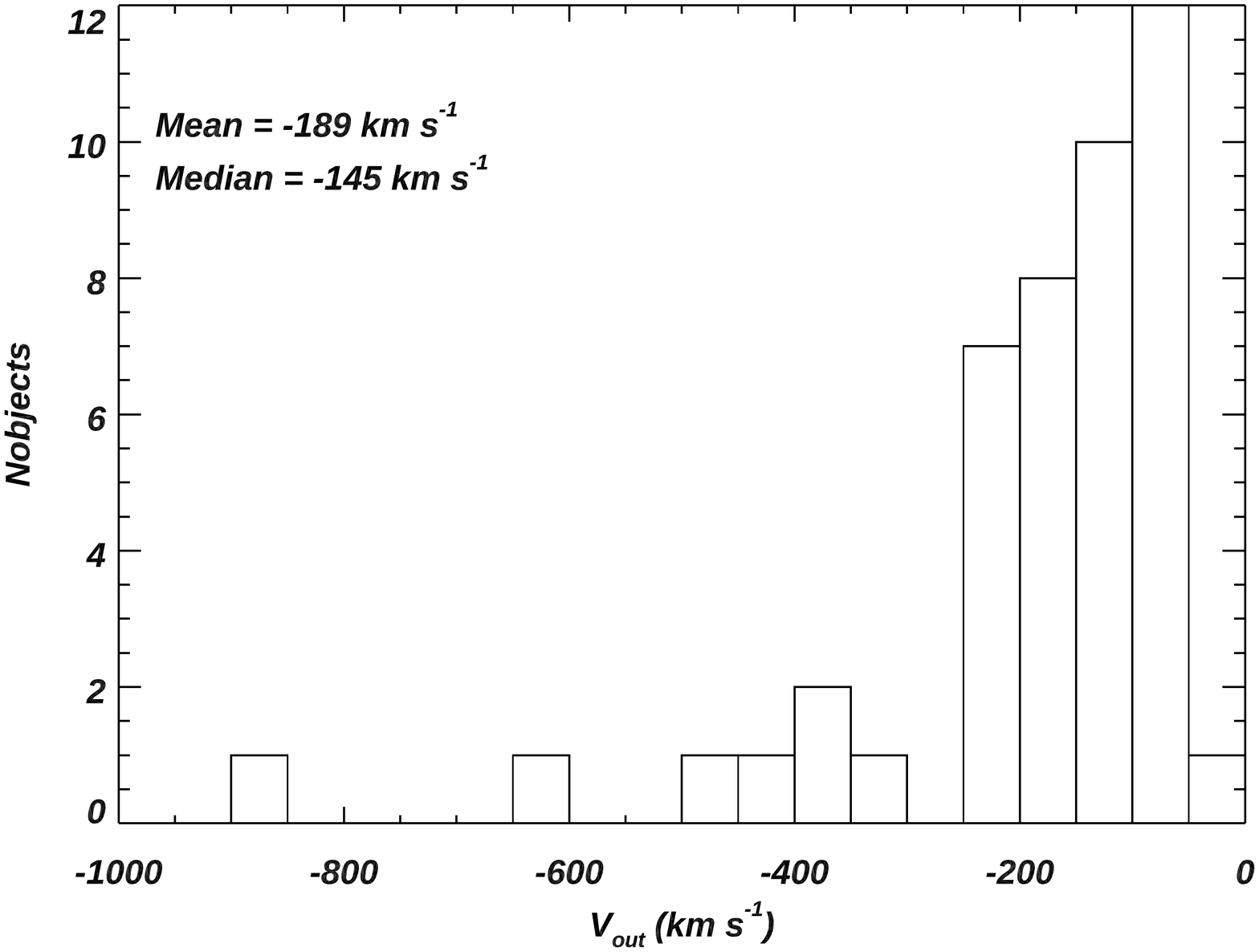}
	\includegraphics[angle=0,trim={2.8cm 0.8cm 0.5cm 0.8cm},clip=true,width=0.45\linewidth,keepaspectratio]{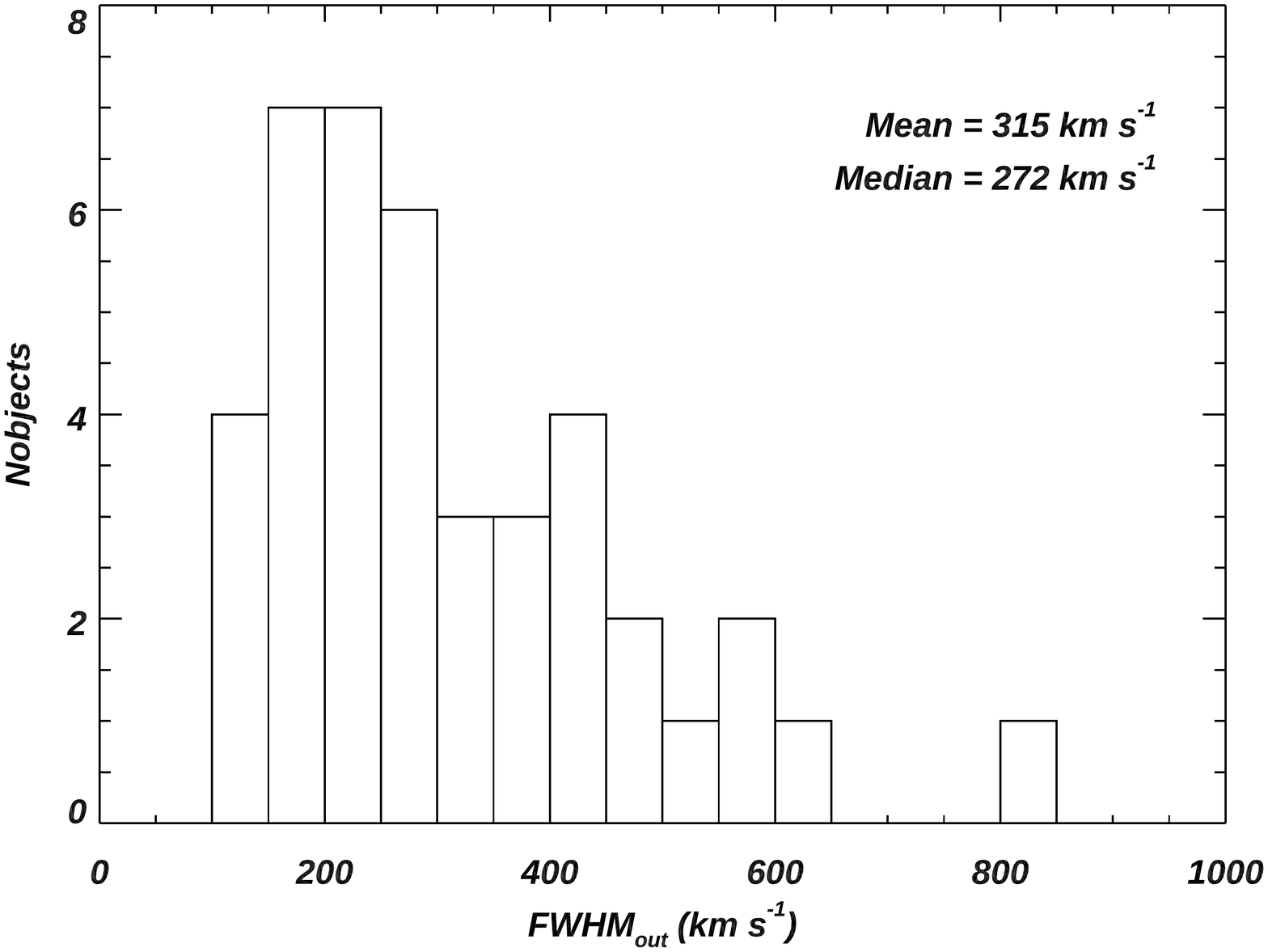}
	
\caption{\normalfont{Distributions for outflow velocity (\textbf{Left}) and the FWHM of the outflow component (\textbf{Right}) for galaxies in our combined sample [CLASSY + \cite{Heckman15}, see Section \ref{sec:reduction}]. The method for estimating these two values are discussed in Section \ref{sec:2GFits}. For \FWHMout, we have already subtracted the contributions from the HST/COS line-spread-functions (LSF).} }
\label{fig:VoutDist}
\end{figure*}

For most galaxies, we find that the effects of the LSFs on the outflow components are relatively small. This is because 1) the LSFs do not alter the measured velocity centers of the fitted outflow components, and 2) the FWHMs of LSF$_{uv}$ are usually $\ll$ 100 km s$^{-1}$, but the FWHMs of the outflow components are usually $>$ 250 km s$^{-1}$. On the contrary, for the narrow static ISM component and galaxies with narrow outflow troughs ($<$ 100 km s$^{-1}$), the HST/COS LSFs have significantly broadened their FWHMs. Therefore, in these cases, our fitting method discussed above is necessary for quantifying the outflow's FWHM (\FWHMout): 
\begin{equation}
   \text{FWHM}_\text{out} = \sqrt{\text{FWHM}_\text{all}^2 - \text{FWHM}_\text{LSF}^2} 
\end{equation}
where $\text{FWHM}_\text{all}$ is the fitted FWHM for a certain trough and $\text{FWHM}_\text{LSF}$ is the FWHM of LSF$_{uv}$ for that galaxy.

For each galaxy, we adopt this method to fit the absorption lines from \oi\ \ly 1302, \cii\ \ly 1334, \Siii\ multiplet (\ly1190, 1193, 1260, 1304, and 1526), \siiii\ \ly 1206, and \siiv\ \ly\ly 1393, 1402, separately. If one individual line falls into a chip gap or is contaminated by Galactic lines (e.g., \siiii\ can be affected by Galactic \lya), we exclude them from the fitting.
For each galaxy, if more than half of the fitted troughs for the different absorption-lines pass the F-test, we label this galaxy as ``hosting outflows". Otherwise, if one galaxy has less than half of its troughs that passes the F-test, ``no outflow" is labelled.

One galaxy's outflow velocity is then calculated from the median value of central velocities (of fitted \Gtwo) from all troughs that have passed the F-test. Similarly, the galaxy's outflow full-width-half-maximum (\FWHMout) is derived from the median value of FWHM from all troughs that pass the F-test. The corresponding errors of \Vout\ (or \FWHMout) are estimated from the standard deviations of \Vout\ (or \FWHMout) from all passed lines.  Note that for these ``no outflow" galaxies, they could still host very low velocity outflows (\Vout $\ll$ FWHM$_\text{ISM}$). However, since we cannot disentangle them from the static ISM component, their \Vout\ and \FWHMout\ are not measurable.

\begin{figure*}
\center
	\includegraphics[angle=0,trim={0.3cm 0.1cm 0.1cm 1.0cm},clip=true,width=0.5\linewidth,keepaspectratio]{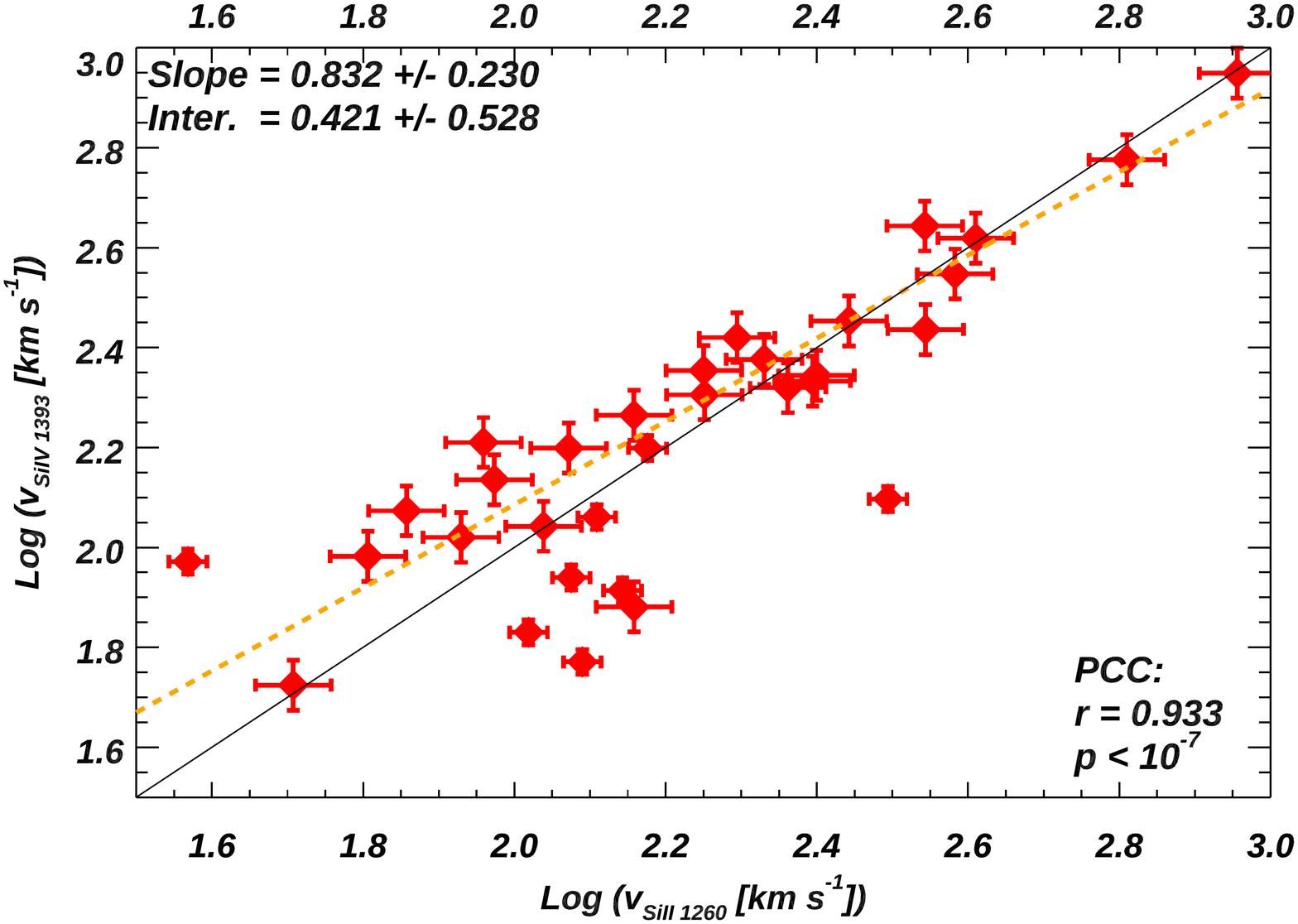}
	\includegraphics[angle=0,trim={0.3cm 0.0cm 0.1cm 1.0cm},clip=true,width=0.5\linewidth,keepaspectratio]{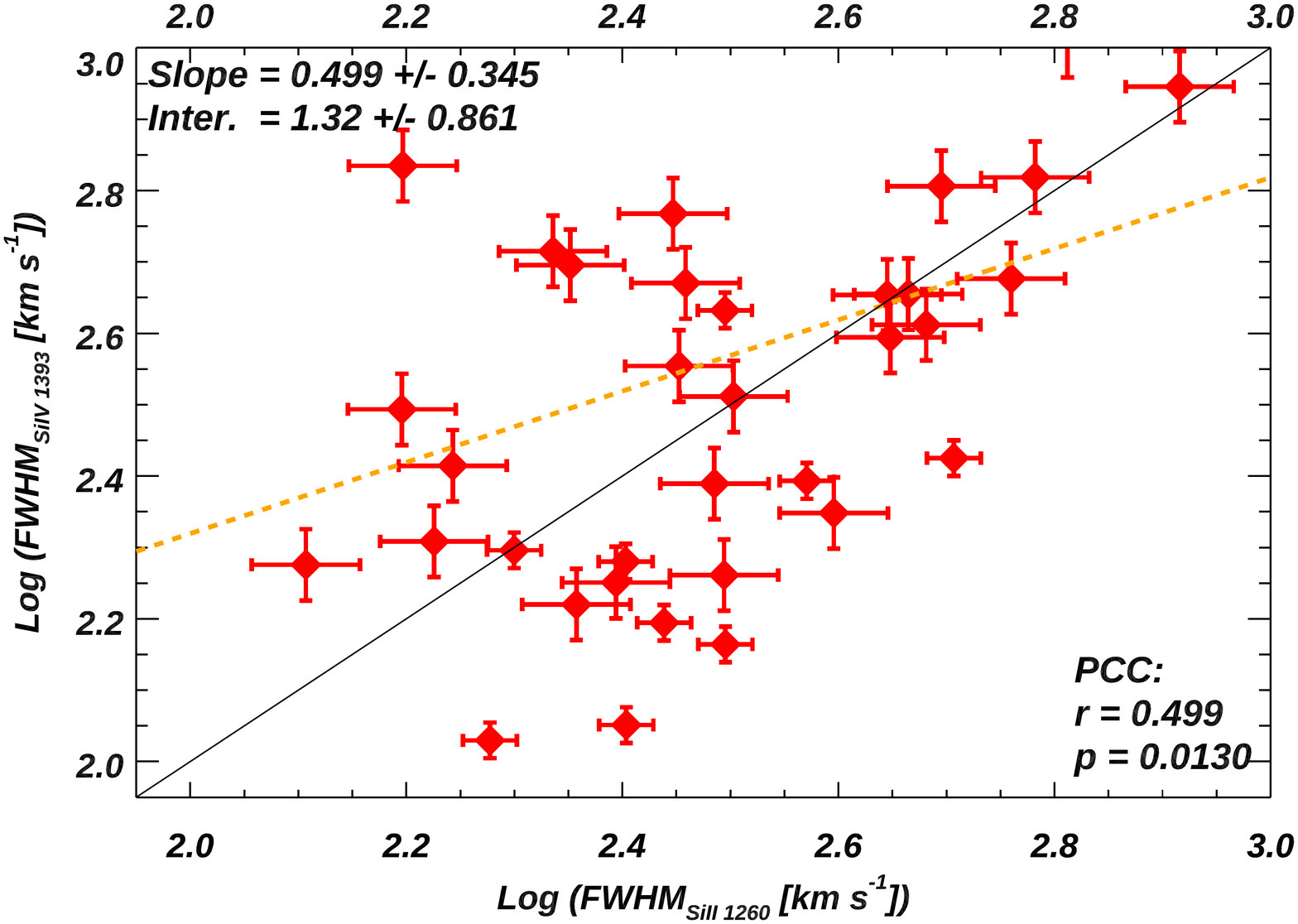}

	\includegraphics[angle=0,trim={0.3cm 0.1cm 0.1cm 1.0cm},clip=true,width=0.5\linewidth,keepaspectratio]{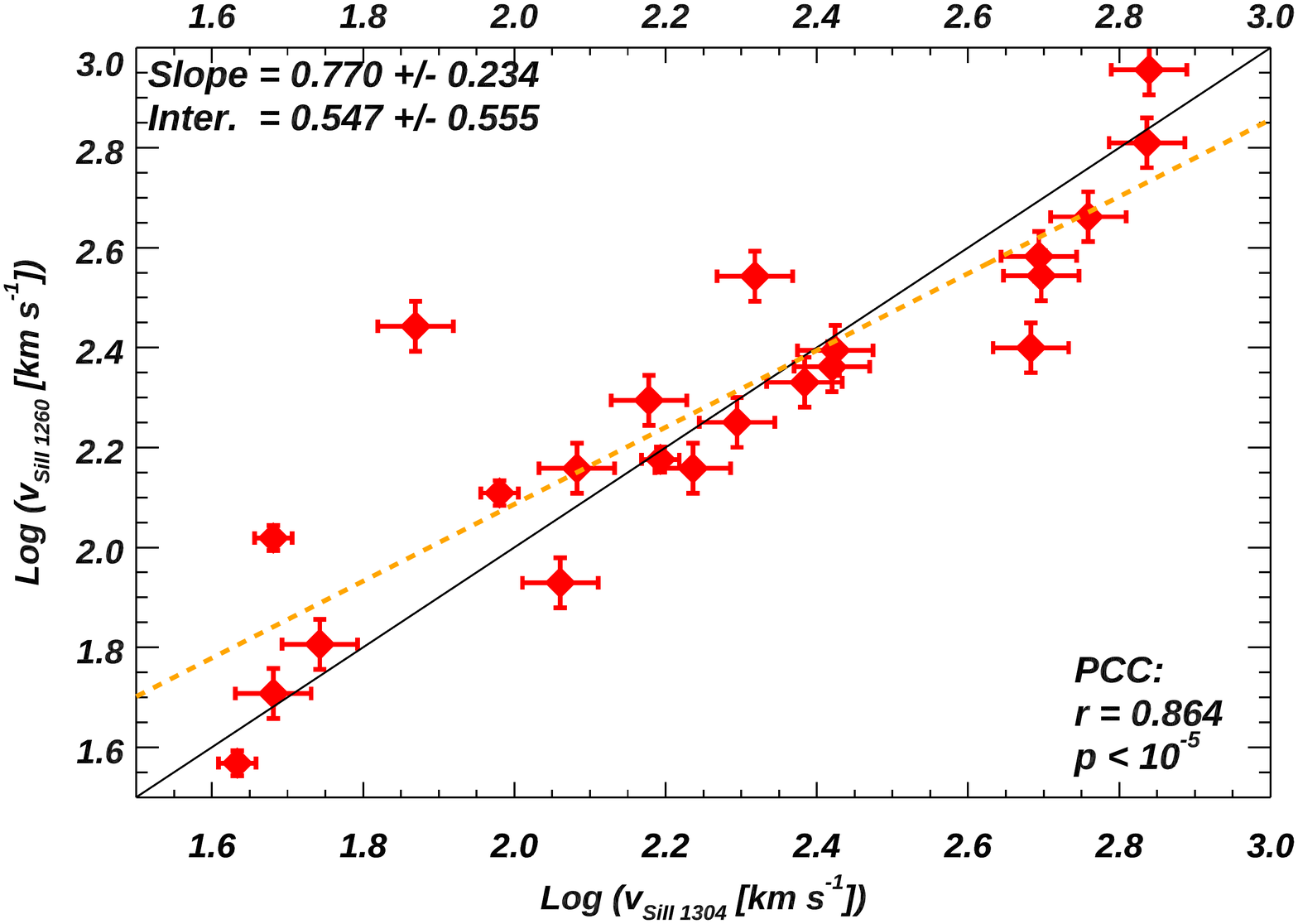}
	\includegraphics[angle=0,trim={0.3cm 0.0cm 0.1cm 1.0cm},clip=true,width=0.5\linewidth,keepaspectratio]{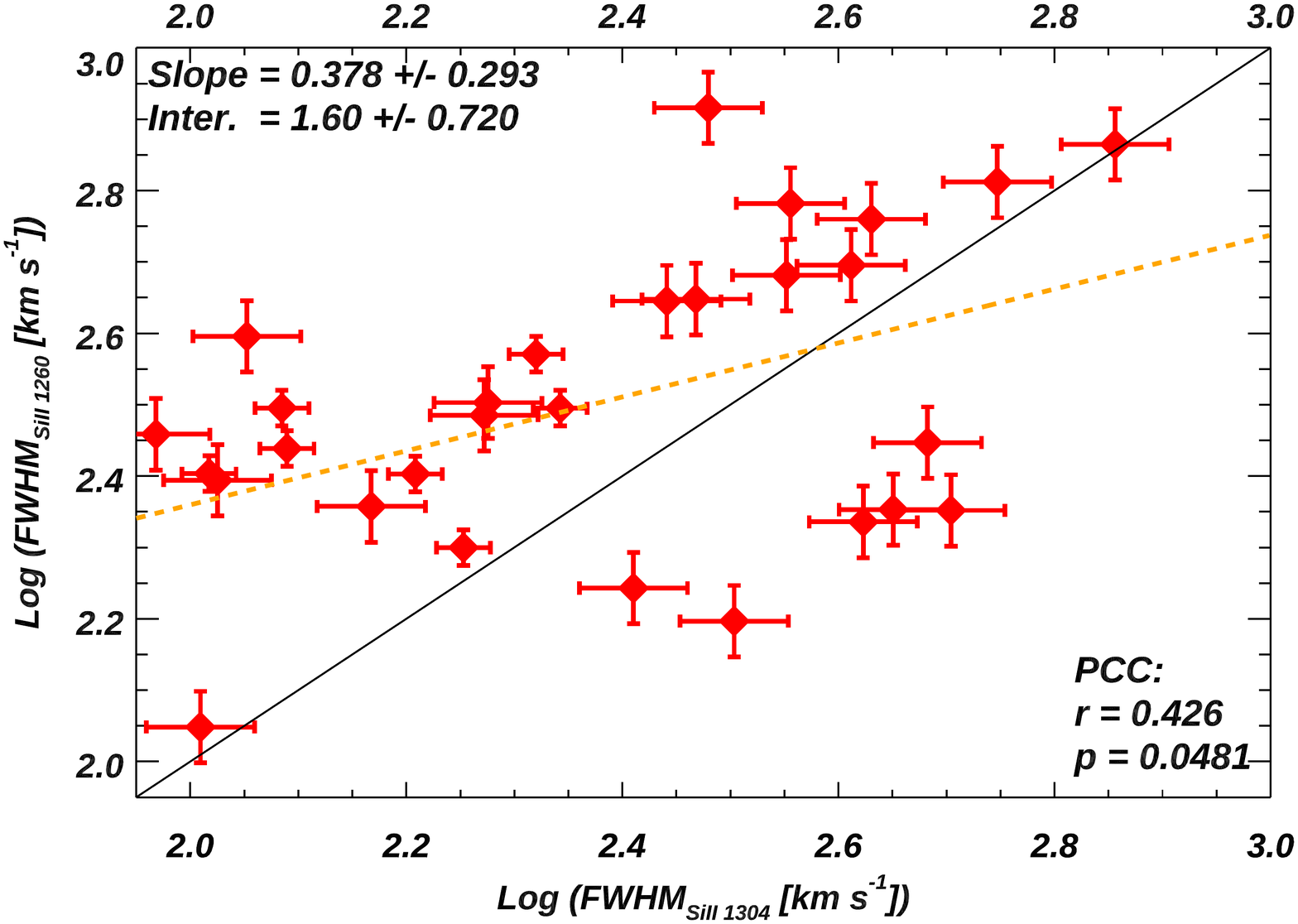}
	
\caption{\normalfont{Comparisons of outflow velocity centers (\Vout) and widths (\FWHMout) from \Siii\ \ly 1260 and \siiv\ \ly 1393 troughs \textbf{(top)} and for \Siii\ \ly1260 and \Siii\ \ly 1304 \textbf{(bottom)} in our combined sample. While all three lines show consistent \Vout, their \FWHMout\ values have substantial scatter. See Section \ref{sec:2GFits} for details. The black lines show the 1:1 relationship, and the orange lines show the best linear fits while the fitted parameters are shown in the left-top corners. The results of Pearson correlation coefficients (PCC) are shown at the bottom-right corners.} }
\label{fig:SiIIvsSiIV}
\end{figure*}

Besides the median values, we also have examined the consistency of the derived outflow velocity and FWHM among the different transitions. Two examples are shown in Figure \ref{fig:SiIIvsSiIV}. We find that the values of outflow velocities are quite consistent, but there is significant scatter in the values of FWHM. The panel showing FWHM for the two \Siii\ transitions is particularly instructive. It shows a systematic trend for the FWHM from \Siii\ \ly 1304 (which is the most optically thin \Siii\ line we measure) to be narrower than that of \Siii\ \ly 1260 (which is the most optically thick \Siii\ line that we measure). We will discuss the implications of this in Sections \ref{sec:ColumnDensity} below.

Overall, 43 out of 50 galaxies ($\sim$ 86\%) in our combined sample are labeled as ``hosting outflows". This indicates that galactic outflows are common in the low-redshift starburst galaxies in the CLASSY sample. The distribution of \Vout\ and \FWHMout\ are shown in Figure \ref{fig:VoutDist}, and the values are presented Table \ref{tab:mea}. 


 
\begin{figure*}
\center
	\includegraphics[angle=0,trim={0.3cm 0.1cm 0.2cm 4.0cm},clip=true,width=0.45\linewidth,keepaspectratio]{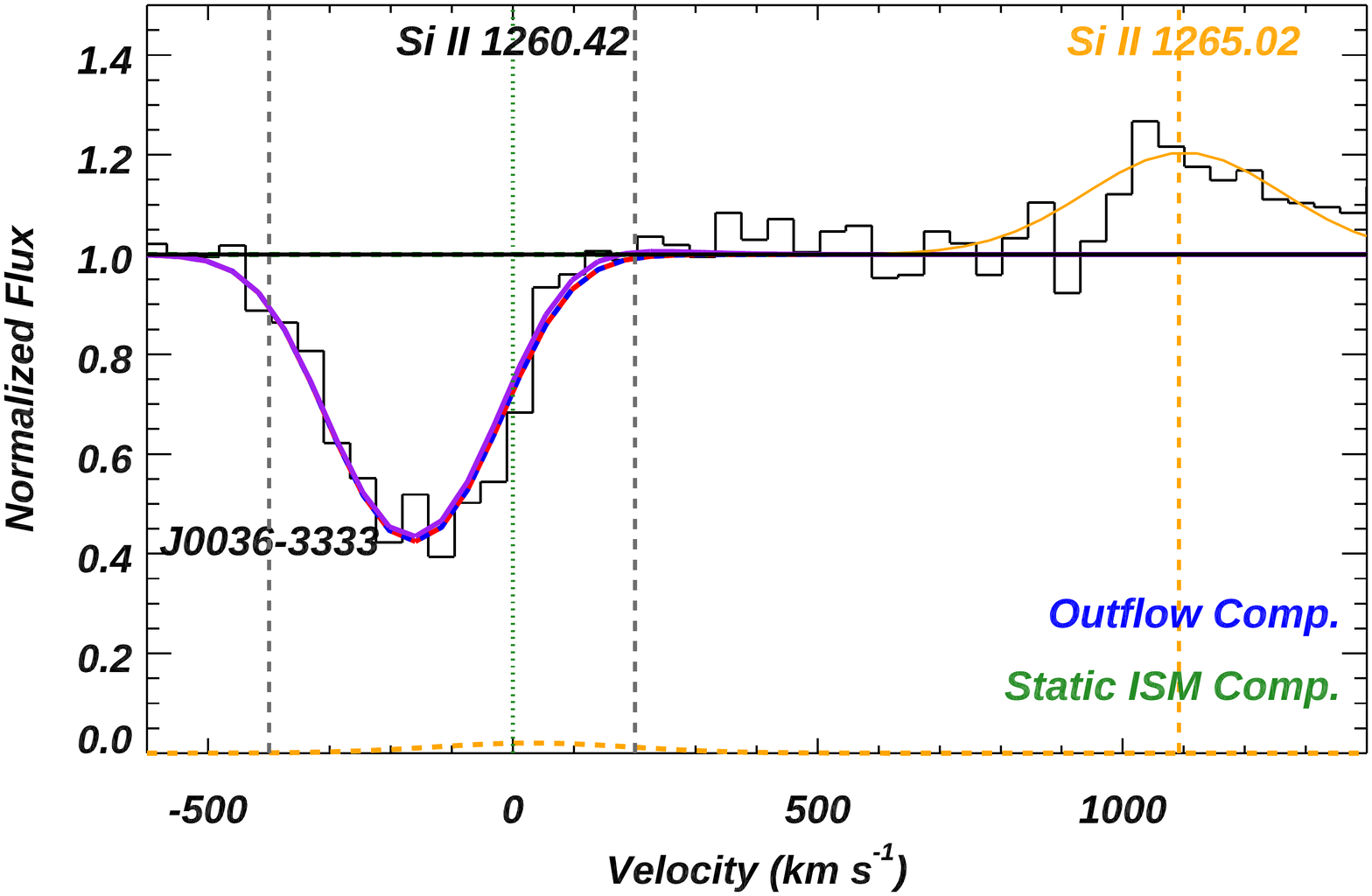}
	\includegraphics[angle=0,trim={0.3cm 0.0cm 0.0cm 4.0cm},clip=true,width=0.45\linewidth,keepaspectratio]{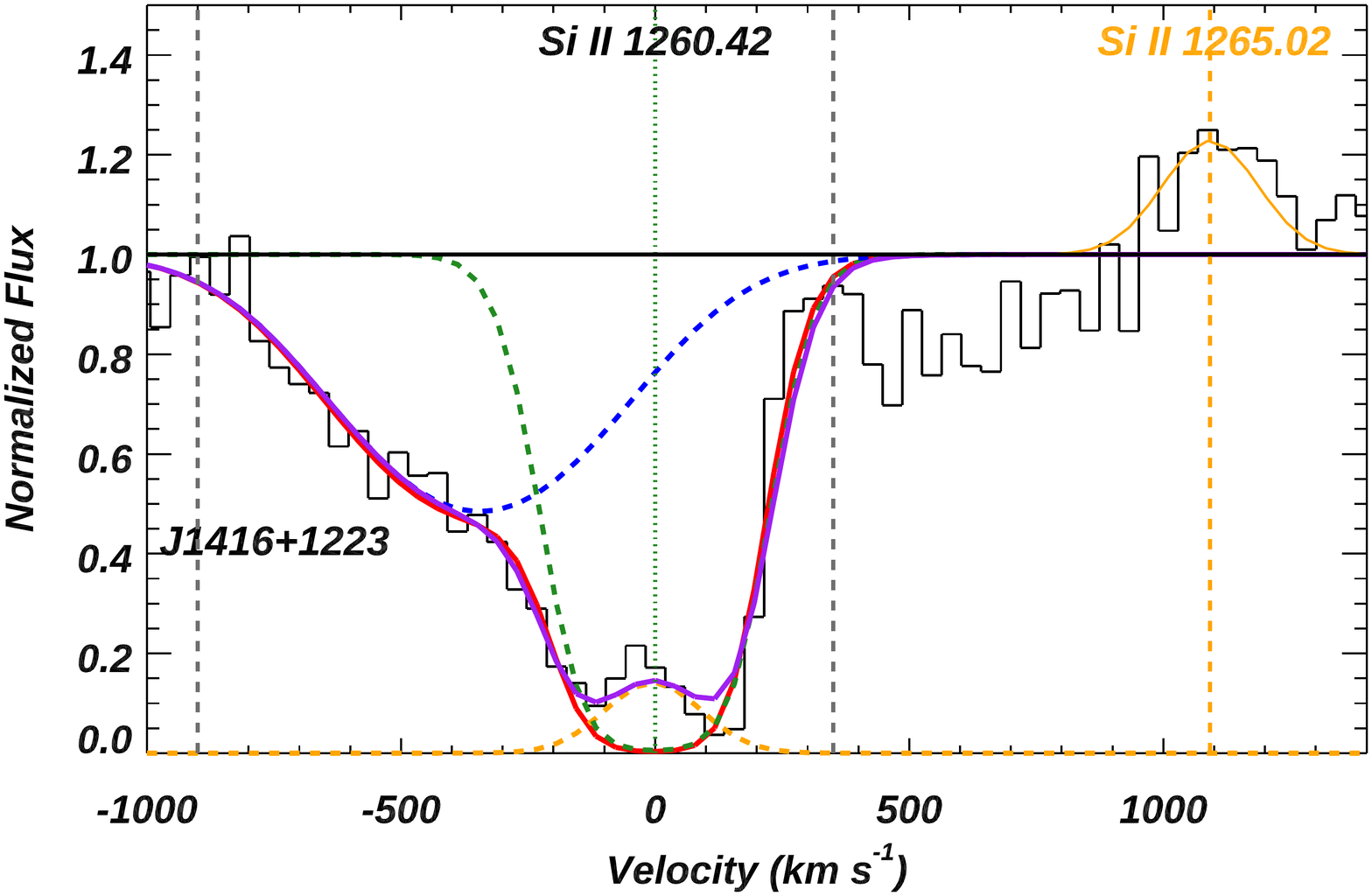}
	
\caption{\normalfont{Examples of checking the amount of infilling by scattered resonance emission based on the properties of the associated fluorescent emission-lines. For double-Gaussian fits, the blue, green, and red lines are for the outflow, static ISM, and summation of both components, respectively. The emission from fluorescent emission-lines are fitted by one Gaussian profile shown as the orange lines. For three-Gaussian fits, the orange dashed lines centered at $v$ = 0 km s$^{-1}$ represent the amount of infilling, which are scaled and shifted from the fluorescent lines. \textbf{Left:} A case when the infilling is negligible. \textbf{Right:} A case when the infilling is obvious at the bottom of static ISM component. This infilling will only affect the fit of the static ISM component, but not the outflow component. See details in Section \ref{sec:infilling}.} }
\label{fig:FEFits}
\end{figure*}

\subsection{Effects of Infilling}
\label{sec:infilling}
As discussed in previous publications \citep[e.g.,][]{Prochaska11, Scarlata15, Alex15, Carr18, Wang20, Carr21}, the absorption of a photon from the ground state may be followed by emission (resonance scattering). While only the gas directly along the line-of-sight to the UV continuum will produce absorption, the emission will come from all the gas within the size of COS aperture. This line emission will infill the absorption trough and could significantly affect our fitted models. Therefore, it is necessary to check the effects of infilling during our fitting processes.


The infilling is difficult to disentangle given only the absorption lines themselves, but fortunately, many of these lines have associated fluorescent emission lines. These are the fine structure transitions that share the same upper energy level as the associated resonance lines, e.g., \Siii*\ \ly 1265 for \Siii\ \ly 1260. Since the fluorescent lines are well displaced in wavelength from the resonance lines, their properties can be easily determined. Furthermore, the physical properties of the infilled emission line is related to that of fluorescent emission lines in various ways. For example, the infilling emission of the \Siii\ \ly 1260 would cover the same velocity range as \Siii*\ \ly 1265, and the line strength ratio of 1260/1265 depends on: 1) their Einstein-A coefficient ratios; and 2) how many times the photon has been scattered \citep[represented by the optical depth of the trough, see, e.g.,][]{Scarlata15}. We mainly test infilling effects on the \Siii\ multiplet since they are the major lines affected and commonly used in our analyses later (see Section \ref{sec:results}). The other major lines are \siiv\ \ly\ly 1393, 1402, but there are no associated fluorescent transitions for them. The steps are as follows:

\begin{enumerate}

    \item For each \Siii\ resonance line, we first fit the corresponding fluorescent emission line adopting one convolved Gaussian (G$^{*}$, see solid orange lines in Figure \ref{fig:FEFits}). 
  
    \item Then we choose the ten objects that have the largest |EW(\Siii*)/EW(\Siii)|, since we expect that these galaxies are affected the most by infilling. We conduct a triple-Gaussian fit to their \Siii\ absorption troughs. Besides the double Gaussians (in absorption) discussed in Section \ref{sec:2GFits}, we add a third Gaussian (\Gthree, in emission) to represent the infilling component. We fix the velocity center and width of \Gthree\ as the same as its associated fluorescent line (fitted in step 1) but allow the amplitude of \Gthree\ to vary between 0 and a constant. For \Siii\ \ly 1260, this constant equals $ A_{1260}/A_{1265} \times H_{1260} \simeq 0.85H_{1260}$  \citep[e.g.,][]{Scarlata15}, where $A$ is the Einstein $A$ coefficient and $H$ is the amplitude of the emission line.
    
    \item Finally, we compare the results from the double-Gaussian fits and triple-Gaussian fits by measuring three parameters, i.e., velocity center, FWHM, and minimum flux of their outflow trough (\Gtwo).

\end{enumerate}

Overall, for these ten galaxies, these measured parameters for \Gtwo\ only have minimal differences ($<$ 5\%) between the double-Gaussian and triple-Gaussian fits. Thus, we conclude that in our sample, the infilling is negligible for the outflow absorption troughs from \Siii\ multiplet \citep[e.g.,][]{Alex15}. Nonetheless, as shown in the right panel of Figure \ref{fig:FEFits}, infilling can more significantly affect the properties of the fit to the static ISM component (centered at v = 0 km s$^{-1}$), which would alter the derived covering fraction, column density, etc., for the ISM component. These conclusions are consistent with extensive studies presented in previous publications of metal absorption lines in galaxies \citep[e.g.,][]{Prochaska11, Martin12, Erb12, Alex15}.
    



\begin{figure*}
\center
	\includegraphics[page = 1, angle=0,trim={0.3cm 3.05cm 0.0cm 2.0cm},clip=true,width=0.5\linewidth,keepaspectratio]{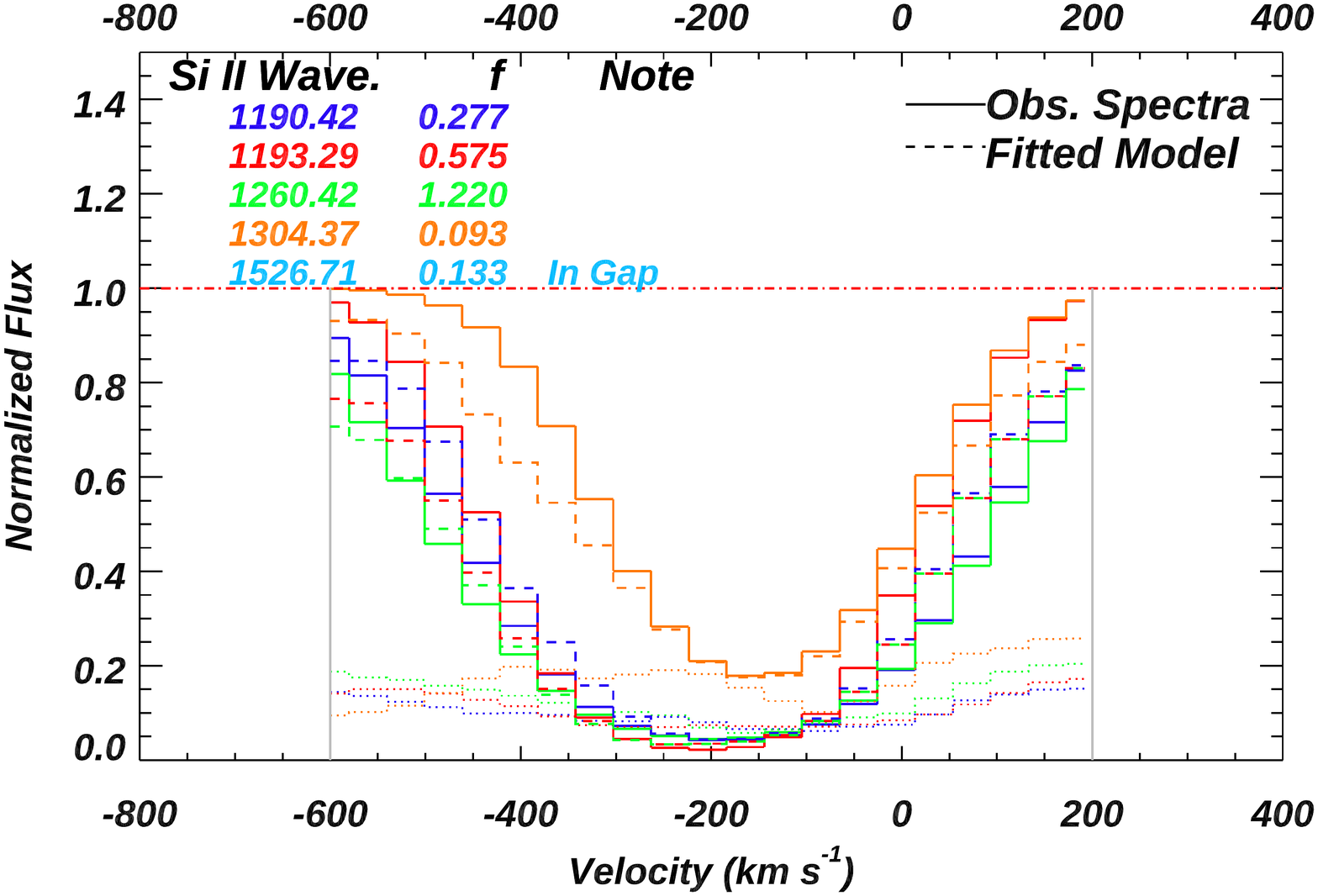}
	\includegraphics[page = 2, angle=0,trim={0.3cm 3.05cm 0.0cm 2.0cm},clip=true,width=0.5\linewidth,keepaspectratio]{./J0055-0021_velocity_N_SiII_FitModelOutflow.pdf}
	
	\includegraphics[page = 3, angle=0,trim={0.2cm 0.0cm 0.0cm 3.3cm},clip=true,width=0.5\linewidth,keepaspectratio]{./J0055-0021_velocity_N_SiII_FitModelOutflow.pdf}%
	\includegraphics[page = 4, angle=0,trim={0.0cm 0.0cm 0.0cm 3.3cm},clip=true,width=0.5\linewidth,keepaspectratio]{./J0055-0021_velocity_N_SiII_FitModelOutflow.pdf}

\caption{\normalfont{An example of the solutions of partial-covering equations and total hydrogen column densities (\Nh) for galaxy J0055-0021. \textbf{Top-Left:} The fitted Gaussian profiles for the four \Siii\ outflow troughs are shown as the solid lines (i.e., \Gtwo, Section \ref{sec:2GFits}). A total of four \Siii\ troughs are clean (shown as different colors), while J0055--0021's \Siii\ \ly 1526 trough is in a detector gap. Then we fit partial-covering models [see equation (\ref{eq:SiII})] to these Gaussian profiles and the results are shown in the dashed lines. The errors of the Gaussian profiles are shown as dotted lines at the bottom of the troughs, and the fitting range is within the two gray lines. Note that the \Siii\ \ly 1304 line is significantly narrower than the other three lines \textbf{Top-Right:} Best fitting covering fraction (\CF) for \Siii\ from solving Equation (\ref{eq:SiII}). \textbf{Bottom-Left:} Best fitting optical depth over velocity ($\tau$(v)) in logarithm for each \Siii\ lines. We only show \Siii\ \ly 1260's error bars to avoid crowding. \textbf{Bottom-Right:} The resulting average column density in the aperture per velocity for \Siii, which is proportional to \CF(v)$\times \tau$(v) [Equation (\ref{eq:SiII})]. In the latter 3 panels, the corresponding errors derived from \textit{mpfit} are shown as vertical lines. This shows that the decline in column density away from the line center is due to drops in both optical depth and covering fraction. See details of the fitting process in Section \ref{sec:ColumnDensity}. } }
\label{fig:SiIIFits}
\end{figure*}

\subsection{Ancillary Parameters}
\label{sec:anc}

Various galaxy properties are used in the rest of the paper. The measurements of them are discussed in Section 4 of \citetalias{Berg22} and listed in their Tables 5 and 6. We take these values and propagate their errors when they are used in our analyses. These include: 1) gas-phase metallicity (12+log(O/H)) and the errors, which are derived using the direct method; 2) \Mstar, SFR, and dust extinction (E(B-V)), and the corresponding errors, which are derived from the SED fitting to the GALEX+SDSS data; 3) redshift of galaxies, which are derived from fitting the optical emission lines (mainly from SDSS spectra, Mingozzi et al. in prep.).

%


%

We also calculate two other ancillary parameters needed for our analyses, including the galaxy circular velocity (\Vcir) and r$_{50}$. Following \citet{Heckman15}, we calculate \Vcir\ based on the galaxy stellar mass. This adopts the empirical calibration in \citet{Simons15} which was derived using spatially-resolved maps of the kinematics of the gas in low-redshift emission-line galaxies. Following \citet{Simons15}, we define \Vcir\ $= \sqrt{2}$ $(V_{rot}^2 + 2 \sigma^2)^{1/2}$, where $V_{rot}$ and $\sigma$ are the rotation speed and mean value of the line-of-sight velocity dispersion, respectively. Finally, we adopt the best-fit relation from \citet{Simons15} as log \Vcir\ = 0.29 \Mstar\ -- 0.78, with root-mean-square (RMS) residuals of 0.1 dex RMS in \Vcir.

For each galaxy, we measure the half-light radius (r$_{50}$) from its HST/COS NUV acquisition image. We accumulate the net photon counts within a certain radius from the galaxy center until it reaches half of the total source counts. However, some galaxies are more spatially extended than the unvignetted region of the acquisition images, and their cumulative counts vs. radius distributions are not flat at large radii. Therefore, for the scaling relations that we discuss in Section \ref{sec:scale}, we adopt the r$_{50}$ values from COS for galaxies with r$_{50}$ smaller than 0.4\arcsec. These galaxies are compact enough, so the vignetting effects are small. For other galaxies that have COS r$_{50}$~$>$~0.4\arcsec, we instead take their SDSS u-band sizes from the SDSS archive, which conducted exponential fits convolved with the measured PSF to the SDSS images.

For three objects (J1129+2034, 1314+3452, J1444+4237) in our sample, the SDSS u-band sizes from the archive only refer to a bright star-formation knot instead of the entire galaxy. Therefore, we remeasure their sizes using the SDSS u-band image and performing the method of aperture photometry discussed above. Furthermore, there are five galaxies (J0036--3333, J0127--0619, J0144+0453, J0405--3648, J0337--0502) that do not have SDSS u-band images. We instead measure sizes from their optical images that are close to the u-band, including Dark Energy Survey (DES) g-band and Pan-STARRS g-band images.


For galaxies with r$_{50}$ $>$ 1.5 \arcsec, the COS aperture does not cover the majority of massive stellar population in the galaxy. This means that their observed absorption-line outflows may be less related to the global properties of the galaxy but instead more related to bright local properties (e.g. individual massive young clusters). Therefore, we will label these galaxies differently (in gray triangles) in all correlation figures in Section \ref{sec:results}. For each galaxy, the final adopted \Vcir\ and r$_{50}$ values along with other important ancillary parameters used in this paper are listed in Table \ref{tab:anc}. More detailed studies of the aperture effects of CLASSY galaxy properties are presented in \PaperV.

%

\section{Results}
\label{sec:results}
In this section, we present detailed analyses and the corresponding results for galaxies in the CLASSY sample. We start with getting robust estimates of the column density and covering fraction (\CF) of outflows from the observed \Siii\ and \siiv\ absorption lines in Section \ref{sec:ColumnDensity}. Then we study the dust depletion of metals in these galaxies from stacked spectra in Section \ref{sec:Dust}. After that, we estimate the total hydrogen column density (\Nh) of outflows from CLOUDY models \citep{Ferland17} in Section \ref{sec:PI}. Finally,  we present various scaling relationships related to outflow kinematics, feedback effects, and galaxy properties in Section \ref{sec:scale}.

\subsection{Column Density and Covering Fraction \\ of \Siii\ and \siiv\ Lines}
\label{sec:ColumnDensity}
Galactic outflows have been found to only partially cover the ionizing source \citep[e.g.,][]{Heckman00,Rupke05,Martin09,Chisholm16a}. In the case of covering fraction (\CF) $<$ 1.0, the column density (\Nion) derived from the absorption lines assuming an apparent optical depth (AOD) can only be viewed as a lower limit. Therefore, to get robust \Nion\ and \CF\ measurements, we adopt the partial-covering (PC) models that were commonly used in analyzing both quasar outflows \citep[e.g.,][]{Arav05, Xu18} and galactic outflows \citep[e.g.,][]{Rupke05, Martin09, Chisholm16a,Rivera-Thorsen15, Rivera-Thorsen17}. These models assume that only a portion of the UV continuum source (in our case, starburst region) is covered by foreground absorbing/outflowing gas. \CF\ and the optical depth ($\tau$) are usually degenerate since they can both affect the depths of a absorption line. However, the degeneracy can be broken given the useful information from doublet and multiplet transitions as follows \citep[e.g.,][]{Arav05}.

For absorption lines from a doublet transition (e.g., \siiv\ \ly\ly 1393, 1402), we have:

\begin{equation}
    \begin{aligned}   
     I_{R}(v)&= 1 - C_\text{f}(v) + C_\text{f}(v) \times e^{-\tau(v)}\\
     I_{B}(v)&= 1 - C_\text{f}(v) + C_\text{f}(v) \times e^{-w \cdot \tau(v)}
    \end{aligned}   
\label{eq:SiIV}
\end{equation}
where $I_{R}(v)$ and $I_{B}(v)$ are the red and blue doublets' flux at velocity $v$, normalized by the stellar continuum (Section \ref{sec:stellarContinuum}), respectively, and $\tau(v)$ is the optical depth of the red line (\siiv\ \ly 1402) that is velocity dependent. The weight $w$ in the bottom equation equals $f_{B}\lambda_{B}$/$f_{R}\lambda_{R}$, where $f$ is the oscillator strength and $\lambda$ is the wavelength of the line (see Table \ref{tab:atomic}). For common doublet lines, such as \nv\ \ly\ly 1238, 1242, \civ\ \ly\ly 1548, 1550, and \siiv\ \ly\ly 1393, 1402, they have $w$ = 2.

For absorption lines from a multiplet transition (e.g., \Siii\ multiplet), we similarly have a set of equations:

\begin{equation}
    \begin{aligned}   
     I_{k}(v)&= 1 - C_\text{f}(v) + C_\text{f}(v) \times e^{-w_{k} \cdot \tau(v)}
    \end{aligned}   
\label{eq:SiII}
\end{equation}
where for \Siii\ multiplet, $k$ stands for \Siii\ \ly 1190, 1193, 1260, 1304, and 1526. We define $w_{k}$ = $f_{k}\lambda_{k}$/$f_{1304}\lambda_{1304}$, and therefore $w_{k}$ = 2.7, 5.7, 12.7, 1.0, 1.7 for the five \Siii\ lines, respectively. If one or more troughs from the multiplet is blended with other lines (e.g., Milky Way absorption lines) or in detector gaps, we exclude them from the equation set.

To solve Equations (\ref{eq:SiIV}) or (\ref{eq:SiII}), we require the detections of absorption troughs from isolated doublet or multiplet transitions, respectively. Given the wavelength coverage of CLASSY data, the major lines discussed in the reminder of this paper are \siiv\ doublet and \Siii\ multiplet. We didn't solve the absorption troughs from \civ\ \ly\ly\ 1548, 1550 doublet. This is because the two troughs blend with each other, and the solutions are usually uncertain \citep[e.g.,][]{Du16}. Furthermore, the \civ\ line is more likely to be affected by nebular emission and stellar P-Cygni absorption.

For each galaxy, our main goal is to solve for the N(\siiv) and N(\Siii) in the galactic outflows. Therefore, we take fitted Gaussian models (i.e., \Gtwo, see Section \ref{sec:2GFits}) which represents the outflow component. Note that, due to possible contamination from infilling at the systemic velocity (Section \ref{sec:infilling}), it is difficult to get robust values for N(\siiv) and N(\Siii) for the static ISM component (i.e., \Gone).

We then adopt \textit{mpfit} \citep{Markward09} to solve $\tau(v)$ and \CFv\ from Equations (\ref{eq:SiIV}) and (\ref{eq:SiII}) for \siiv\ and \Siii, respectively.  An example for the fitted \Siii\ troughs and the resulting \CFv, $\tau(v)$, and \Nion\ are shown in Figure \ref{fig:SiIIFits}. For each galaxy, we also list the outflow mean covering fraction (\barCFv) in Table \ref{tab:mea}. This is calculated from the average of \CFv\ in the range of [\Vout\ - $\sigma_\text{Vout}$,\Vout\ + $\sigma_\text{Vout}$] for both the \Siii\ and \siiv\ multiplet/doublet lines. This choice is because most of the outflowing material (i.e., \Nion) has velocities around the center of the troughs. Therefore, \barCFv\ represents the covering fraction for the bulk of outflowing material we observe from \Siii\ and \siiv.

Finally, we calculate \Nion\ for \Siii\ and \siiv\ as below \citep[e.g.,][]{Savage91, Edmonds11}.

\begin{equation}
\label{Eq:GaussianFit2}
	\begin{aligned}
	\bar{N}_{ion}(v)  &=\frac{3.8 \times\ 10^{14}}{f\cdot\lambda}\cdot C_\text{f}(v) \cdot \tau(v)\\
	N_{ion}     &= \int {\bar{N}_{ion}(v)} dv  
	\end{aligned}
\end{equation}
where $\bar{N}_{ion}(v)$ is the average column density per velocity over the aperture at  $v$, and \Nion\ is the integrated column density over all velocity bins. The derived N(\Siii) and N(\siiv) values for each galaxy is presented in Table \ref{tab:mea}. 


For other singlet transitions such as \siiii\ \ly\ 1206, even though we have the absorption troughs observed in most of our galaxies, the corresponding N(\siiii) is difficult to measure due mainly to 3 reasons: 1) \siiii\ \ly\ 1206 is close to \lya, where the damped \lya\ absorption from Milky Way (and sometimes also from the observed galaxy) can cause significant contamination and/or make the continuum hard to determine; 2) \siiii\ \ly\ 1206 is a strong line and usually saturated (see Section \ref{sec:PI}). In this case, the measured N(\siiii) from the AOD method will be much smaller than the actual N(\siiii) considering the PC effects of outflows; 3) \siiii\ \ly\ 1206 is only a single line. It is not possible to solve the PC equations [i.e., Equations (2) or (3)] and determine robust N(\siiii). Overall, we do not present direct measurements of N(\siiii) from the spectra. We instead constrain N(\siiii) from the photoionization models discussed in Section \ref{sec:PI}.

In Figure \ref{fig:SiIIFits}, we show in the bottom-right panel the resulting N(\Siii)(v) for one of our galaxies (J0055--0021). The column density peaks near the line center and declines towards both higher and lower velocities. While the low velocity gas in the outflow will be affected by the accuracy to which the static ISM component is removed, the decline in column densities towards higher outflow velocity is robust. The top-right and lower-left panels show that the decline in N(\Siii)(v) is driven by a decline in both the optical depth and the covering fraction. 

We also see that the \Siii\ absorption troughs with smaller $f$ (i.e., \Siii\ \ly 1304) have narrow widths (see top-left panel), which is consistent with \cite{Wang20}. This is as expected from the curve of growth, i.e., while N(\Siii)(v) drops in the line wings, stronger lines are more optically thick and, therefore, show wider troughs (see the third panel of Figure \ref{fig:SiIIFits}). We further compare the \Vout\ and \FWHMout\ values from \Siii\ \ly 1260 and 1304 for all our galaxies in the bottom two panels of Figure \ref{fig:SiIIvsSiIV}. We find troughs from \Siii\ \ly 1260 indeed commonly have larger \FWHMout\ than that of \Siii\ \ly 1304, while their \Vout\ are quite consistent. These results imply that the \FWHMout\ parameter is quite sensitive to the distribution of column densities of the outflows (unlike \Vout), and there is a relatively narrow range around the characteristic outflow velocity where the bulk of the outflowing gas resides. This gas is traced most clearly by the least optically-thick lines.



\begin{figure}
\center
	\includegraphics[page = 2, angle=0,trim={0.3cm 0.0cm 0.0cm 1.0cm},clip=true,width=1\linewidth,keepaspectratio]{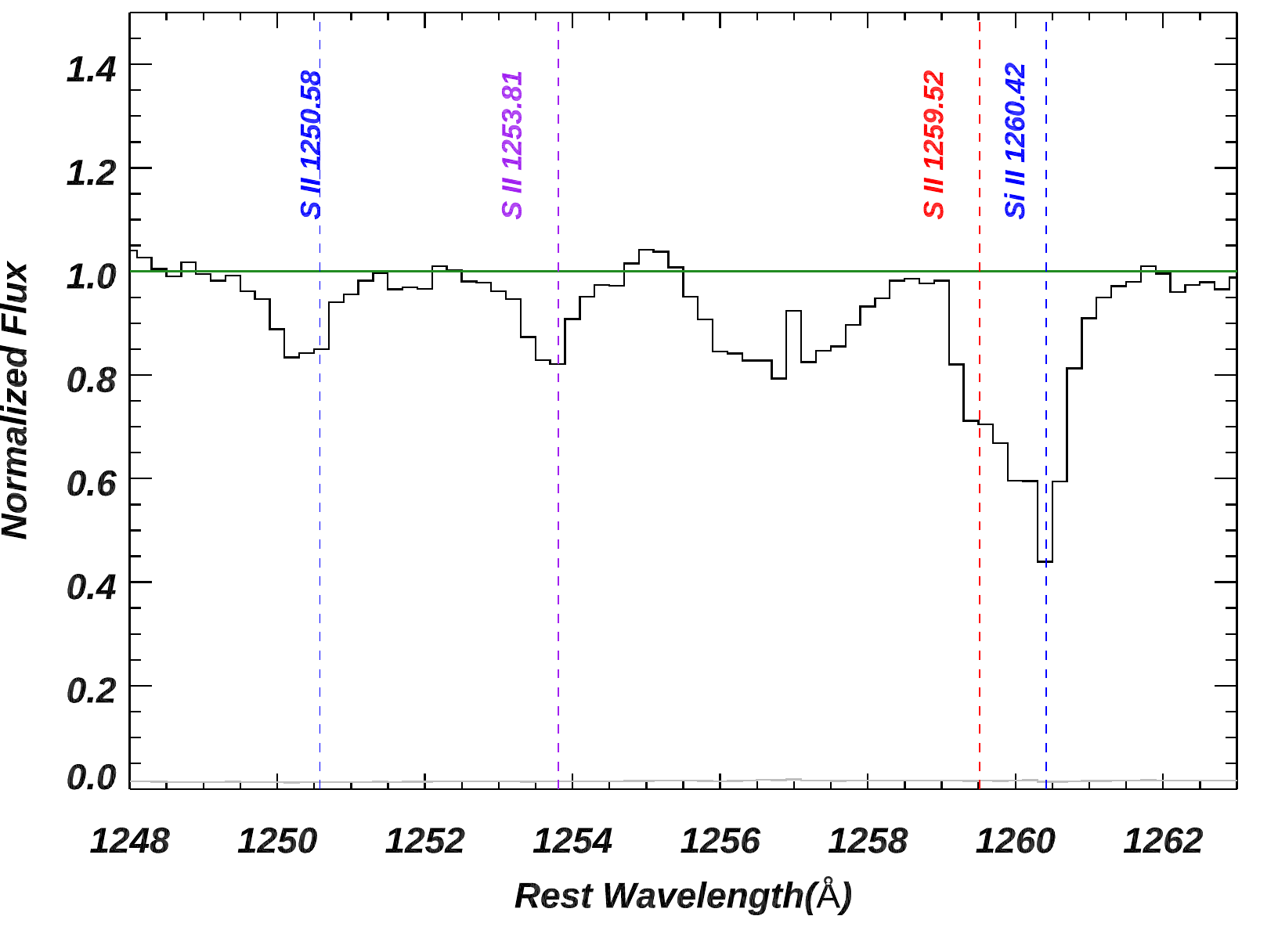}

\caption{\normalfont{Zoom-in on the spectral regions around  \sii\ \ly\ly\ly 1250, 1253, 1259 from the stacked spectra of 31 galaxies that cover all these lines. The normalized spectra are shown in black while the errors are shown in gray at the bottom. The trough from 1259 is commonly blended with \Siii\ \ly 1260. Therefore, we exclude it when calculating the column density of \sii\ (see Section \ref{sec:Dust}).} }
\label{fig:stack}
\end{figure}

\subsection{Dust Depletion of Gas-Phase Metals}
\label{sec:Dust}

To estimate the total hydrogen column density (\Nh) of outflows from the measured N(\Siii) and N(\siiv), we need first to determine the amount of depletion of gas-phase silicon onto dust. As discussed in \cite{Savage96,Jenkins09}, sulfur has been treated as a standard for an element with very little depletion. Therefore, in this subsection, we compare the measured silicon column densities to those of sulfur to estimate the dust depletion of silicon.

For $>$ half of our galaxies, the HST/COS spectra cover weak lines from \sii\ \ly\ly\ly 1250, 1253, 1259. To measure these weak features in our COS sample, we create a high S/N stacked spectrum that covers these lines \citep[e.g.,][]{Alex15}. A total of 31 galaxies are coadded following the methodology in \cite{Thomas19}. After de-redshifting and re-sampling wavelengths for all galaxies into the same wavelength array, we calculate SNR$^{2}$ weighted normalized flux for each bin. We also conduct a 3-sigma clipping in the weighting process, i.e., for one bin, we iteratively remove galaxies with flux that is out of the range [$med$ -- 3$\sigma$, $med$ + 3$\sigma$], where $med$ and $\sigma$ are the median and standard deviation of all flux at this bin in the coadding sample, respectively. This process helps remove outliers in each wavelength bin. The \sii\ region from the stacked spectra are shown in Figure \ref{fig:stack}.


\begin{table}
	\centering
	\caption{Column Densities for the Stacked Spectra}
	\label{tab:stack}
	\begin{tabular}{ccc|cc} 
		\hline
		\hline
		Labels  	    & \# of Galaxy$^{(1)}$     &	N(\Siii)$_\text{obs}$  & N(\sii)$_\text{obs}$  & N(\sii)$_\text{pred}$$^{(2)}$   \\
        \hline
        \sii\ stack    &31		& 774$^{+12}_{-12}$	    & 168$^{+35}_{-35}$    & 203$^{+5}_{-5}$ \\

		\hline
		\hline
		  	    &          &	N(\siiv)  & N(\siv)$_\text{obs}$  & N(\siv)$_\text{pred}$$^{(2)}$   \\
        \hline
        \siv\ stack             &24		    & 500$^{+3}_{-3}$	    & 193$^{+39}_{-39}$	    & 175$^{+2}_{-2}$     \\

		\hline
		\hline
	\multicolumn{5}{l}{%
  	\begin{minipage}{8cm}%
	Note. --\\
	    (1)\ \ The number of galaxies stacked. \\ 
    	(2)\ \ The predicted column densities are scaled from N(\Siii) or N(\siiv) assuming solar ratios of S/Si and typical ionization corrections (see Equation (\ref{eq:SII})).\\
    	
  	\end{minipage}%
	}\\
	\end{tabular}
	\\ [0mm]
	
\end{table}

We follow the same methodology in Section \ref{sec:ColumnDensity} to measure the column densities of \sii\ [i.e., N(\sii)$_\text{obs}$]. For \sii\ \ly 1259, since it is too close to \Siii\ \ly 1260, we exclude it from the calculation. In Table \ref{tab:stack}, we compare the measured N(\sii) from the stacked spectra (4th columns) to the predicted values (5th columns) by scaling N(\Siii) as follows:
\begin{equation}
    \begin{aligned}   
    \text{ N(S II})_\text{pred} &=\overbar{\text{N}}\text{(Si II})\cdot(\text{S}/\text{Si})\cdot\frac{\text{ICF(S II})}{\text{ICF(Si II})}
    \end{aligned}   
\label{eq:SII}
\end{equation}
where S/Si is the abundance ratio, and we assume solar abundance; \Nbar(\Siii) is the mean value of measured N(\Siii) for all stacked galaxies (Section \ref{sec:ColumnDensity}); and ICF(\sii) and ICF(\Siii) are the ionization corrections for \sii\ and \Siii, respectively. For typical parameters of our galaxies (see Section \ref{sec:PI}), we get ICF(\sii)/ICF(\Siii) $\sim$ 0.6 from CLOUDY models \citep{Ferland17}. This is consistent with \cite{Hernandez20} for similar star-forming galaxies. Our choice of a solar abundance ratio for sulfur-to-silicon is a good one because both are alpha elements \citep[created in core-collapse supernovae,][]{Steidel16,Kobayashi20}.  


From Table \ref{tab:stack}, we show that N(\sii)$_\text{obs}$ and N(\sii)$_\text{pred}$ from our stacked spectra are consistent within 1$\sigma$.  Similarly, we conduct another stack which includes all galaxies covering \siv\ \ly 1062 region, and find N(\siv)$_\text{obs}$ is also consistent with N(\siv)$_\text{pred}$ within 1$\sigma$ (see the second part of Table \ref{tab:stack}). In this case, N(\siv)$_\text{pred}$ is scaled from \Nbar(\siiv) given a solar S/Si ratio and ICF(\siv)/ICF(\siiv) $\sim$ 0.8 . This suggests that the dust depletion of silicon is similar to sulfur in our galaxies, implying both are primarily in the gas-phase. Qualitatively-similar depletion patterns have been observed in the Milky Way halo \citep{Savage96,Jenkins09} and in shocked regions in the disk \citep[]{Welty02}.

Overall, we conclude that silicon is mostly undepleted into the dust for galaxies in our sample. Therefore, it is viable to estimate \Nh\ of outflows from the measured N(\Siii) and N(\siiv), which is discussed next.




\subsection{Column Densities from CLOUDY Models}
\label{sec:PI}

\subsubsection{Methodology}

Given the low dust depletion of silicon and the measured N(\Siii) and N(\siiv), we can estimate two important properties of the observed outflows, i.e., the total silicon column density (\NSi) and total hydrogen column density (\Nh). We do so by running a variety of grid models adopting CLOUDY [version c17.01, \citep{Ferland17}]. The fixed parameters are: 1) Starburst99 \citep{Leitherer99} models as the input spectral energy distribution (SEDs), where we assume Geneva tracks with high mass loss and constant star-forming history (SFH) with age = 5 Myr. We also assume a standard Kroupa IMF \citep[with slopes of 1.3 and 2.3 in mass ranges of 0.1 -- 0.5 and 0.5 -- 100 \Msun, respectively, see][]{Kroupa01}; 2) an electron number density (\ne) = 10 cm$^{-3}$ for typical starburst galaxies; and 3) for each galaxy, we adopt the GASS10 abundance \citep{Grevesse10} scaled by the measured O/H values discussed in Section \ref{sec:anc}. The conversion from \NSi\ to \Nh\ assumes that the outflow has the same metallicity as measured from the nebular emission lines.\footnote{In principle, the metallicity of the outflow could be larger than in the HII regions, if the former is strongly contaminated by metals ejected by the supernovae \citep{Chisholm18,Hogarth20}. We will show in Section \ref{sec:SiliconRate} that this is unlikely.} We have tested models for \ne\ = 100 cm$^{-3}$, and find that the resulting \NSi\ and \Nh\ only have minor changes ($<$ 3 -- 5\%). In our grid models, the two varied parameters are: 1) the logarithm of ionization parameter, log(\Uh), in the range between -4.0 and 0.0 with a step size of 0.05 dex, and 2) the logarithm of \Nh\ in the range between 18.0 and 23.0 [log(cm$^{-2}$)] with a step size of 0.02 dex.

For each velocity bin, we solve the best-fit model, i.e., a set of \Uh($v$) and \Nh($v$) values, by comparing the CLOUDY model predicted N(\Siii)($v$) and N(\siiv)($v$) to the measured ones at this velocity (see Section \ref{sec:ColumnDensity}). This is done through $\chi^2$-minimizations of the difference between the model predicted and the measured column densities \citep[e.g.,][]{Borguet12, Xu19}. Then for the whole outflow, we integrate \Nh($v$) over all velocity bins to get \Nh. Hereafter, we use \Nion($v$) to represent the column density per velocity at $v$ [in unit of cm$^{-2}$/(km s$^{-1}$)], while \Nion\ stands for the integrated column density over all velocity bins in the outflow trough [in unit of cm$^{-2}$]. The best-fit \Nh\ values are listed in Table \ref{tab:mea}, and its distribution is shown in Figure \ref{fig:NhDist}, where we find that the mean \Nh\ for the observed outflows is 10$^{20.70}$ cm$^{-2}$. 
Based on this best-fit model, we can then predict the column densities for other ions , e.g., N(\siiii)($v$) and N(\hi)($v$). We also show their integrated values in Table \ref{tab:mea}. We find that, in all galaxies, \siiii\ is the dominant ion of silicon for the observed outflows, so we estimate the total column density of silicon as \NSi\ =  N(\Siii) + N(\siiii) + N(\siiv).

Furthermore, the mean value of N(\siiii) is 10$^{15.53}$ cm$^{-2}$. For an average FWHMout $\sim$ 300 km s$^{-1}$ (see Figure \ref{fig:VoutDist}), and assuming the trough is flat, we can calculate the average optical depth of \siiii\ by solving Equation (\ref{Eq:GaussianFit2}) as:

\begin{equation}
\label{Eq:tau}
	\tau\ = \frac{ N_\text{Si III}\cdot f_\text{Si III}\cdot\lambda_\text{Si III}}{3.8 \times\ 10^{14}\cdot FWHM_{out}}
\end{equation}
where we have $f_\text{Si III}$ = 1.67 and $\lambda_\text{Si III}$ = 1206.51\angstrom. This leads to $\tau$(\siiii) $\sim$ 60. Combined the fact that the observed troughs of \siiii\ are usually non-black (i.e., I $>$ 0), this suggests that \siiii\ troughs of our galaxies are strongly non-black saturated and affected by PC effects. This is consistent with our claims in Section \ref{sec:ColumnDensity}.  

We also see that only a few outflows are consistent with a unit covering factor (with a median value $0.7$). The outflows are therefore somewhat patchy. We see no evidence for any systematic difference between the values of the covering fractions for \Siii\ and \siiv, implying that the two ions likely trace similar regions in the outflow \citep[see also][]{Chisholm16a}.

\subsubsection{The Neutral Phase of the Outflow}

For our galaxies, the mean N(\hi) from CLOUDY models is 10$^{18.28}$ cm$^{-2}$, which is only $\sim$ 0.4\% of the mean \Nh\ value (10$^{20.70}$ cm$^{-2}$). This suggests that our outflows detected in UV absorption lines are mostly ionized gas, and the neutral gas is only a minor part of the outflows in our galaxies.

This also has interesting implications for the origin of the gas traced by low ionization transitions of \Siii\ and \cii. These are sometimes used as proxies for neutral hydrogen \citep[e.g.,][]{Jones13, Alex15, Gazagnes18}. However, the ratios of the observed column densities in \Siii\ vs. those derived via CLOUDY in \hi\ are much too large for the \Siii\ to arise primarily in the \hi\ gas. More quantitatively, we find that typically only 1 to 10\% of the observed \Siii\ comes from the \hi\ phase, so \Siii\ is a poor proxy for the neutral gas. This means that the \oi\ \ly 1302 line should be used instead, since the nearly identical ionization potentials of \oi\ and \hi\ ensure that the two species arise in the same phase.




\subsection{Scaling Relationships}
\label{sec:scale}
In this subsection, we present various empirical scaling relationships between the outflow  and galaxy properties. These correlations for low-redshift galaxies (z $\lesssim$ 0.4) have already been discussed in previous publications in the literature \citep[e.g.,][]{Heckman00, Martin05,Rupke05, Heckman15, Chisholm16a}. However, we have conducted more robust analyses for the CLASSY sample. Namely: 1) For each galaxy, our outflow velocity and FWHM are determined from up to 10 lines where both low- and high-ionization transitions are considered (see Section \ref{sec:2GFits}). In contrast, most previous publications usually have access to only 1 -- 2 low-ionization lines \citep[e.g., \nai\ D studied in][]{Heckman00, Martin05, Rupke05}. 2) We solve for the outflow's column density  given a partial covering (PC) model (see Sections \ref{sec:ColumnDensity} and \ref{sec:PI}). In contrast, previous publications are usually aware of the PC property of outflows, but did not include it in their analyses \citep[except in a few cases, e.g.,][]{Chisholm16b, Chisholm17,Chisholm18}. This was usually due to the inability to solve the PC equations without doublet or multiplet transitions being well-detected, 3) We have used CLOUDY models to calculate the total H column densities based on the measured ionic column densities. These ionization corrections were not possible in the prior work based on only low-ionization lines, e.g., \nai\ D and \mgii.

Finally, the CLASSY sample was specifically designed to span maximum ranges in the fundamental galaxy properties of stellar mass, star-formation rate, and metallicity. This makes it ideal for exploring how the outflow properties correlate with these other properties. Overall, it is clearly important to revisit these scaling relationships.

\begin{figure*}
\center
	\includegraphics[angle=0,trim={2.8cm 0.8cm 0.5cm 0.8cm},clip=true,width=0.45\linewidth,keepaspectratio]{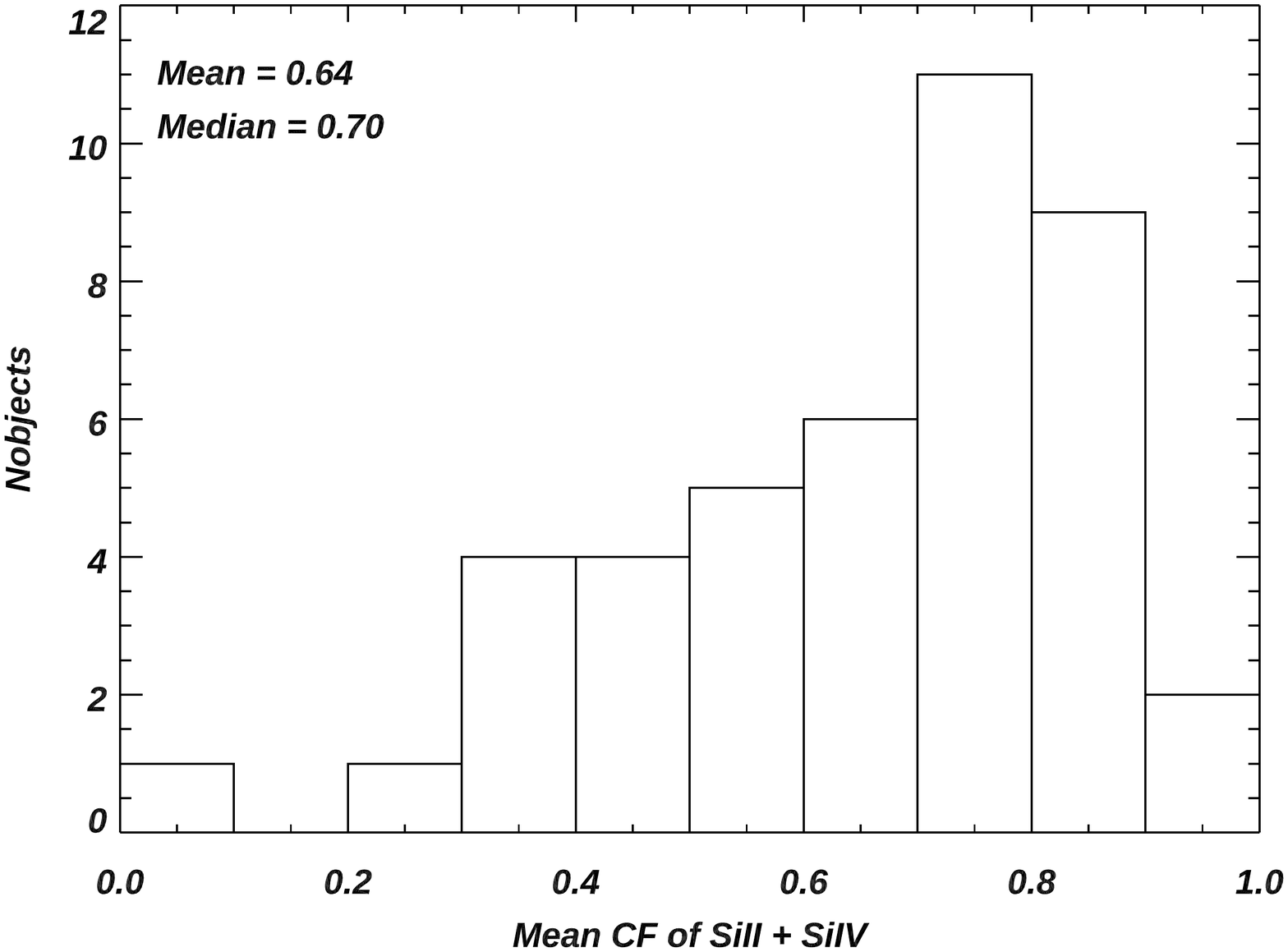}
	\includegraphics[angle=0,trim={2.8cm 0.8cm 0.5cm 0.8cm},clip=true,width=0.45\linewidth,keepaspectratio]{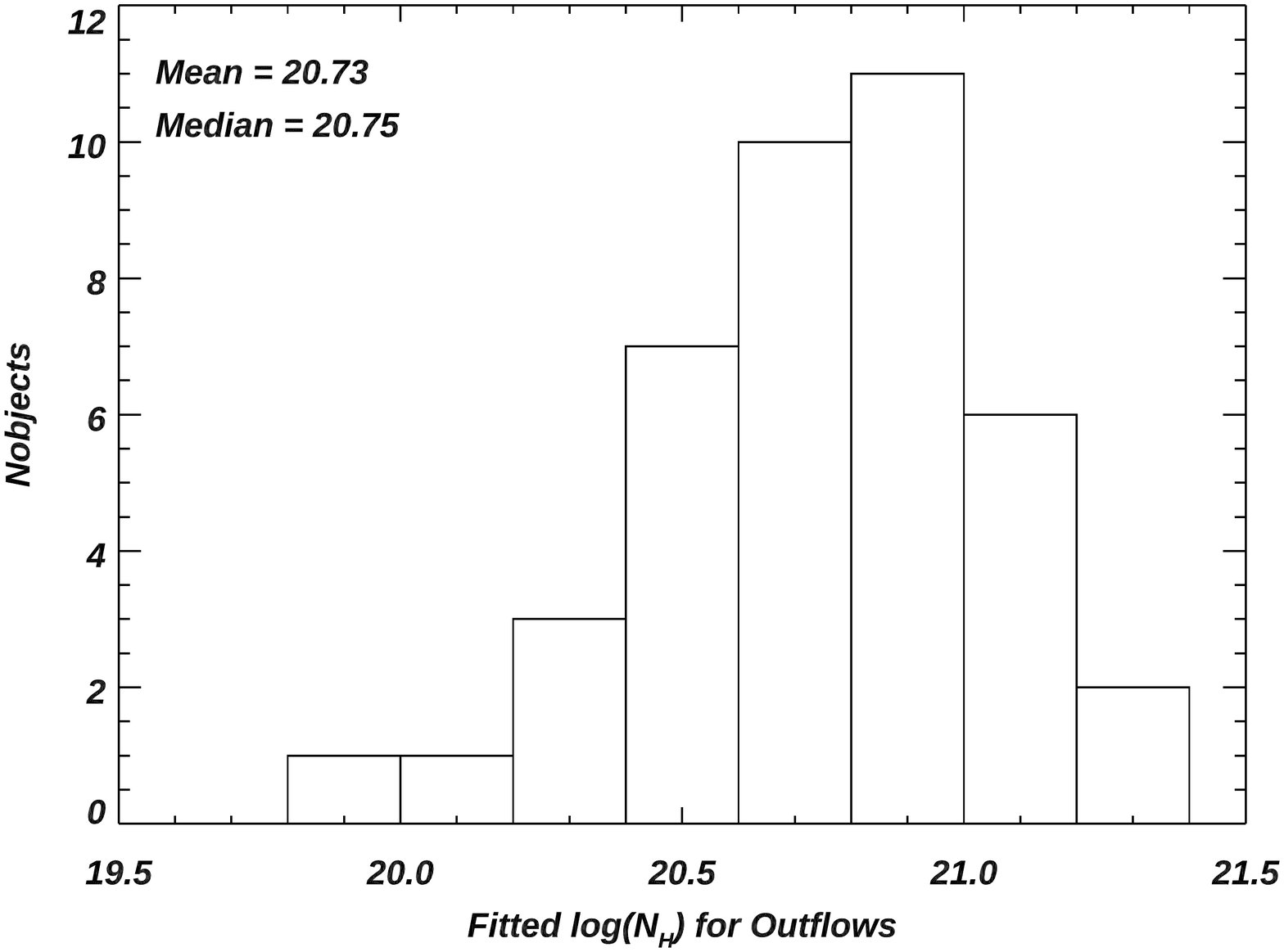}

\caption{ \normalfont{Two distributions for galaxies from our combined sample of CLASSY and \cite{Heckman15}. \textbf{Left:} Distribution for the mean covering fraction of outflows. For each object, this is calculated as the average of the covering fractions for \Siii\ and \siiv\ around the measured outflow velocity, i.e., \Vout\ $\pm$ $\sigma_\text{Vout}$ (see Section \ref{sec:ColumnDensity}). \textbf{Right:} Distributions for the total column density of outflows (\Nh, in units of cm$^{-2}$), which are derived from the photoionization models of CLOUDY (see Section \ref{sec:PI}). Note that for both panels, galaxies labelled as ``No outflow" are \textit{not} included.} }
\label{fig:NhDist}
\end{figure*}

As discussed in Section \ref{sec:reduction}, for all figures in this subsection, we have added 5 LBA galaxies from \cite{Heckman15} that have HST/COS observations but were not part of the CLASSY sample. These LBA galaxies satisfy all selection criteria of the CLASSY ones, but they can provide additional starbursts with relatively larger values of SFR and \Mstar. To be consistent, we follow the same methodology as discussed in Sections \ref{sec:analyses} and  \ref{sec:results} to measure the required quantities for these 5 galaxies. In all figures, galaxies from CLASSY and \cite{Heckman15} are in red and blue colors, respectively.


\begin{figure*}
\center
	\includegraphics[page = 1, angle=0,trim={0.2cm 0.8cm 0.20cm 0.3cm},clip=true,width=0.5\linewidth,keepaspectratio]{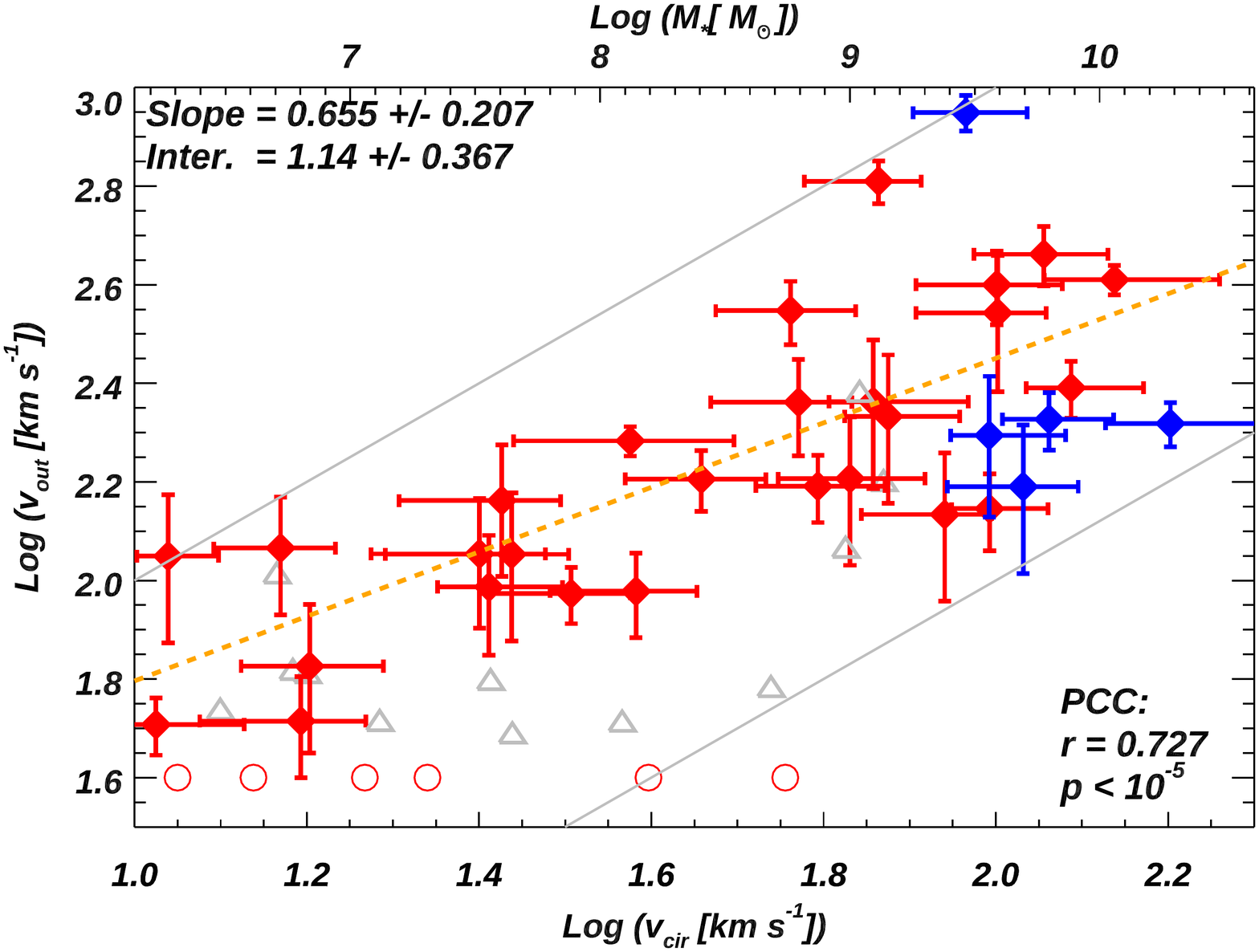}
	\includegraphics[page = 1, angle=0,trim={0.2cm 0.8cm 0.20cm 1.2cm},clip=true,width=0.5\linewidth,keepaspectratio]{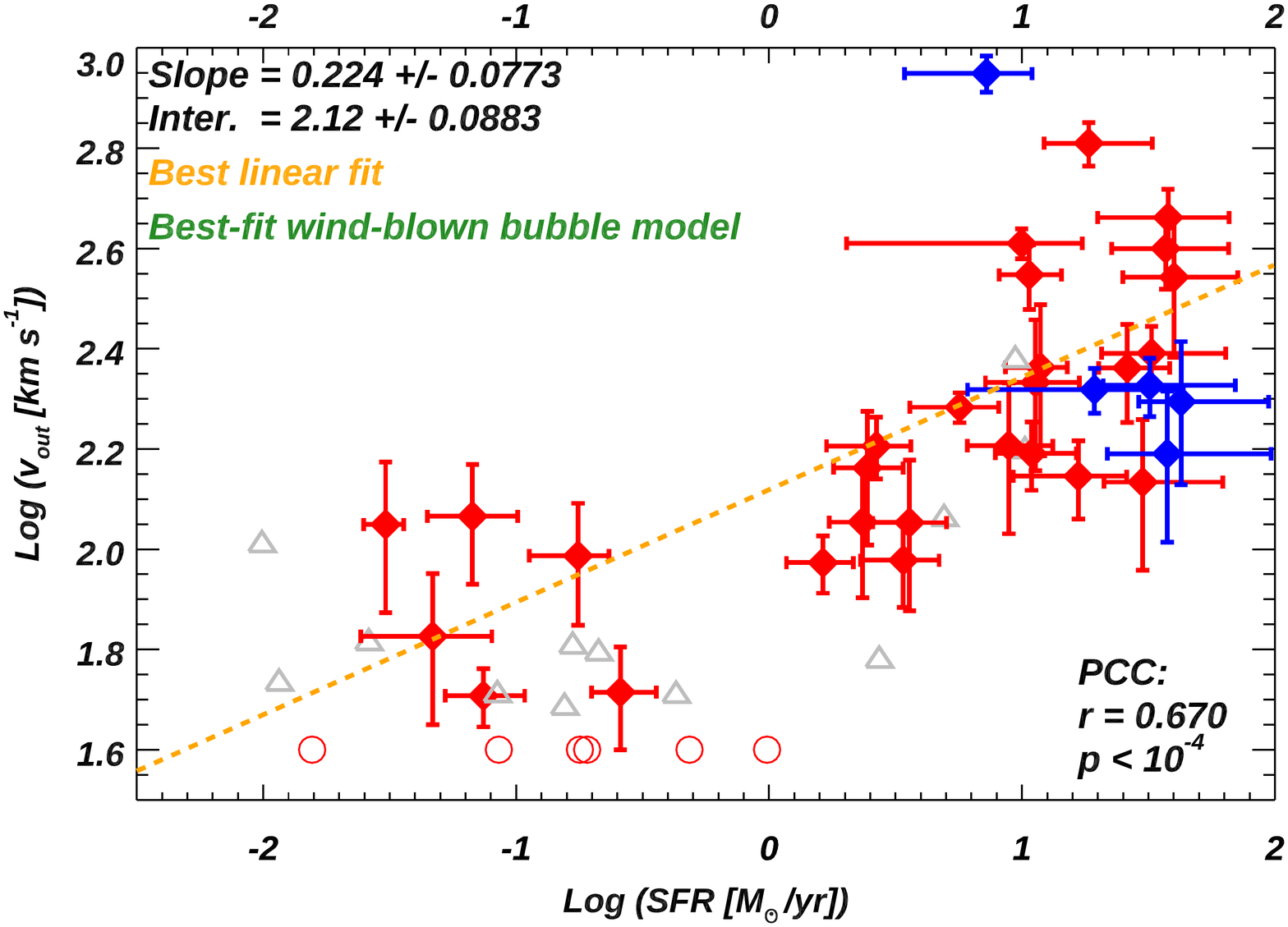}
	
	\includegraphics[page = 1, angle=0,trim={0.2cm 0.8cm 0.20cm 0.3cm},clip=true,width=0.5\linewidth,keepaspectratio]{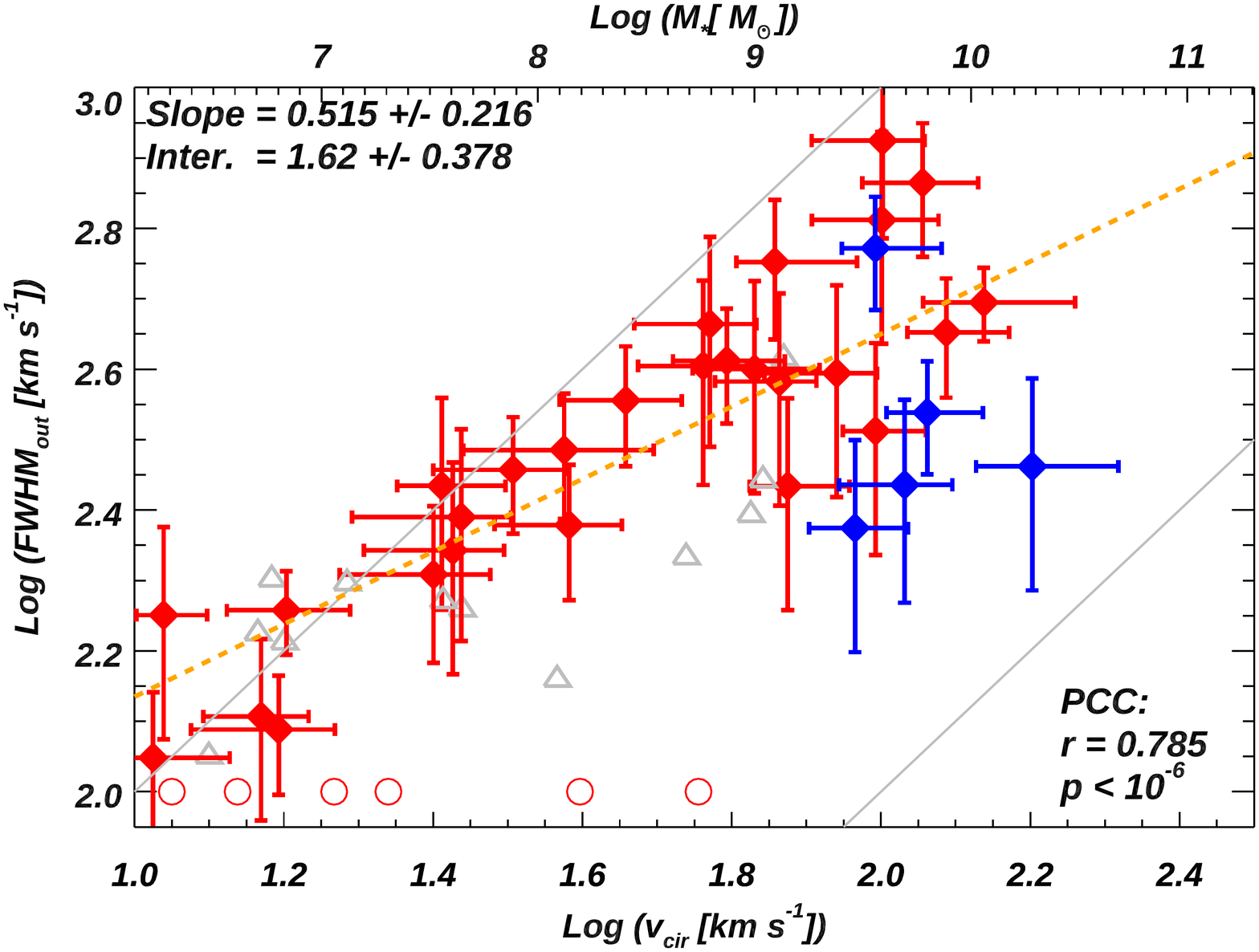}
	\includegraphics[page = 1, angle=0,trim={0.2cm 0.8cm 0.20cm 1.2cm},clip=true,width=0.5\linewidth,keepaspectratio]{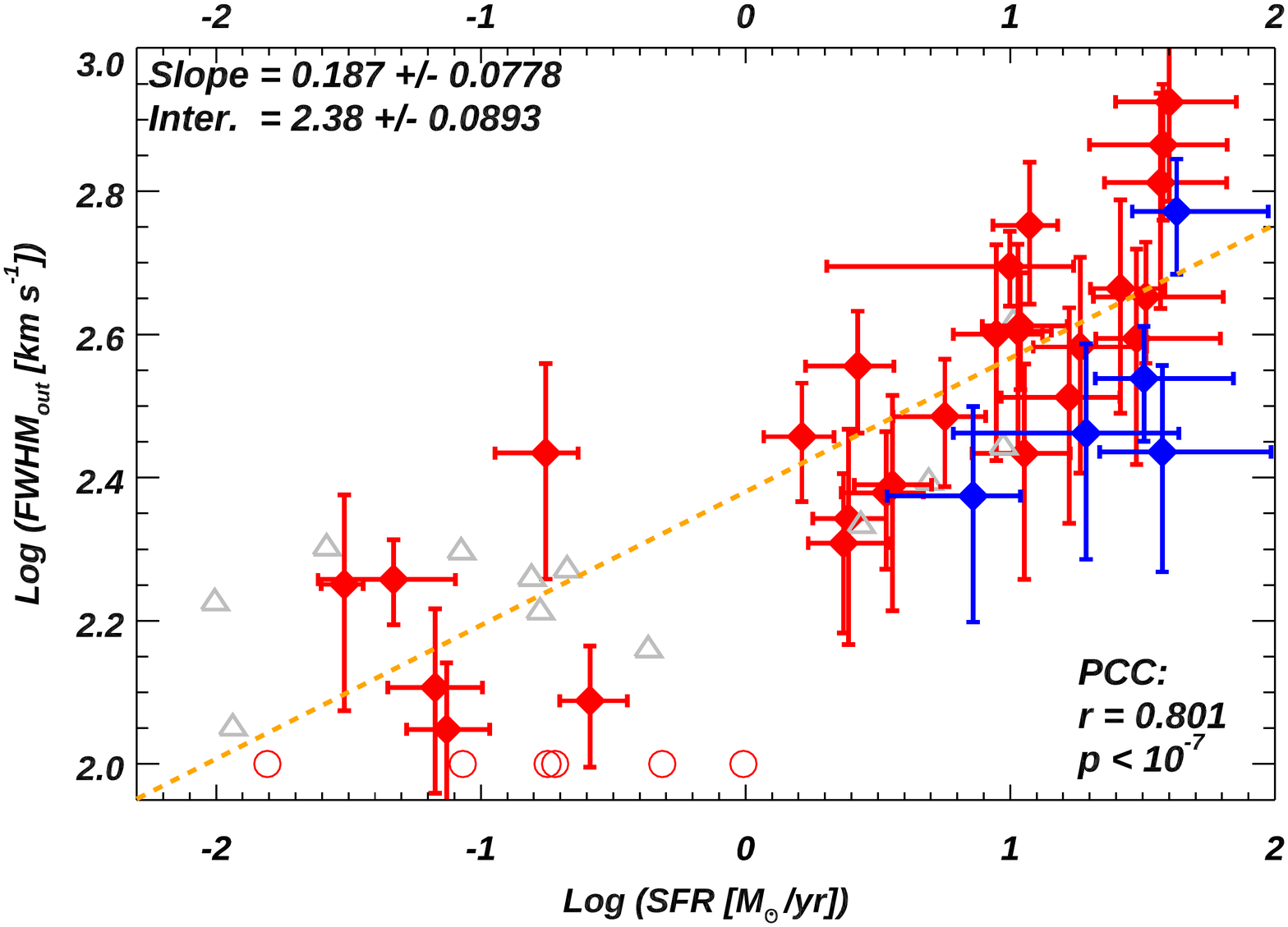}
	
\caption{\normalfont{The log of the outflow velocity (\Vout) and Full-Width-Half-Maximum (\FWHMout) plotted as a function of the basic properties of the star-forming galaxies. From top-left to bottom-right are log(\Vout) vs. circular velocity, star formation rate, and log(\FWHMout) vs. circular velocity, star formation rate, respectively. Galaxies with HST/COS observations from CLASSY and \cite{Heckman15} are shown in red and blue, respectively. The corresponding error bars are shown as crosses. The gray solid lines in the left panels indicate \Vout $= V_{cir}$ and \Vout $= 10 V_{cir}$ (and the same for \FWHMout). Galaxies with non-detections of outflows are shown as hollow circles. We set their \Vout\ = 40 km s$^{-1}$ and log(\FWHMout) = 100 km s$^{-1}$ (only to include them in the figures). Galaxies with large UV sizes compared to COS aperture are labeled as hollow gray triangles. For these galaxies, the COS aperture covers only part of the starburst and the data may reflect local rather than global outflow properties. These galaxies' spectra will also be more affected by vignetting. We therefore exclude them when calculating the correlation coefficients. The results for Pearson correlation coefficients (PCC) are shown at the bottom-right corner of each panel. The linear fit of the y to x values is shown as the orange dashed line, and the fitted slope and intercept are shown in the top-left corner. See discussion in Section \ref{sec:Vout}. For the top-right panel, we also show the best-fit wind-blown bubble model in green assuming momentum-conserving, which matches the data well [see Section \ref{sec:compModel} and Equation (\ref{eq:bubble})]. } }
\label{fig:VoutCorr}
\end{figure*}

\begin{figure*}
\center
	\includegraphics[page = 1, angle=0,trim={0.2cm 0.8cm 0.0cm 0.3cm},clip=true,width=0.5\linewidth,keepaspectratio]{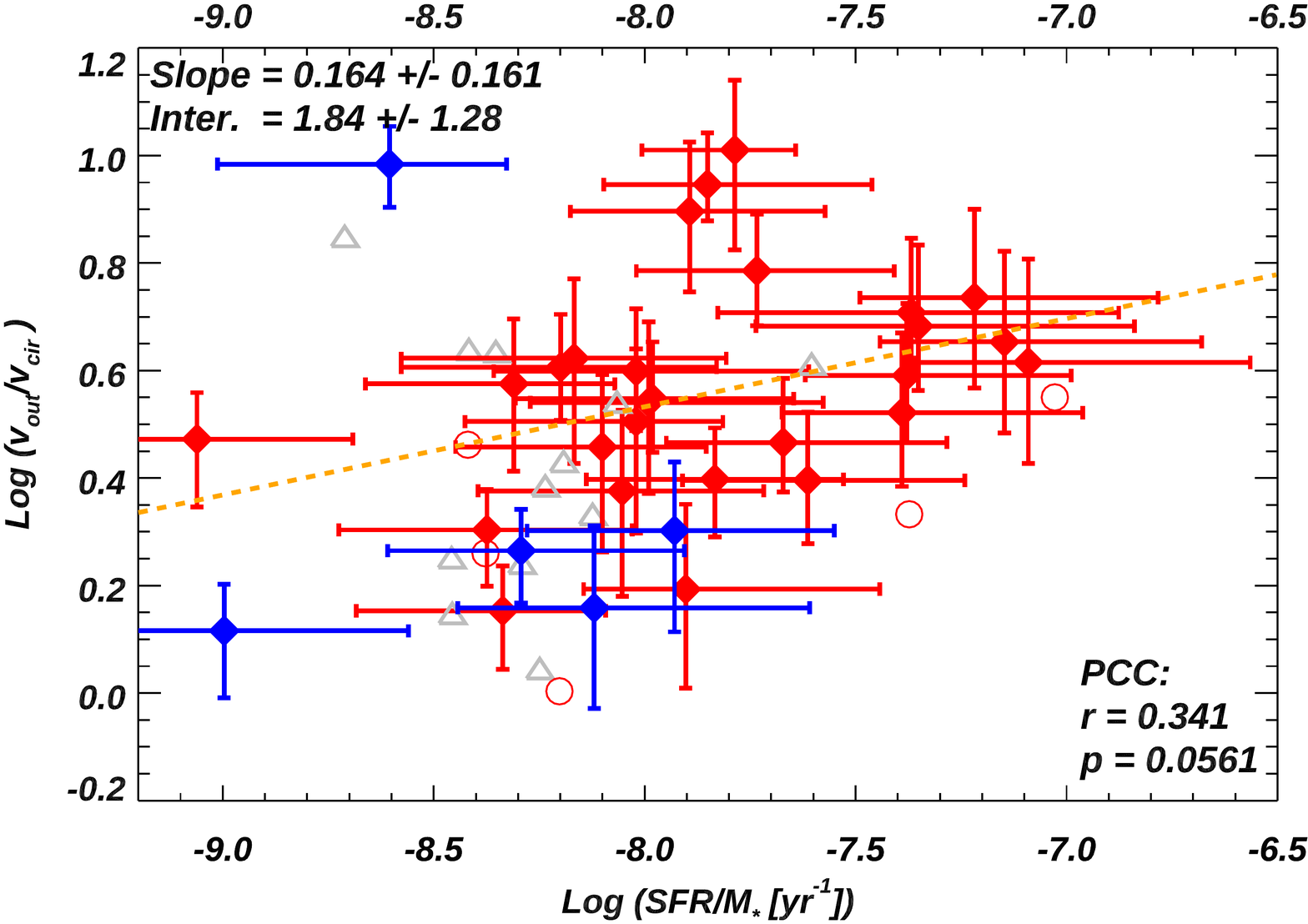}
	\includegraphics[page = 1, angle=0,trim={0.2cm 0.8cm 0.0cm 1.2cm},clip=true,width=0.5\linewidth,keepaspectratio]{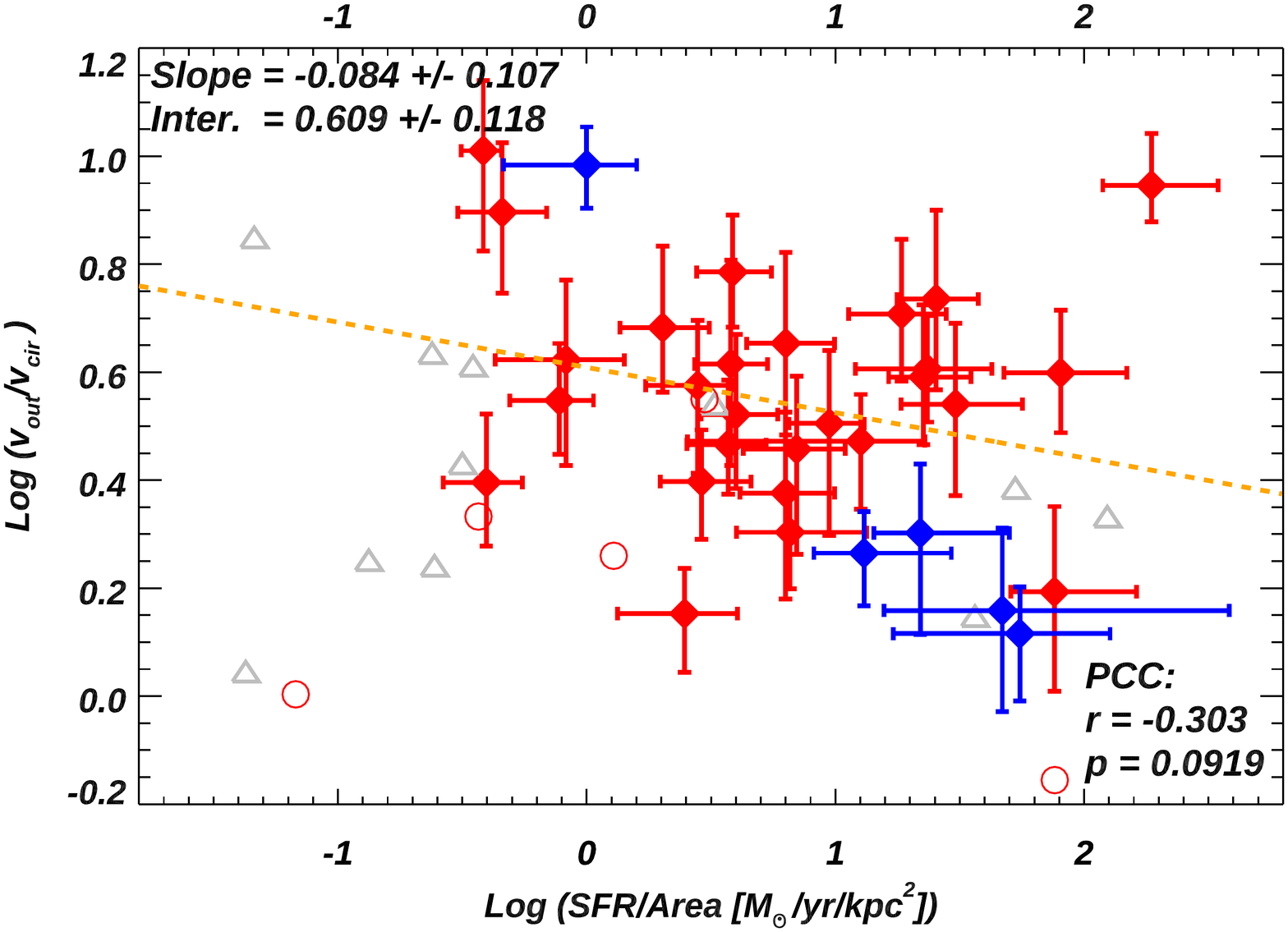}

\caption{\normalfont{Comparisons of normalized outflow velocity (\Vout/\Vcir) with normalized star formation rate (SFR/\Mstar\ and SFR/area). The captions and labels are the same as Figure \ref{fig:VoutCorr}. Due to the small dynamic range of \Vout/\Vcir\ in our sample, the correlations in these two panels are weak (L) to insignificant (R). See discussion in Section \ref{sec:Vout}. } }
\label{fig:VoutNormCorr}
\end{figure*}

\subsubsection{Outflow Velocity and FWHM}
\label{sec:Vout}
We begin by examining the correlations of \Vout\ and \FWHMout\ (Section \ref{sec:2GFits}) with the principal properties of the galaxy, including the circular velocity (\Vcir), stellar mass (\Mstar, in units of \Msun) and star formation rate (SFR, in units of \Msun/yr). We show these correlations in logarithm scale in Figure \ref{fig:VoutCorr}. The corresponding error bars are shown as crosses. Galaxies with non-detections of outflows are shown as hollow circles, and we arbitrarily set their values to \Vout\ = 40 km s$^{-1}$ and \FWHMout\ = 100 km s$^{-1}$ (in order to show them in the figures). Note that the non-detections are preferentially in galaxies with low stellar masses and SFRs (for such galaxies, the fraction of detected outflows is smaller). 

For each panel, the results of Pearson correlation coefficients (PCC) are shown at the bottom-right corner. The linear fit of the y to x values is shown as the orange dashed line, and the fitted slope and intercept are shown in the top-left corner. We have also conducted the Kendall $\tau$ test to assess the statistical significance of each correlation. Both the Kendall and PCC coefficients are listed in Table \ref{tab:corr}. Objects with UV radii r$_{50} >$ 1.5\arcsec\ are labeled as hollow triangles. As discussed in Section \ref{sec:anc}, most of the UV light in these galaxies lies outside of the COS aperture, so their observed absorption outflows may not be representative of the global properties of the galaxy. This is further confirmed by the fact that these galaxies do not lie upon the locus defined by other more centrally concentrated galaxies in Figure \ref{fig:VoutCorr}. Therefore, we exclude galaxies with UV radii r$_{50} >$ 1.5\arcsec\ when calculating the correlation coefficients.

As shown in the left column of Figure \ref{fig:VoutCorr}, there are statistically significant correlations between \Vout\ (or \FWHMout) with \Mstar\ (and hence with \Vcir). It is noteworthy that the slopes of the relationships between \Vout\ and \FWHMout\ with \Vcir\ are sub-linear, meaning that their ratios decrease with increasing \Vcir. We will discuss this further in Section \ref{sec:discussion} below. In the right column of Figure \ref{fig:VoutCorr}, we see strong positive correlations between \Vout\ (or \FWHMout) with SFR. The strengths of the correlations with SFR and \Mstar\ are very similar. This is in part because SFR and \Mstar\ are themselves well-correlated in this sample \citepalias{Berg22}. In contrast, the sample studied in \citet{Heckman15} included galaxies observed by Far Ultraviolet Spectroscopic Explorer (FUSE) that had significantly lower values of SFR/\Mstar\ (typically 10$^{-10}$ to 10$^{-9}$ M$_{\odot}$) compared to CLASSY. With the inclusion of these galaxies, they found a stronger correlation of \Vout\ with SFR than with \Mstar.


We have also noticed that the correlations with \FWHMout\ are slightly stronger than the ones with \Vout\ in Figure \ref{fig:VoutCorr}. As we explained in Section \ref{sec:ColumnDensity}, unlike \Vout, \FWHMout\ is sensitive to both the bulk kinematics of the outflow and to the distribution of column density (as reflected in the extent of the broad and shallow wings of the outflow profiles). These wings represent lower column densities in outflows and are more sensitive to the variations of column densities (see Figure \ref{fig:SiIIFits}). Therefore, differences in galaxy properties could affect \FWHMout\ somewhat differently than \Vout.

In Figure \ref{fig:VoutNormCorr}, we also test the correlations between the normalized outflow velocity (\Vout/\Vcir) and normalized measures of the star formation rate (SFR/area and SFR/\Mstar). Previous studies found positive correlations for them \citep[e.g.,][]{Martin12, Heckman15}. Our galaxies in these figures show large scatter, and the correlations are weak or insignificant. As noted above, the main difference between our sample and the \citet[]{Heckman15} sample is that we do not have galaxies with the relatively low values of SFR/area and SFR/\Mstar\ represented by their galaxies with FUSE data. In a future paper, we will re-analyze the FUSE data in the same way as we have done for the current sample and then revisit these potential correlations.

Similar results relating \Vout\ to SFR, \Mstar, and \Vcir\ have been found previously in low-z starbursts \citep[]{Martin05, Rupke05, Chisholm15, Chisholm16a, Heckman16}. A direct comparison to our results is not straightforward, largely because few of the studies defined \Vout\ the same way we have done. In some cases, a ``maximum'' velocity was used rather than a line centroid \citep[e.g.,][]{Rupke05,Heckman16}. In other cases, the entire absorption feature was treated as a single component \citep[e.g.,][]{Heckman15,Chisholm15,Chisholm16a}. The only study that used something similar to our double-Gaussian approach was \citet[]{Martin05}. Another difference is that \citet[]{Martin12} and \citet[]{Rupke05} used the \nai\ D doublet to probe the outflows. This traces a dusty \hi\ component in the outflow, while the UV lines used in CLASSY, \citet[]{Chisholm15}, \citet[]{Chisholm16a}, and \citet{Heckman16} trace warm ionized gas.

With these caveats in mind, we summarize the results from these papers and compare them to CLASSY in Table \ref{tab:scale}. Given the differing definitions of \Vout, we only list the log-log slopes (and not their normalizations). All the studies are rather consistent. The main advantage of the CLASSY sample is an improved sample size at low values of SFR, \Vcir, and \Mstar.

\begin{figure*}
\center

	\includegraphics[page = 1, angle=0,trim={0.1cm 0.8cm 0.1cm 0.3cm},clip=true,width=0.50\linewidth,keepaspectratio]{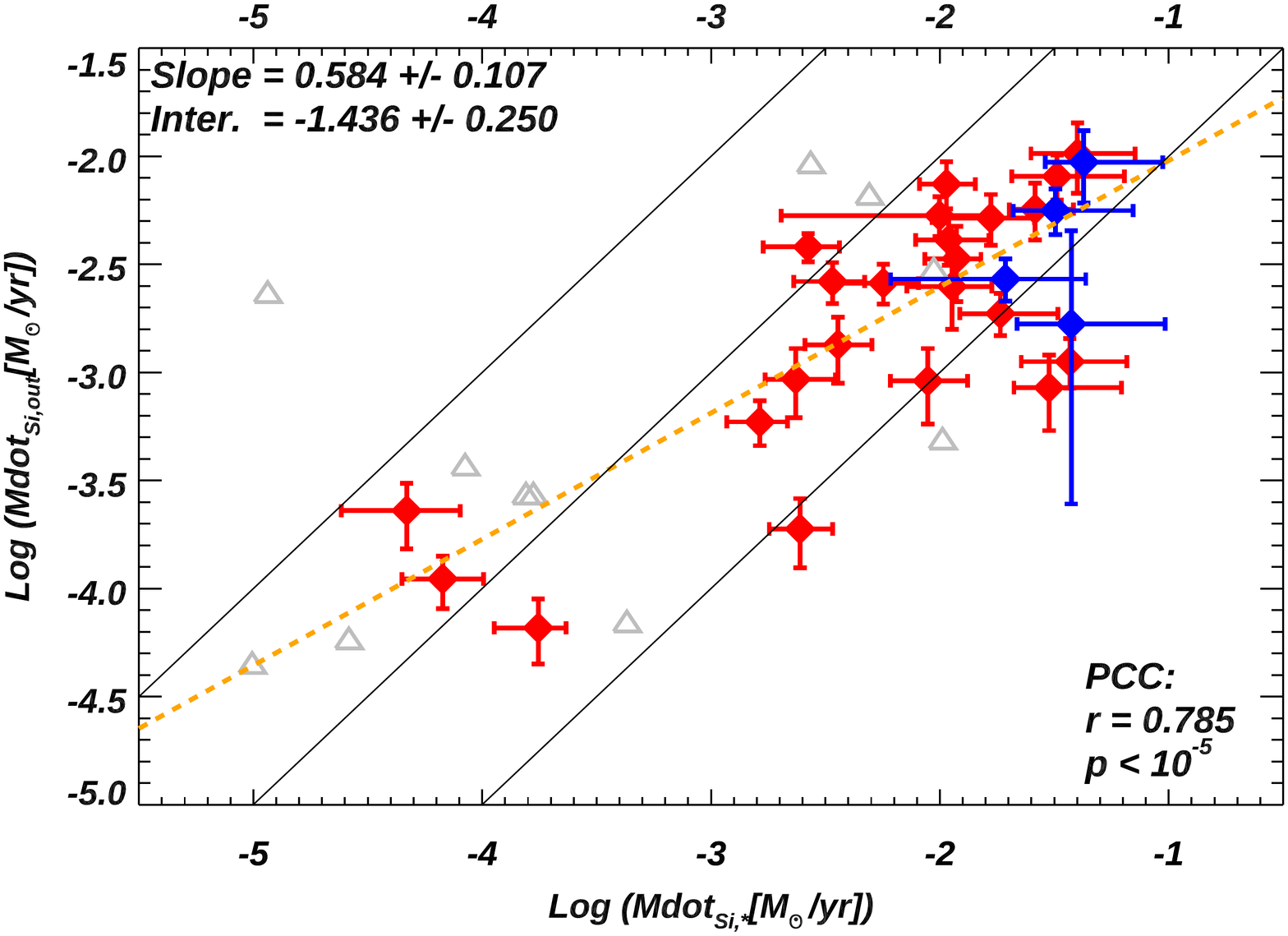}
	\includegraphics[page = 1, angle=0,trim={0.1cm 0.8cm 0.1cm 0.3cm},clip=true,width=0.50\linewidth,keepaspectratio]{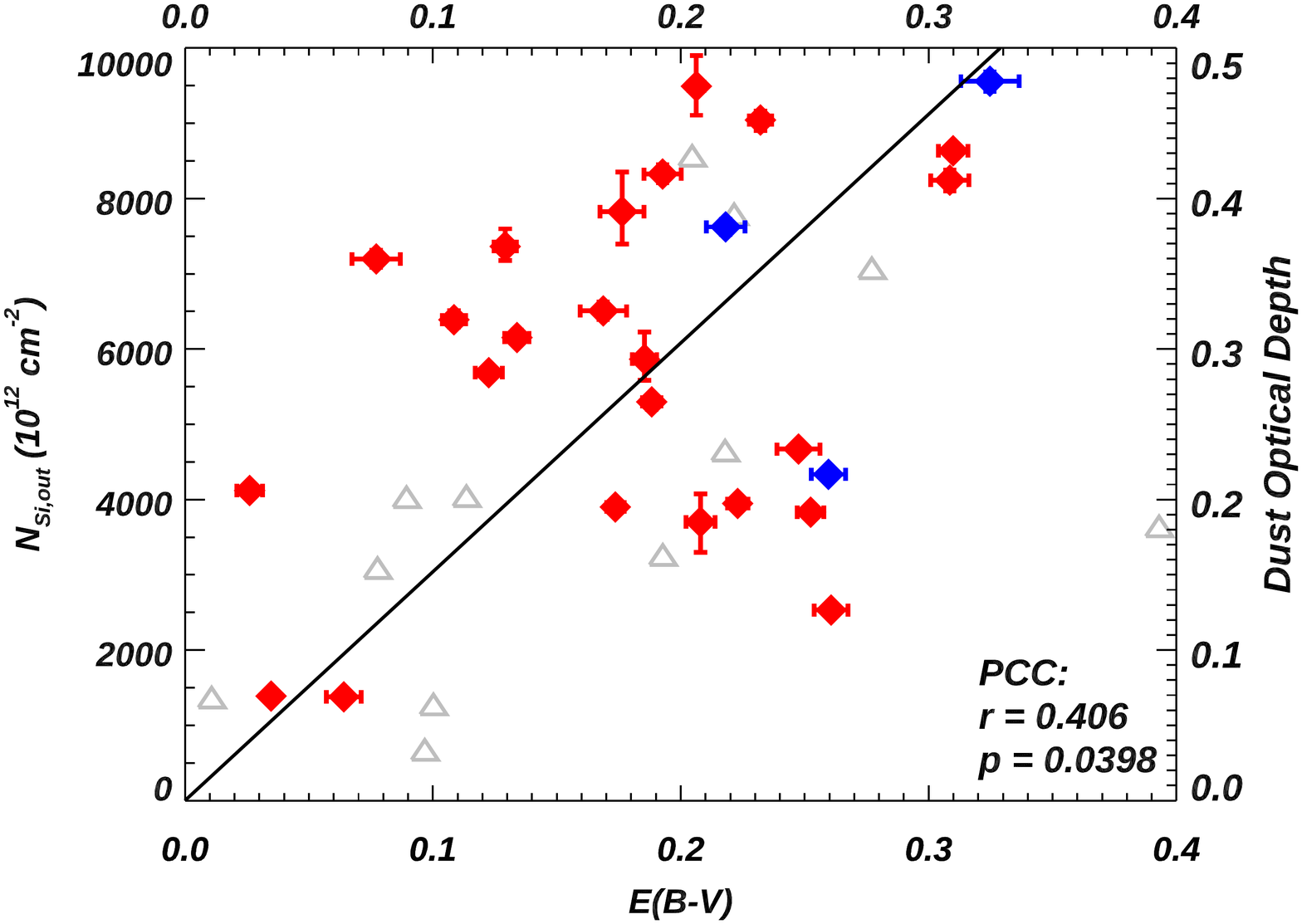}

\caption{\normalfont{Correlations related to the silicon mass flow rate (\MdotSi) and total silicon column density (\NSi). Labels and captions are the same as Figure \ref{fig:VoutCorr}. \textbf{Left:} We compare \MdotSi\ from the outflow to the amount that is provided from the starburst. The solid lines from left to right represent where y = 10, 1, 0.1 x, separately. \textbf{Right:} We compare \NSi\ to the dust extinction calculated from fitting the stellar continuum (Section \ref{sec:anc}). We also show the prediction (in black line) from \cite{Draine11} assuming a standard Milky Way dust/metals ratio and the observed median dust extinction in outflow $\sim$ 16\% of the total dust measured. The right y-axis represents the dust optical depth at 1200\angstrom\ assuming \cite{Reddy15} extinction law. See discussion in Section \ref{sec:SiliconRate}.} }
\label{fig:MdotSiCorr}
\end{figure*}

\begin{table*}
	\centering
	\caption{Comparisons of \Vout\ Correlations with Published Star-forming/Starburst Galaxy Samples}
	\label{tab:scale}
	\begin{tabular}{ccccccc} 
		\hline
		\hline
		References  & Redshift & Lines$^{(1)}$ & Def. of \Vout$^{(2)}$   &SFR  &\Vcir     & \Mstar \\
		\hline
		This paper     & 0.002 -- 0.23    & Far-UV lines$^{(3)}$   & Double Gaussian    & 0.22$\pm$0.08 &0.66$\pm$0.21   & 0.19$\pm$0.06\\
		\cite{Martin05}     & 0.04 -- 0.16    & \nai\ \ly\ly 5890, 5896   & Double Gaussian    & 0.35$\pm$0.06 &\dots\   & \dots\\
        \cite{Rupke05}      & 0.01 -- 0.50    & \nai\ \ly\ly 5890, 5896   & Max velocity    & 0.21$\pm$0.05 &\dots\   & \dots\\
        \cite{Chisholm15}$^{(4)}$      & 0.0007 -- 0.26    & \Siii\ 1190, 1193, 1260, 1304   & Single Gaussian    & 0.22$\pm$0.04 &0.87$\pm$0.17   & 0.20$\pm$0.05\\
        \cite{Chisholm16a}$^{(4)}$      & 0.0007 -- 0.26    & Far-UV lines$^{(3)}$   & Max velocity    & 0.12$\pm$0.02 &\dots\   & 0.15$\pm$0.02\\
        \cite{Heckman16}         & 0.4 -- 0.7    & UV lines$^{(3)}$   & Max velocity    & 0.32$\pm$0.02 &1.16$\pm$0.37   & 0.34$\pm$0.11\\        
		\hline
	\multicolumn{7}{l}{%
  	\begin{minipage}{18cm}%
	Note. -- We compare the slopes of the scaling relationships with other known samples of star-forming/starburst galaxies in the literature. The log-log slopes between \Vout\ and SFR, \Vcir, and \Mstar\ are shown in the 5th, 6th, and 7th columns, respectively.\\\
	    \textbf{(1).}\ \ The lines adopted to measure the outflow velocity (\Vout). \\
	    \textbf{(2).}\ \ Definitions for \Vout. In this paper, we derive \Vout\ from the double-Gaussian fitted results (see Section \ref{sec:analyses}), which is similar to \cite{Martin05}. Other publications adopted either single-Gaussian fitting or took the maximum velocity (or $v_{90}$) of the troughs as \Vout. \\ 
	    \textbf{(3).}\ \ From HST/COS spectra, the major rest-frame Far-UV lines adopted to estimate \Vout\ are: \oi, \cii, \Siii\ multiplet, \siiii, \siiv\ \citep[used in this paper, see Section \ref{sec:analyses}, and][]{Chisholm16a, Heckman16}. For works that also considered FUSE data \citep{Heckman16}, additional Far-UV lines are adopted, including \ciii\ \ly 977, \cii\ \ly 1036, and \nii\ \ly 1084. \\	 
	    \textbf{(4).}\ \ These did not exclude the cases in which the COS aperture did not cover at least 50\% of the starburst FUV continuum (see Section \ref{sec:anc}). \\
  	\end{minipage}%
	}\\
	\end{tabular}
	\\ [0mm]
	
\end{table*}


\begin{figure*}
\center
	\includegraphics[page = 1, angle=0,trim={0.1cm 0.8cm 0.1cm 0.3cm},clip=true,width=0.5\linewidth,keepaspectratio]{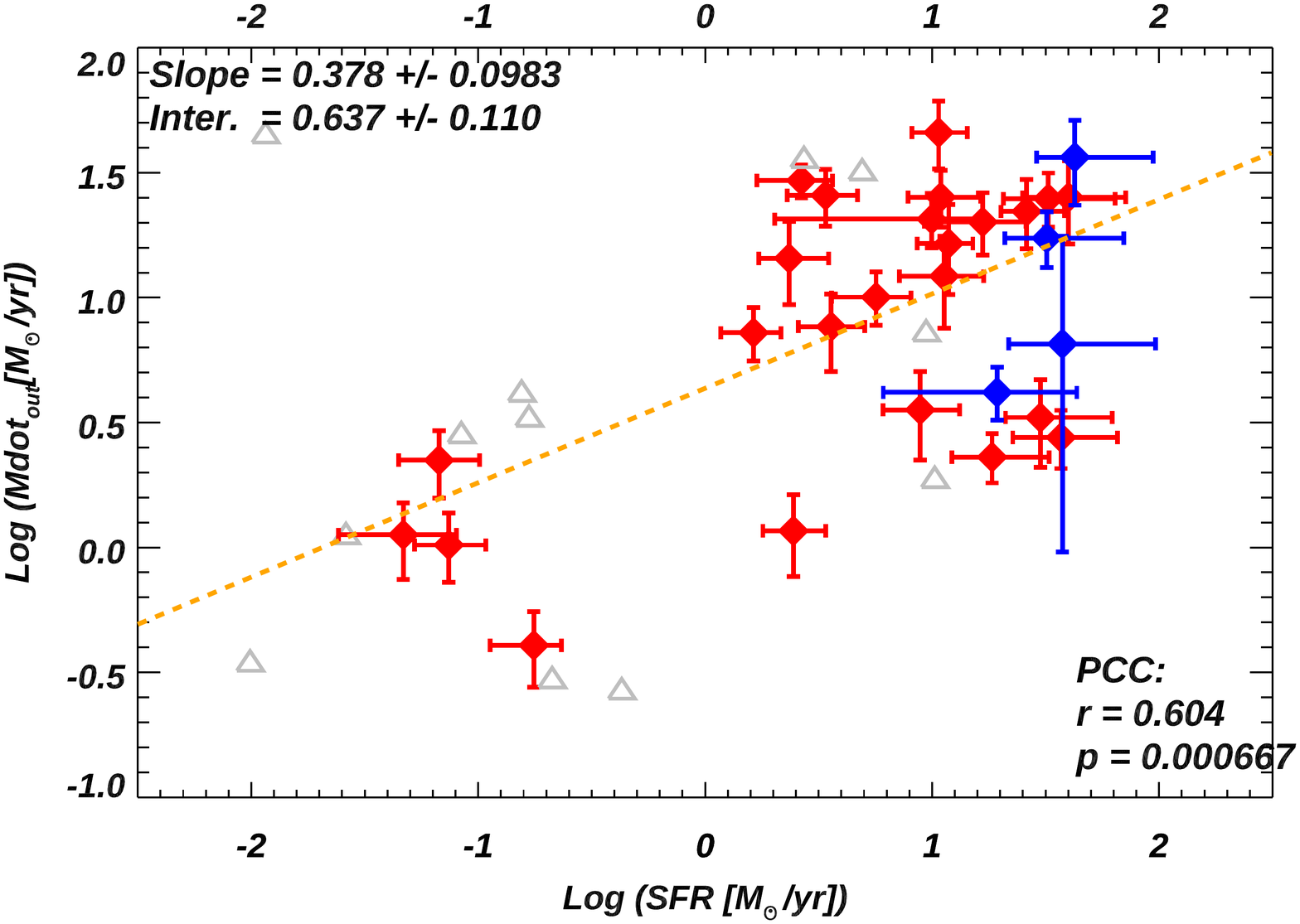}
	\includegraphics[page = 1, angle=0,trim={0.1cm 0.8cm 0.1cm 0.3cm},clip=true,width=0.5\linewidth,keepaspectratio]{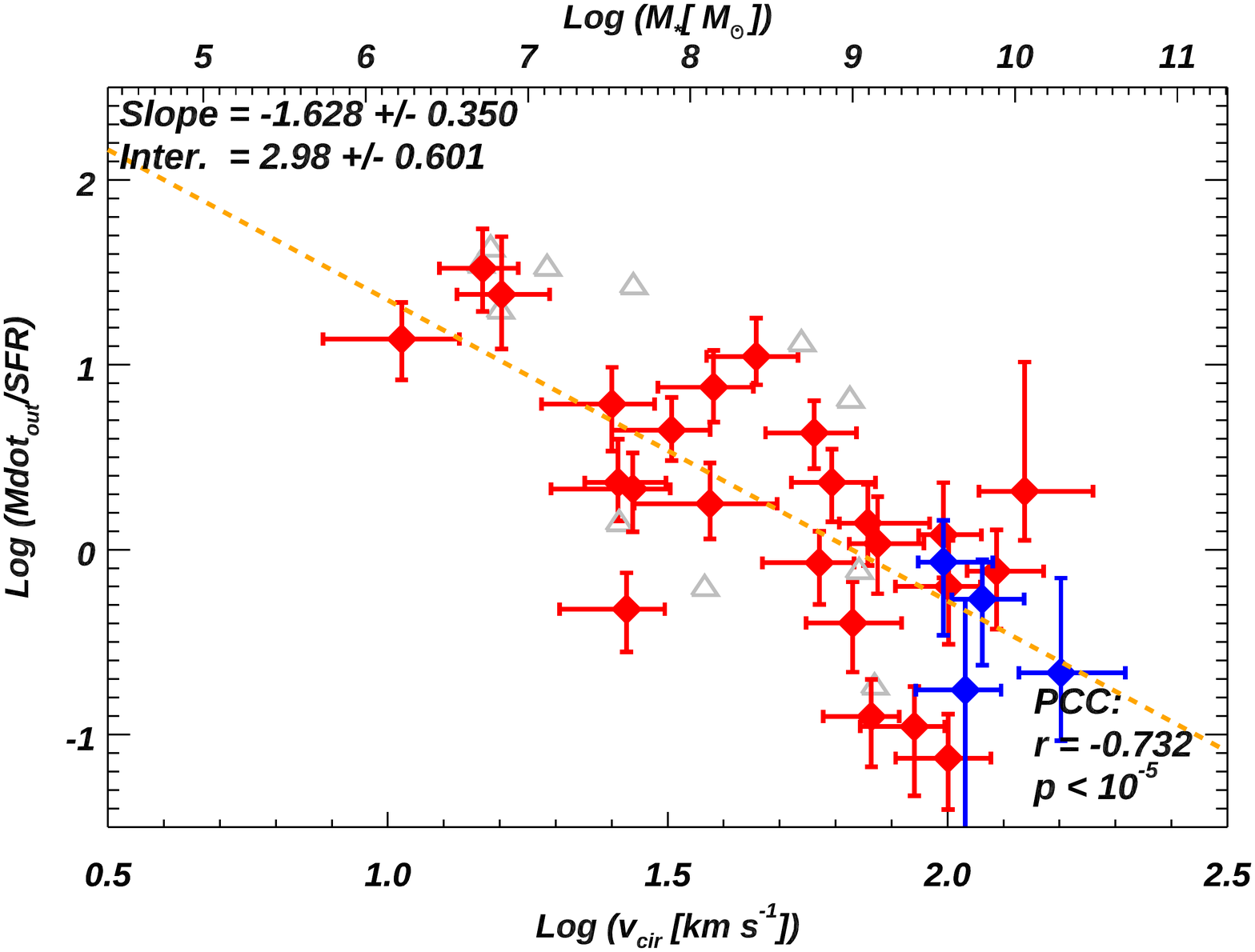}
	
	\includegraphics[page = 1, angle=0,trim={0.1cm 0.8cm 0.1cm 0.3cm},clip=true,width=0.5\linewidth,keepaspectratio]{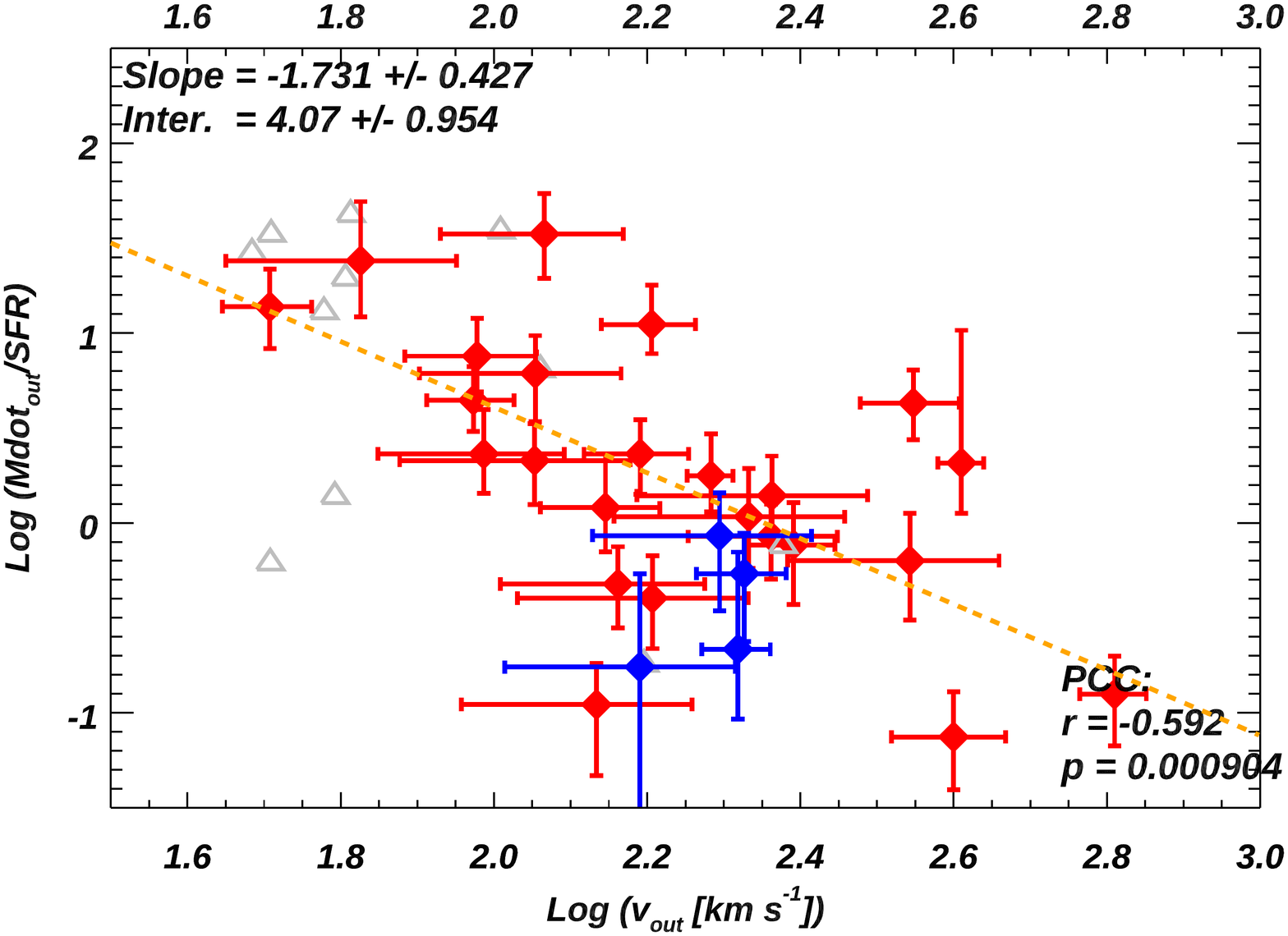}

\caption{\normalfont{Correlations related to total mass flow rate (\Mdot) and mass loading factor (\Mdot/SFR). Labels and captions are the same as Figure \ref{fig:VoutCorr}. The top left panel shows that the mass outflow rate has sub-linear dependence on the SFR. The top right and lower left panels show how the ratio of \Mdot/SFR scales with the galaxy circular velocity and with the outflow velocity. In both cases, the ratio increases rapidly at lower velocities (slopes $\sim$ --1.6 to --1.7). See further discussion in Section \ref{sec:TotalMassRate}.} }
\label{fig:MdotCorr}
\end{figure*}

\subsubsection{Outflow Rates: General Considerations}
\label{sec:OutflowGeneral}
Before using these data to estimate outflow rates in the galaxies, it is useful to briefly consider the general methodology and resulting uncertainties. We will use the mass outflow rate as a specific example, but these general considerations will apply to all the outflow rates discussed later.

Simple dimensional analysis tells us that the average mass outflow rate ($\dot{M}_{out}$) will just be the mass of outflowing gas ($M_{out}$) within a radius $R_{out}$ divided by the time it takes the flow to traverse this distance, i.e., $R_{out}/V_{out}$. This leads to $\dot{M}_{out} = M_{out} V_{out}/R_{out}$, where \Vout\ is the outflow velocity. The strength of the observed absorption-lines depends on the column density of the outflowing gas (particles per unit area).

Let us consider a few idealized cases. In the simplest case of an expanding thin shell with mean velocity $V_{shell}$, radius $R_{shell}$, and solid angle $\Omega$ we have:

\begin{equation}\label{eq:Mdots}
    \begin{aligned}
        \dot{M}_{out}\simeq \Omega N_\text{H} \mu m_p R_\text{shell} V_\text{shell} 
    \end{aligned}
\end{equation}
where $\mu$ is the mean mass per proton ($\sim 1.4$) and $m_p$ is the proton mass. The difficulty in using this equation is that we do not know the value of $R_\text{shell}$. We will return to consideration of shells in the context of a wind-blown bubble model in Section \ref{sec:discussion} below.

For now, we consider cases of continuous mass-conserving outflows with different radial profiles of density and velocity. The simplest such case is an outflow with constant velocity \Vout\ and a density profile $n(R) = n_0 (R/R_{0})^{-2}$, where $R_{0}$ is the radius at which the outflow begins. This simple case is broadly consistent with both numerical simulations \citep[e.g.,][]{Schneider20} and analytic models \citep[e.g.,][]{Fielding21} of multi-phase galactic outflows. In this situation (hereafter, case 1), it is straightforward to show that 

\begin{equation}\label{eq:Mdot2}
    \begin{aligned}
        \dot{M}_{out}\simeq \Omega N_\text{H} \mu m_p R_\text{0} V_\text{out} 
    \end{aligned}
\end{equation}

However, based on the analysis of the UV emission-line properties of outflows \citep[e.g.,][]{Wang20,Burchett21}, we also consider a shallower radial density profile $n(R) = n_0 (R/R_0)^{-1}$, which implies $V(R) = V_{max} (R/R_0)^{-1}$ for mass-conservation (and also means that the outflow's maximum velocity is at $R_0$). In this case (hereafter, case 2):
\begin{equation}\label{eq:Mdot1}
    \begin{aligned}
        \dot{M}_{out}\simeq \Omega N_\text{H} \mu m_p R_0 V_{max}/ln(R_{max}/R_0)
    \end{aligned}
\end{equation}
where $R_{max}$ is the maximum radius which the outflow reaches. 

The key point here is that for reasonable choices for the relevant parameters, the values of $\dot{M}_{out}$ will be the same to better than a factor of two in two radial density laws described  above. More explicitly, if we define $V_{max}$ $\sim$ \Vout + 0.5 \FWHMout, then Figure \ref{fig:FEFits} implies $V_{max}$ $\sim$ 2 \Vout. The mass outflow rates will be exactly the same in case 1 and case 2 above if $ln(R_{max}/R_0$) = 2 (i.e. $R_{max}$ = 7.3 $R_0$). Since this involves the log of the ratio, its value depends only weakly on $R_{max}/R_0$.  

\subsubsection{Metal Mass Outflow Rates and Dust Extinction}
\label{sec:SiliconRate}
With these considerations in mind, we calculate the metal mass outflow rates of silicon (\MdotSi) for galaxies in our sample. One advantage of \MdotSi\ over the total (i.e., hydrogen) mass outflow rate (\Mdot) is that we do not need to know the metallicity of the outflows, for which we don't have direct measurements.


Let us consider the simple cases of an outflow with a constant velocity and a $R^{-2}$ density profile [see Equation (\ref{eq:Mdot2})] for silicon:

\begin{equation}\label{eq:Mdot3}
    \begin{aligned}
        \dot{M}_\text{Si,out}\simeq \Omega \mu m_p R_\text{0} \int \frac{dN_\text{Si}}{dv}\times vdv
    \end{aligned}
\end{equation}
where $dN_\text{Si}/dv$ = N$_\text{Si}(v)$ is the silicon column density per velocity, and the integration is for the velocity range of the observed outflow trough. We have three unknowns (i.e., $\Omega$, \NSi, and $R_0$) required to calculate \MdotSi. For $\Omega$, as shown in Section \ref{sec:2GFits}, in the CLASSY sample, $\sim$ 85\% of galaxies are identified as "hosting outflows". Given this high detection rate and the possible existence of outflows with directions in the sky-plane (which are undetectable), we take $\Omega$ = 4$\pi$. 

The value for \NSi\ is derived from the CLOUDY models discussed in Section \ref{sec:PI}. For $R_0$, previous studies either assume a fiducial radius \citep[e.g., 1 -- 5 kpc in][]{Martin05, Rupke05, Martin12}, or relate it to the starburst radius \citep[e.g.,][]{Chevalier85, Heckman15}. We take the second method and assume $R_0 = 2$ $\times$ r$_{50}$ (the outflow begins at twice the half-light radius of the starburst). This choice is consistent with \cite{Heckman15}, and corresponds to an outflow that begins at a radius enclosing 90\% of the starburst (for an exponential disk model). It is also consistent with the radius at which the outflow of ionized gas is observed to begin in M 82 \citep{Shopbell98}. As noted above, the outflow rates scale linearly with the adopted radius, so this choice is important.

In the left panel of Figure \ref{fig:MdotSiCorr}, we compare \MdotSi\ with the rate at which silicon is created and injected by SNe in the starburst, i.e., \MdotSiStar\ (both in units of \Msun/yr). We approximate \MdotSiStar\ = 1$\times$10$^{-3}$ SFR assuming Starburst99 models for the Si yield \citep{Leitherer99, Heckman15}. While there is a good correlation, the median ratio of \MdotSi\ and \MdotSiStar\ is only $\sim$ 0.2. This implies inefficient incorporation of the Si ejected by SNe into the warm ionized outflow that we observe.  We will discuss this further in Section \ref{sec:semi-analytic} below.



In the right panel, we show the relationship between $N_{Si}$ and $E(B-V)$. The latter is the dust extinction derived from stellar continuum fits to the FUV spectra for each galaxy (see Section \ref{sec:anc}). There is a positive correlation, but with considerable scatter. A correlation is not surprising since both \NSi\ and dust extinction are related to the total column density of metals. The difference is that \NSi\ only measures metals in the outflow, while the extinction includes dust in both the outflow and the static ISM. Furthermore, the scatter is also not surprising. For one thing, \NSi\ is derived from partial covering fraction models, while E(B-V) is for a screen that covers the starburst.

To evaluate this further, we can estimate the maximum amount of dust extinction in the outflow using \NSi, assuming a standard Milky Way (MW) dust/metals ratio \citep{Draine11} and assuming the solar ratio of silicon to metals mass. This then implies that the total dust extinction is related to \NSi\ as \citep[see][]{Draine11}:

\begin{equation}\label{eq:NSi}
   N_\text{Si} > E(B-V)_\text{tot} \times 1.9 \times 10^{17} cm^{-2}
\end{equation}

In our sample, we then check the distribution of \NSi/E(B-V)$_\text{tot}$/(1.9 $\times$ 10$^{17}$), and the resulting median is $\sim$ 0.16. This suggests that the dust in our observed outflows only represents $\sim$ 16\% of E(B-V)$_\text{tot}$. Therefore, most of the dust responsible for the observed extinction must reside in the static ISM. This relationship [Equation (\ref{eq:NSi}) with multiplication factor of 0.16 on the right side] is shown as the black line in Figure \ref{fig:MdotSiCorr}. On the right y-axis, we also show the implied upper limits on the dust optical depths in the outflow ($\tau_{1200}$) at 1200\angstrom\ assuming the \cite{Reddy15} extinction law and Equation~(\ref{eq:NSi}). This leads to a typical value of $<\tau_{1200}>$ $\sim$ 0.3 due to dust in the outflow. 




\subsubsection{Total Mass Outflow Rates and Mass Loading Factors}
\label{sec:TotalMassRate}
Similarly to the calculations of \MdotSi\ [see Equation (\ref{eq:Mdot3})], we can derive the total mass outflow rate (\Mdot) as:
\begin{equation}\label{eq:Mdot}
    \begin{aligned}
        \dot{M}_{out}\simeq \Omega \mu m_p R_0 \int \frac{dN_\text{H}}{dv}\times vdv
    \end{aligned}
\end{equation}
where $v$ is the outflow velcity, $dN_\text{H}/dv$ = \Nh(v)\ is the total hydrogen column density per velocity and has been derived in Section \ref{sec:PI}. We use the same values described above for $R_0$ and $\Omega$. In Figure \ref{fig:MdotCorr}, we present the correlations related to \Mdot. 

In the first panel, we show a strong correlation between \Mdot\ and SFR, which has been found in previous publications for low-redshift galaxies \citep[e.g.,][]{Martin05, Rupke05, Heckman15}. For our combined sample, \Mdot\ ranges from 0.3 -- 40 \Msun/yr, and the best linear fit yields that \Mdot\ = 3.8~$\times$~SFR$^{0.41}$ (a sub-linear slope).

In the next two panels, we show the correlations between the mass loading factor (i.e., \Mdot/SFR) and \Vcir\ (and \Vout). Both figures show strong inverse correlations, with the correlation between \Mdot/SFR and \Vcir\ being stronger. The best fit slopes for both figures are $\sim -1.6$ to $-1.7$, intermediate between the expectations for a so-called ``momentum-driven'' and ``energy-driven'' outflows. We will discuss this in Section \ref{sec:discussion}. It is striking that the mass-loading factors reach values of $\sim 10$ in the lowest mass galaxies.


\begin{figure*}
\center
	\includegraphics[page = 1, angle=0,trim={0.1cm 0.8cm 0.2cm 0.3cm},clip=true,width=0.5\linewidth,keepaspectratio]{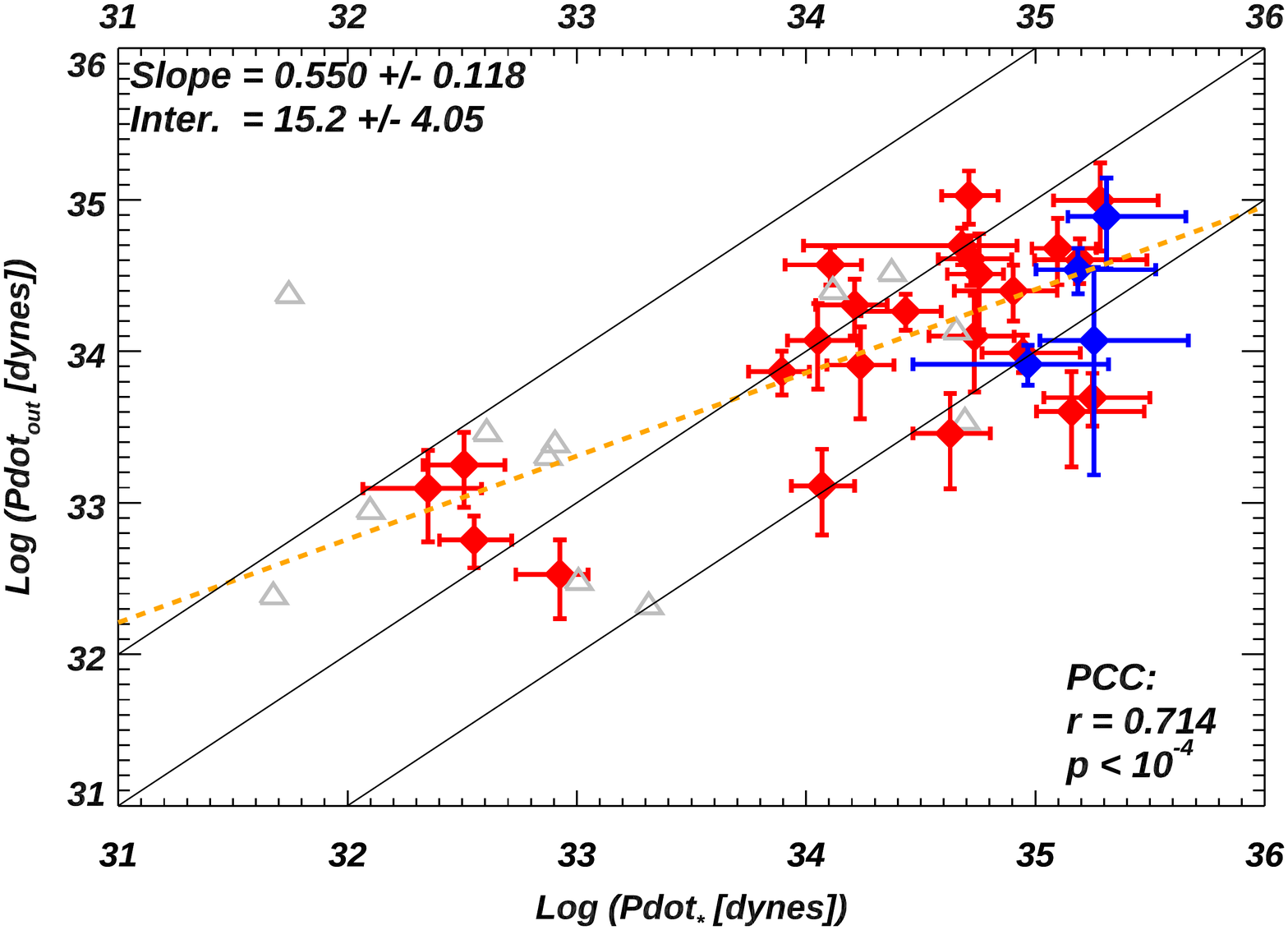}
	\includegraphics[page = 1, angle=0,trim={0.1cm 0.8cm 0.2cm 0.3cm},clip=true,width=0.5\linewidth,keepaspectratio]{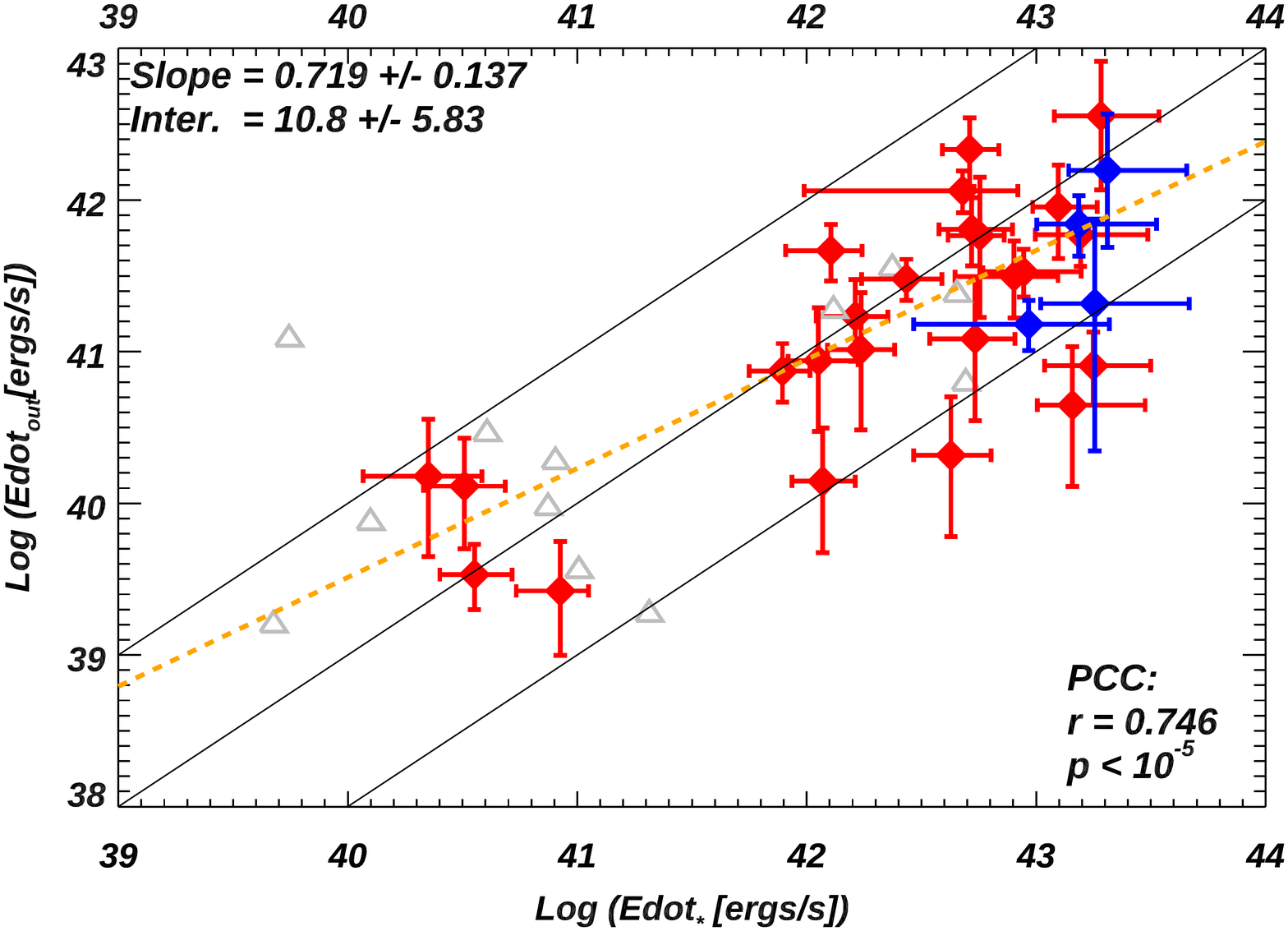}

	\includegraphics[page = 1, angle=0,trim={0.1cm 0.8cm 0.2cm 0.3cm},clip=true,width=0.5\linewidth,keepaspectratio]{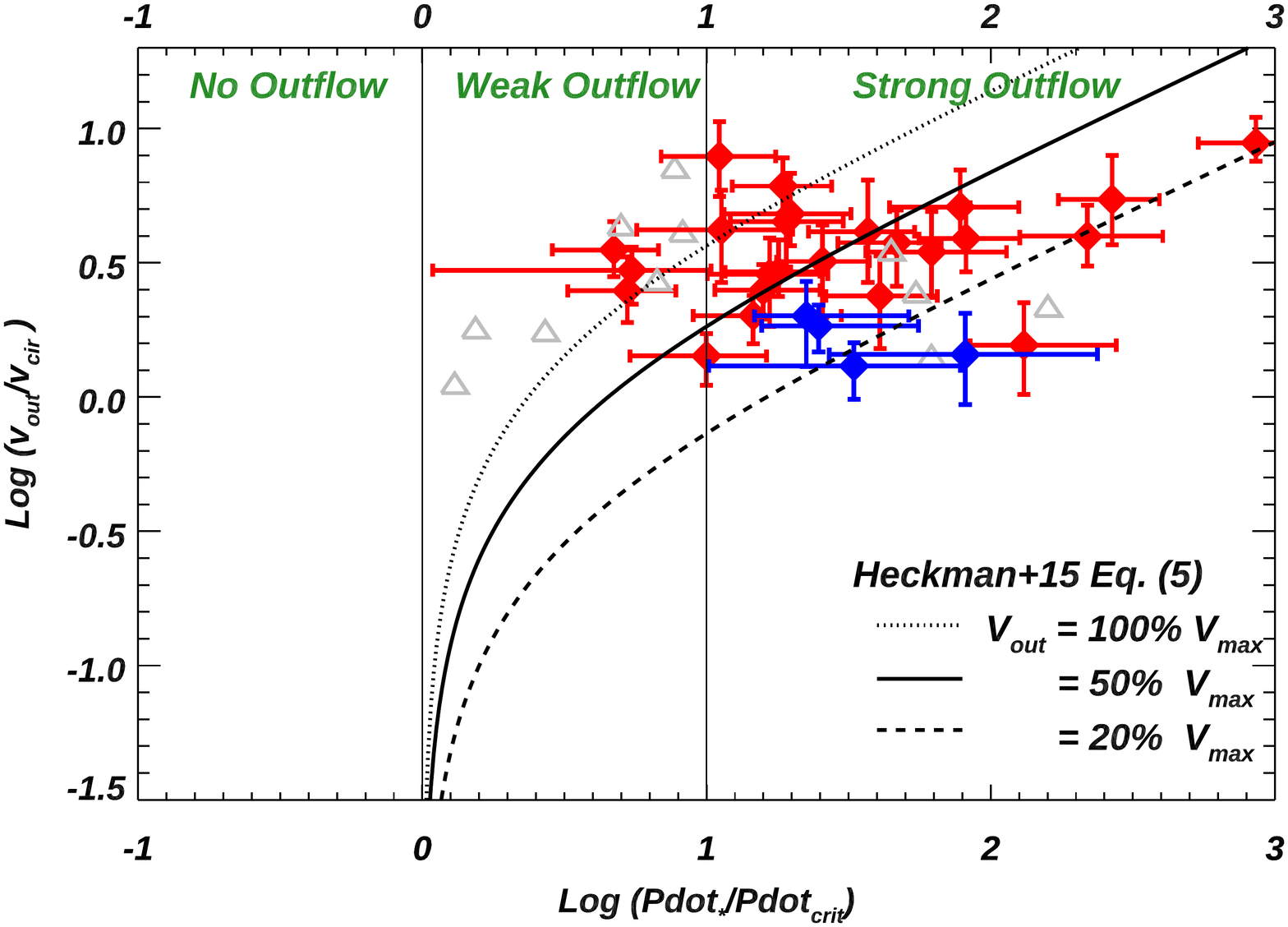}
	
\caption{\normalfont{Correlations comparing \textbf{(top left)} the estimated momentum outflow rate (\Pdot) and \textbf{(top right)} the estimated kinetic energy outflow rate (\Edot) to the rates of momentum and kinetic energy supplied by the starburst. Labels and captions are the same as Figure \ref{fig:VoutCorr}. The three black lines refer to Y = 0.1, 1, 10 X in the left panel and Y = 0.01, 0.1, 1 X in the right panel. See discussion in Section \ref{sec:TotalPERate}.} In the last panel, we present comparisons to a simple analytic model of momentum-driven outflows from \cite{Heckman15} (see Section \ref{sec:compModel}).}
\label{fig:XdotCorr}
\end{figure*}

\subsubsection{Momentum and Kinetic Energy Outflow Rates}
\label{sec:TotalPERate}

Given the calculated \Mdot, we can derive the momentum and kinetic energy outflow rates as:

\begin{equation}\label{eq:PEdot}
    \begin{aligned}
       \dot{p}_{out} &= \int \dot{M}_{out}(v) v = \Omega \mu m_p R_0 \int \frac{dN_\text{H}}{dv}\times v^{2}dv\\
       \dot{E}_{out} &= \int \frac{1}{2}\dot{M}_{out}(v) v^{2} = \frac{1}{2}\Omega \mu m_p R_0 \int \frac{dN_\text{H}}{dv}\times v^{3}dv\\
    \end{aligned}
\end{equation}

The integration over $v$ in \Pdot\ and \Edot\ is necessary since both functions are strongly dependent on $v$. We can then compare \Pdot\ and \Edot\ with the momentum and kinetic energy supplied from starburst, i.e., \PdotStar\ and \EdotStar. As discussed in \cite{Heckman15}, \PdotStar\ is a combination of a hot wind fluid driven by thermalized ejecta of massive stars \citep{Chevalier85} and radiation pressure \citep{Murray05}. In Starburst99 models \citep{Leitherer99}, this leads to \PdotStar\ = 4.6$\times$10$^{33}$ SFR dynes, where SFR is in units of \Msun/yr. Similarly, \EdotStar\ = 4.3$\times$10$^{41}$ SFR erg$/$s. For each galaxy, these values are listed in Table \ref{tab:mea}. 

We present the comparisons in the first two panels of Figure \ref{fig:XdotCorr}, where we find strong positive correlations for both. We draw several solid black lines to represent where Y = 10, 1, 0.1 X in the left panel and Y= 1, 0.1, and 0.01 X in the right panel. For most of galaxies in our combined sample, we have \Pdot\ = 10 -- 100\% \PdotStar, and \Edot\ = 1 -- 20\% \EdotStar. These ranges are similar to the ones reported in \cite{chisholm17}, but both several times smaller than the typical values derived by \cite{Heckman15} based on a much simpler analysis.

In both cases, the slopes are sub-linear, implying that as the SFR increases, less of the available momentum and kinetic energy supplied by stars is being carried by the warm ionized phase of the outflow that we are probing. This trend is stronger for the momentum than for the kinetic energy (see more discussion in Section \ref{sec:discussion}).

We see that there are no cases in which \Edot $>$ \EdotStar. However, there are three galaxies in which \Pdot $>$ \PdotStar. This can be understood in the situation in which the wind fluid that accelerates the gas we see has been substantially mass-loaded but conserves kinetic energy. We will return to this in Section \ref{sec:discussion}.

\begin{table*}
	\centering
	\caption{Fitted Parameters for Scaling Relationships}
	\label{tab:corr}
	\begin{tabular}{l|ll|ll|ll} 
		\hline
		\hline
		Correlations & r$^{(1)}_\text{k}$ & p$^{(1)}_\text{k}$ &r$^{(2)}_\text{PCC}$ & p$^{(2)}_\text{PCC}$& Slope$^{(3)}$ & Intercept$^{(3)}$\\
		\hline
		
\FWHMout\ vs. \Vcir\ 	&	0.55	&	8.88E-06	&	0.79	&	1.19E-07	&	 0.52$\pm$0.22 	&	 1.62$\pm$0.38\\
\FWHMout\ vs. SFR 	&	0.62	&	7.15E-07	&	0.80	&	0.00E+00	&	 0.19$\pm$0.08 	&	 2.38$\pm$0.09\\
\Vout\ vs. \Vcir 	&	0.53	&	2.15E-05	&	0.73	&	2.50E-06	&	 0.65$\pm$0.21 	&	 1.14$\pm$0.37\\
\Vout\ vs. SFR 	&	0.49	&	8.70E-05	&	0.67	&	2.77E-05	&	 0.22$\pm$0.08 	&	 2.12$\pm$0.09\\
\Vout/\Vcir\ vs. sSFR 	&	0.30	&	1.64E-02	&	0.34	&	5.61E-02	&	 0.16$\pm$0.16 	&	 1.84$\pm$1.28\\
\Vout/\Vcir\ vs. SFR/Area 	&	 --0.18	&	1.44E-01	&	 --0.30	&	9.19E-02	&	 --0.08$\pm$0.11 	&	 0.61$\pm$0.12\\
\Mdot\ vs. SFR 	&	0.29	&	3.29E-02	&	0.60	&	6.67E-04	&	 0.38$\pm$0.10 	&	 0.64$\pm$0.11\\
\Mdot/SFR vs. \Vout\ 	&	 --0.37	&	5.03E-03	&	 --0.59	&	9.04E-04	&	 --1.73$\pm$0.43 	&	 4.07$\pm$0.95\\
\Mdot/SFR vs. \Vcir 	&	 --0.52	&	9.16E-05	&	 --0.73	&	9.66E-06	&	 --1.63$\pm$0.35 	&	 2.98$\pm$0.60\\
\Pdot\ vs. \PdotStar 	&	0.38	&	4.44E-03	&	0.71	&	1.97E-05	&	 0.55$\pm$0.12 	&	 15.17$\pm$4.05\\
\Edot\ vs. \EdotStar 	&	0.46	&	6.78E-04	&	0.75	&	5.13E-06	&	 0.72$\pm$0.14 	&	 10.77$\pm$5.83\\
\MdotSi/\MdotSiStar\ vs. \Vout\ 	&	 --0.18	&	1.75E-01	&	 --0.38	&	4.75E-02	&	 --0.88$\pm$0.47 	&	 1.45$\pm$1.06\\
\MdotSi/\MdotSiStar\ vs. \Vcir 	&	 --0.38	&	5.56E-03	&	 --0.57	&	1.95E-03	&	 --0.97$\pm$0.39 	&	 1.16$\pm$0.68\\
N(Si) vs. E(B-V) 	&	0.19	&	1.65E-01	&	0.41	&	3.98E-02	&	 --0.97$\pm$0.39 	&	 1.16$\pm$0.68\\
\MdotSi\ vs. \MdotSiStar 	&	0.47	&	5.82E-04	&	0.79	&	1.19E-06	&	 0.58$\pm$0.11 	&	 --1.44$\pm$0.25\\

		\hline
	\multicolumn{7}{l}{%
  	\begin{minipage}{13cm}%
	Note. --\\
    	\textbf{(1).}\ \ Coefficients for Kendall $\tau$\ test.\\
    	\textbf{(2).}\ \ Coefficients for Pearson Correlation (PCC).\\
    	\textbf{(3).}\ \ Fitted slope and intercept assuming linear correlations between the X and Y values. Note that except for ``N(Si) vs. E(B-V)'', all other correlations are fitted in log-log scale. See Figures \ref{fig:VoutCorr} -- \ref{fig:XdotCorr}.\\
  	\end{minipage}%
	}\\
	\end{tabular}
	\\ [0mm]
	
\end{table*}

\section{Discussion}
\label{sec:discussion}
We begin the discussion by summarizing the most widely used simple model for galactic winds. We then describe several straightforward analytical models for the outflows. This is then followed by a discussion of the scaling relations often adopted for galactic winds in semi-analytic models and numerical simulations of cosmic volumes and we contrast them to our results. Finally, we compare our results to a recent analytic models and high-resolution numerical simulations of multi-phase galactic winds.

\subsection{Theoretical Background}
\label{sec:standardModel}
A very influential model for galactic winds driven by a population of massive stars was that of \citet[]{Chevalier85}. It was simplified, with an assumption of spherical symmetry and neglected gravity. It was also a single-phase model treating only the thermalized ejecta of massive stars (supernovae and stellar winds). As such, it served as a important foundation for more elaborate models to come. The model predicts a very hot region of gas within the starburst. This gas then passes through a sonic point at the starburst radius and then transitions into a supersonic galactic wind. The inputs to the model are the mass and kinetic energy injection rates from the massive stars ($\dot{M}_\star$ and $\dot{E}_\star$) and the radius of the spherical starburst. For the purposes of this paper, the most important outputs of the model are the terminal velocity of this wind, and the outflow rates (of mass, momentum, and kinetic energy, all of which are conserved with radius). These quantities are given as:

\begin{equation}\label{eq:vwind}
\begin{aligned}
v_{wind} = (2 \dot{E}_{wind}/\dot{M}_{wind})^{1/2} 
\end{aligned}
\end{equation}

\begin{equation}\label{eq:outwind}
\begin{aligned}
\dot{M}_{wind} = \beta \dot{M}_\star\\
\dot{E}_{wind} = \alpha \dot{E}_\star\\
\dot{p}_{wind} = (\alpha \beta)^{1/2} \dot{p}_\star
\end{aligned}
\end{equation}

In the above equations $\beta$ accounts for ambient gas that is mixed into the stellar ejecta (so that the total outflow rate in the wind fluid is a factor $\beta$ larger than the rate at which stellar ejecta are created).\footnote{The $\beta$ term is not to be confused with mass-loading factor $\dot{M}_{out}/SFR$, where $\dot{M}_{out}$ is the total mass outflow rate in all phases. For a standard IMF, $\beta = 1$ implies that $\dot{M}_{\star} \sim 0.2$ SFR.} The term $\alpha$ accounts for the effects of radiative losses that drain away the energy carried by the wind fluid. Thus, $\beta \geq$ 1 and $\alpha \leq$ 1. 

{\it It is crucial to emphasize that the outflowing gas described above is not the gas we observe in absorption. In the rest of the discussion we will make this distinction by referring to the former as the wind fluid and the latter as the warm outflow.} 

In this simple standard model, the warm outflow traces pre-existing gas clouds that are being accelerated via momentum and/or kinetic energy transferred from the wind fluid to the clouds \footnote{There is an alternative class of models in which the wind fluid suffers strong radiative cooling and the warm outflow forms directly out of the wind fluid \citep{Thompson16,Schneider18,Lochhaas21}. We will compare these models to the data in Section \ref{sec:Cooling} below.}. The initial idea \citep[]{Chevalier85} was that the clouds were accelerated by the ram pressure of the wind fluid. A challenge has been understanding how clouds survive being shocked by the wind fluid long enough to be accelerated to the observed velocities in the warm outflows \citep{Heckman17a}. 

Note that radiation pressure from the starburst coupled to dust in the clouds can also transfer momentum to the clouds \citep[]{Murray05}. However,  for the choices $\alpha = \beta = 1$, the wind fluid's ram pressure is about four times larger than the radiation pressure ($L_{bol}/c$, where $L_{bol}$ is the bolometric luminosity). We have seen in Section \ref{sec:SiliconRate} above that any dust in the outflow is optically-thin in the far-UV ($\tau_{1200} \sim 0.3$). This means that only $\sim$ 25\% of the momentum provided by UV radiation can be transferred to dust in the outflow. Thus, $P_{ram}/P_{rad} > 16$, and radiation pressure is most likely negligible in driving these outflows.

One relatively new and promising idea is that the momentum transfer from the wind fluid to the clouds occurs in turbulent mixing layers at the interface between the cool cloud and the wind fluid, and (more generally) that mass and momentum can be exchanged in both directions between the clouds and the wind fluid \citep[e.g.,][]{Gronke20,Fielding21}. We will explore such models in Section \ref{sec:semi-analytic} below.

We can now examine some of the scaling relations we have measured in the context of the simple standard model described above of an outflow driven by a fast-moving wind fluid. Observations of the hard X-ray emission in the M 82 starburst favor the choice $\alpha \sim \beta \sim 1$ \citep[]{Strickland09}. In this case, $V_{wind} \sim$ 2700  km s$^{-1}$. This is over an order-of-magnitude higher than the warm outflow velocities we see, but again, we are not directly observing the wind fluid itself.

One key prediction of this model is that the kinetic energy carried by the warm outflow cannot exceed the kinetic energy supplied by the starburst. Our results in Figure \ref{fig:MdotCorr} are fully consistent with this, as the typical ratios of \Edot/\EdotStar\ are 1 to 20 \%. We note that this does not imply a small value for $\alpha$ (which pertains to the wind fluid rather than the warm outflow). Indeed the models and simulations discussed in Sections \ref{sec:ScaCos} and \ref{sec:semi-analytic} below find that the great majority of the kinetic energy is carried by the wind fluid rather than the warm outflow.

The momentum flux in the warm outflow reflects the momentum transferred to it from the wind fluid. We see in Figure \ref{fig:XdotCorr} that the measured momentum fluxes are generally similar to (but smaller than) the predicted momentum flux in the wind fluid, with a median ratio of $\sim$ 30\%. This suggests an efficient transfer of momentum. Interestingly, there are several cases in which $\dot{p}_{out} >$ \PdotStar. This is not in conflict with the model of an outflow driven by the wind fluid. That is, in the case where $\alpha \times \beta > 1$, the momentum flux in the wind fluid can exceed the momentum flux from the starburst by a factor $(\alpha \times \beta)^{1/2}$ (see Equation set (\ref{eq:outwind})). This would correspond to a case in which radiative losses in the wind fluid are negligible ($\alpha \sim $ 1) but the wind fluid is significantly contaminated by ambient gas mixed into it ($\beta >>1 $). This would of course require an efficient transfer of this momentum from the wind fluid to the clouds in the warm outflow. 

\subsection{Comparison to a Simple Analytic Models of Momentum-Driven Outflows}
\label{sec:compModel}

We first compare our observations with a model of a population of clouds acted upon by a combination of outward momentum from the starburst and the inward force of gravity, as discussed in \cite{Heckman15}. They have derived a critical momentum flux required to drive the outflow (see their Equation 3):

\begin{equation}\label{eq:Pcrit}
    \begin{aligned}
       \dot{p}_{crit} &= 4\pi R_0 N_\text{H} \mu m_{p} v^{2}_\text{cir} \\
                      &= 10^{32.9}\text{dynes} \times \frac{R_0}{kpc} \times \frac{v_\text{cir}}{100 km s^{-1}} \times \frac{N_\text{H}}{10^{20} cm^{-2}}
    \end{aligned}
\end{equation}

The resulting \PdotCrit\ values are listed in Table \ref{tab:mea}. In the last panel of Figure \ref{fig:XdotCorr}, we compare the normalized outflow velocity, i.e., \Vout/\Vcir, versus the normalized outflow momentum flux, i.e., \Pdot/\PdotCrit. The two black vertical lines split the figure into three different regimes for outflows: (1) \Pdot/\PdotCrit\ $<$ 1, where no outflow is expected to be driven due to the lack of momentum inputs. This is consistent with the fact that no outflows from our combined sample fall in this region. (2) 1 $<$ \Pdot/\PdotCrit\ $<$ 10, where we expect relatively weak outflows are driven in this regime. A few of our observed outflows fall into this region. (3)  \Pdot/\PdotCrit\ $>$ 10, where relatively strong outflows should be driven. We find most of our observed outflows fall into this last region of \Pdot/\PdotCrit. This is as expected since weaker outflows are harder to detect or more easily marked as ``no outflows'' due to various issues mentioned in Section \ref{sec:2GFits}. 

In in last panel of Figure \ref{fig:XdotCorr}, we also show the expectations from Equation (5) of \cite{Heckman15}, where they predict the maximum velocity of an outflowing cloud (\VMax). This velocity corresponds to the radius at which the outward force due to ram and radiation pressure equals the inward force due to gravity (and beyond which the cloud begins to decelerate). In dotted, solid, and dashed curves, we overplot this equation on our data where we assume the observed outflow velocity \Vout\ = 100\%, 50\%, and 20\% of \VMax, respectively. We find that most of our confirmed outflows are consistent with the \Vout\ = 50\%~\VMax\ curve. This can be explained as a ``natural choice'', i.e., \VMax\ = \Vout~+~0.5~\FWHMout\ $\sim$ 2\Vout\ (see Section \ref{sec:OutflowGeneral}). Unlike \citet{Heckman15}, we do not probe the regime of low \Pdot/\PdotCrit\ very well, and so do not see its correlation with \Vout/\Vcir. In a future paper, we will analyze the \citet{Heckman15} FUSE data using the methods in this paper and revisit this plot.

Next we consider the simple model of wind-blown bubble driven into the ISM/CGM by the momentum flux of the wind fluid.\footnote{We have also considered the case in which the bubble expansion is driven by the kinetic energy of the wind fluid. This model is a poorer fit the data, so we omit a discussion here.} In this model,  the gas we see in absorption is swept-up ambient gas at the surface of the expanding bubble. For simplicity, we use the simple spherically-symmetric model described in \citet{Dyson89} in which the ISM is treated as having a uniform density and gravity is neglected. The latter is supported by the large values of \Pdot/\PdotCrit\ seen in Figure \ref{fig:XdotCorr} above.

In this case, the predicted radius and expansion velocity are given by:

\begin{equation}
    \begin{aligned}\label{eq:bubble}
    R_s &= 2.16~\dot{p}_{35}^{1/4} n_0^{-1/4} t_7^{1/2} kpc\\
        &= 1.77~SFR_{10}^{1/4} n_o^{-1/4} t_7^{1/2} kpc\\
    V_s &= 106~\dot{p}_{35}^{1/4} n_0^{-1/4} t_7^{-1/2} km/s\\
        &= 87~SFR_{10}^{1/4} n_0^{-1/4} t_7^{-1/2} km/s
    \end{aligned}
\end{equation}
Here $\dot{p}_{35}$ is the momentum flux in the wind in units of 10$^{35}$ dynes, $SFR_{10}$ is the star formation rate in units of 10 M$_{\odot}$ yr$^{-1}$) (the median value for the outflow sample), $n_o$ is the H density (cm$^{-3}$) in the region into which the bubble expands, and t$_7$ is the time since the expansion began in units of 10$^7$ years, similar to the ages derived from fits of SB99 models to the COS data (Senchyna et al. 2022, in preparation).

It is intriguing that the predicted dependence of $V_s \propto \dot{p}_{35}^{1/4}$ (i.e., $V_s \propto SFR^{1/4}$) is very similar to the measured slope between \Vout\ and SFR ($\sim$ 0.24, see Figure \ref{fig:VoutCorr}). In the second panel of Figure \ref{fig:VoutCorr}, we show the best-fit relation with $n_0 t_7^2$ treated as a free parameter (green dashed line). We find a best-fit value of $n_0 t_7^2$ = 2.8 $\times 10^{-2}$ cm$^{-3}$ yr$^{2}$. Assuming $t_7 \sim$ 1, this implies low-density gas (i.e., 2.8 $\times 10^{-2}$ cm$^{-3}$), which should be located well outside the starburst region.

We can go further with this model. The predicted column density through the bubble wall is given by $n_0 R_s/3$. For the above value of $n_0$ and $t_7$, Equation (\ref{eq:bubble}) yields $N_H = 1.5 \times 10^{20} \dot{p}_{35}^{1/4}$ cm$^{-2}$, which is only a factor of two to three below the typical values we measure (see Figure \ref{fig:NhDist}). However, we do not see the predicted decline in $N_H$ with decreasing $\dot{p}_{*}$ in the data.

Finally, we can compute the mass-outflow rate in this model (the time averaged rate at which ambient gas has been swept-up by the expanding bubble). Using the equations above, and taking $n_0 = 2.8 \times 10^{-2}$ cm$^{-3}$, we get:

\begin{equation}
    \begin{aligned}
    \dot{M} &= 54~\dot{p}_{35}^{3/4} t_{7}^{1/2} M_{\odot} yr^{-1}\\
             &= 31~SFR_{10}^{3/4} t_7^{1/2} M_{\odot} yr^{-1}
    \end{aligned}
\end{equation}

These rates are about three times larger than we have estimated, and would lead to a steeper slope in the dependence of $M_{\odot}$ on SFR (0.75) than we observe (0.4, see Figure \ref{fig:MdotCorr}). 
 
We conclude that a simple wind-driven bubble model has some success in matching the data.
However, it cannot be a complete model: for an expanding bubble, the observed absorption-lines will only come from the part of the bubble located directly along the line-of-sight to the starburst. This will result in a narrow blueshifted absorption-line with \FWHMout $<<$ \Vout. This is inconsistent with the data (see Figures \ref{fig:2GFits}, \ref{fig:VoutDist}, and \ref{fig:FEFits}).

\subsection{Theoretical Scaling Relations in a Cosmological Context}
\label{sec:ScaCos}
There is a substantial literature in which the properties of galactic winds in a cosmological context are modeled using simple physically-motivated prescriptions \citep[e.g.,][]{Somerville15,Naab17}. These include both semi-analytic models and numerical simulations in which the winds are not modeled {\it ab initio} (due to insufficient spatial resolution), but rather, are implemented using various sub-grid recipes.

Here we briefly summarize these popular prescriptions and compare them to the data. For the most part, these models assume that there is a linear proportionality between \Vout\ and \Vcir\ \citep[e.g.,][]{Guo11,Dave13}. This is not fully consistent with the results shown in Figure \ref{fig:VoutCorr}, which imply \Vout$ \propto $ $V_\text{cir}^{0.6}$. At face value, this would suggest that the outflowing gas is more likely to escape from the low-mass galaxies.

There are several different prescriptions for how the mass-loading factor $\dot{M}_{out}/SFR$ should vary with \Vout. Since these models assume a linear relationship between \Vout\ and \Vcir, the mass-loading factor will scale the same with both velocities. In one case the prescription is that outflows all carry the same fraction of the momentum supplied by the starburst. In this case, $\dot{M}_{out}/SFR \propto V_{out}^{-1} \propto V_{cir}^{-1}$ \citep[]{Oppenheimer08,Dutton10}. An analogous assumption is that the outflows instead carry a fixed fraction of the kinetic energy supplied by the starburst \citep[]{Baugh05,Somerville08}, leading to  $\dot{M}_{out}/SFR \propto V_{out}^{-2} \propto V_{cir}^{-2}$. Finally, some models assume that the mass-loading factor is a constant \citep[e.g.,][]{Dave13,Ford16}.

As seen in Figure \ref{fig:MdotCorr}, the observed scaling relations of the mass-loading factor with \Vout\ and \Vcir\ are both intermediate between these two cases (slope $\sim$ --1.6 to --1.7).  Note that \citet{Heckman15} found a shallower slope. The main difference is that the CLASSY sample reaches much lower values of \Vcir\ (providing better dynamic range).

\subsection{Comparisons to an Semi-Analytic Model of a Multi-phase Outflow}
\label{sec:semi-analytic}
In this section, we compare our results to a new semi-analytic model for multi-phase galactic winds \citep[]{Fielding21}. This model starts with the \citet{Chevalier85} model for the wind fluid, but adds radiative cooling and gravity. More importantly, the model incorporates the bi-directional exchange of mass, metals, momentum, and energy between the wind fluid and a population of much denser clouds. These transfers take place in turbulent mixing layers arising as the wind fluid flows along the cloud surface. It is the interaction between the wind fluid and the clouds that lead to the warm outflow we observe.

The case considered is based on M 82, with \Vcir\ = 150 km s$^{-1}$, SFR $ = 20$ M$_{\odot}$ year$^{-1}$, and a starburst radius of 300 pc. The free parameters that are explored are what we have called $\beta$ for the wind fluid, which is varied between 1 and 5, the {\it initial} mass-loading factor (\Mdot/SFR) of the warm phase arising as the wind fluid interacts with ambient clouds, which is varied between 0.2 and 0.5, and the mass of an individual cloud, which varies from $10^2 - 10^6$ M$_{\odot}$. 

The outflow velocity in the model depends most strongly on the cloud mass, with the most massive clouds being accelerated to the lowest velocity. Our results in Figure \ref{fig:VoutCorr} imply that \Vout\ will be 260 km $^{-1}$ based on the SFR in the model and 360 km s$^{-1}$ based on \Vcir\ in the model. These velocities are most consistent with the most massive clouds in the models (10$^6$ M$_{\odot}$), which reach \Vout\ $ =$ 350 (450) km s$^{-1}$ at radii of 1 (10) kpc.

The final mass-loading factor for clouds in the model depends on both its initial value, and the subsequent transfer of mass between the wind fluid and clouds as they flow out. Our results in Figure \ref{fig:MdotCorr} imply that an M82-like galaxy would have a mass-loading factor of $\sim$ 0.6 based on SFR and $\sim$ 0.3 based on \Vcir. This is in the range adopted by the models for the initial mass-loading factor. In the models, the mass-loading factor in the warm outflow only drops significantly with radius for the case in which $\beta = 1$ (uncontaminated wind fluid) {\it and} the cloud masses are small ($<10^4$ M$_{\odot}$). In these cases, the clouds are shredded and incorporated into the wind fluid. These particular models are not consistent with the data.

The momentum flux carried by the clouds in the models shows little radial dependence and ranges from $\sim 30$ to 100\% of \PdotStar\ for $\beta$ from $\sim$ 1 to 5, respectively. The $\beta = 1$ model is therefore a better fit to our data (Figure \ref{fig:XdotCorr}). Finally, the kinetic energy flux in the clouds depends most strongly on $\beta$, being in the range 20 to 40\% \EdotStar\ for $\beta = 5$ {\it vs.} only 4 to 10\% for $\beta = 1$. Our median value of $\sim$5\% (see Figure \ref{fig:XdotCorr}) again favors the $\beta =1$ models.

Next, we can consider the question of the mixing of the metals created and dispersed by SNe and carried by the wind fluid with the material observed in the warm outflow. The \citet{Fielding21} models do not explicitly calculate that quantity, but they do compute the separate radial dependences of the metallicity of the wind fluid and of the clouds for the different models. They assume an initial metallicity of twice solar in the wind fluid and solar in the clouds. For strong metal mixing, the wind and cloud metallicities converge at large radii. This is the case for all cloud masses in the $\beta =5$ model, but only for the lower mass clouds in the $\beta =1$ model. In Section \ref{sec:SiliconRate}, we have found that the observed outflow rate of Si in the warm gas has a median value of only about 20\% of the Si creation rate by SNe (consistent with most of the newly-created Si residing in the wind fluid). This result is at least most qualitatively consistent with the $\beta = 1$ model with massive clouds.

These models also make different predictions for the total hydrogen column density (\Nh) in clouds along the line of sight (D. Fielding, private communication). For the $\beta =1$ models, $N_H$ increases from about 2 $\times 10^{20}$ to 8 $\times 10^{20}$ cm$^{-2}$ as the individual cloud masses increase from $10^2$ to $10^6$ M$_{\odot}$. For the $\beta = 5$ model the corresponding range is $\sim$ 1 to 2 $\times 10^{21}$ cm$^{-2}$. For our sample we find a median value of $N_H = 5 \times 10^{20}$ cm$^{-2}$, which agrees best with the $\beta =1$ model and massive clouds.

In summary, at least for parameters appropriate to an M82-level starburst, our data are most consistent with the models with $\beta =1$ (a fast wind that has not been contaminated) and with relatively massive clouds ($\sim 10^5$ to 10$^6$ M$_{\odot}$). In the future, it will be interesting to compare our data to new models that have been tuned to cover the ranges of the CLASSY sample in \Vcir, SFR, and starburst size.

\subsection{Comparisons to a Numerical Simulation of a Multi-phase Outflow}
\label{sec:numerical}
First, we compare our results to the highest-resolution numerical simulations of galactic winds currently available \citep{Schneider20}. Their simulations are designed to roughly correspond to M 82 in terms of SFR and \Vcir, but the starburst is modeled as a population of massive star-forming clumps distributed within a radius ($R_*$) of 1 kpc. The results of the simulations are shown for gas in three temperature regimes: 1) a hot phase, i.e., $T > 5 \times 10^5$~K, 2) an intermediate phase with 5 $\times 10^5 > T > 2 \times 10^4$ K, and 3) a cool phase with $T < 2\times 10^4$ K. We focus here on the results shown for a time of 35 Myr (since they significantly drop the input SFR at later times).

The predicted values for \Vout\ for the cool phase rise rapidly with radius to a value of $\sim$ 500 km s$^{-1}$ at 4 kpc (4 $R_*$) and then gradually increase to about 750 km s$^{-1}$ at 10 kpc. Our results in Figure \ref{fig:VoutCorr} imply that \Vout\ will be 260 km $^{-1}$ based on their assumed SFR and 360 km s$^{-1}$ based on their assumed \Vcir, roughly half the values from the simulations.

The outflow rates presented in the figures in \citet{Schneider20} refer only a bi-conical region with an opening half-angle of 30$^{\circ}$. They find that integrating over $4\pi$ ster leads to total outflow rates that are three times larger. We use this scale factor to compare to our results.
This yields a mass-loading factor in the cool phase that peaks at about 10\% at radii between 2 and 5 kpc and then drops at larger radii. This is significantly smaller than our results in Figure \ref{fig:MdotCorr}, which imply that an M82-like galaxy would have a mass-loading factor of 0.6 based on SFR and 0.3 based on \Vcir. 

Similarly, the momentum flux in the cool phase in the simulations is about 10 to 20\% of \PdotStar\ (falling at large radii). This is a bit smaller than the median value of $\sim$ 30\% in the data (Figure \ref{fig:XdotCorr}). Finally, the kinetic energy flux in the cool phase in the simulations is in the range of $\sim$ 2 to 4\% of \EdotStar, compared to a median value of 5\% in our data (Figure 11). As in the analytic models by \citet{Fielding21}, the lion's share of the kinetic energy is in the hot phase (i.e., the wind fluid).

Overall, the results in the simulations of \cite{Schneider20} are in reasonable agreement when compared with the galaxies studied here, but overall produce outflows of the cool gas that are too fast and do not carry enough mass, momentum, and kinetic energy (all by factors of $\sim 2$). The simulations most resemble the analytic models of \citet{Fielding21} in the regime of low cloud masses. 

\begin{figure}
\center
	\includegraphics[page = 1, angle=0,trim={0.1cm 0.8cm 0.1cm 0.3cm},clip=true,width=1.0\linewidth,keepaspectratio]{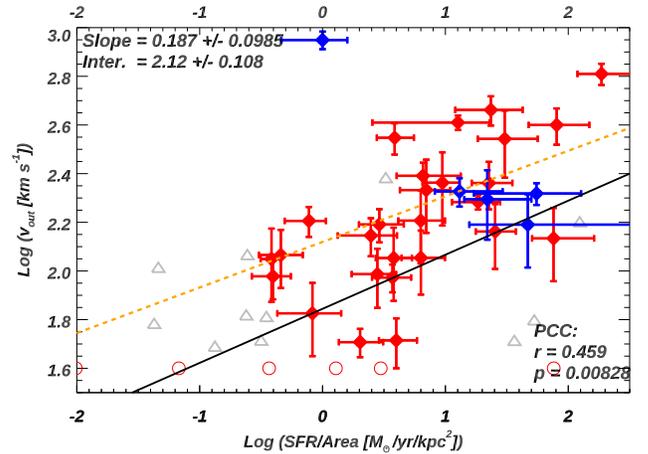}

\caption{\normalfont{Correlations between \Vout\ and the SFR surface density for galaxies in our combined sample. Labels and captions are the same as Figure \ref{fig:VoutCorr}. The black solid line represents the predictions from multiphase outflow simulations presented in \cite{Kim20}. The slopes match, but the model velocities are about a factor of 3 -- 5 smaller than the measured ones (see Section \ref{sec:numerical}).} }
\label{fig:Vout_sSFR}
\end{figure}

There are several other publications that present simulations of multiphase galactic winds and predict various scaling relationships. We briefly discuss two of them below, which are both built on Athena MHD code \citep{Stone08}: 1). \cite{Tanner17} adopts 3D hydrodynamical simulations assuming similar models as Equations (\ref{eq:vwind}) and (\ref{eq:outwind}) to predict the outflow velocities for various silicon ions (\Sii\ -- \sixiii). They find velocities increase significantly ($\gtrsim$ a factor of 2) from \Siii\ to \siiv, which is not seen in our galaxies (Figure \ref{fig:SiIIvsSiIV}). They also predict that \Vout\ versus SFR correlation has an abrupt flattening above SFR = 5 -- 20 \Msun/yr defined by the hot wind velocity. We do not see any statistically significant evidence of this flattening in SFR > 10 \Msun/yr in Figure \ref{fig:VoutCorr}. 2). \cite{Kim20} presents a suite of parsec-resolution numerical simulations for multiphase outflows and shows that \Vout\ correlates with the surface density of SFR with a slope $\sim$ 0.2, which is similar to what we get in our combined sample (Figure \ref{fig:Vout_sSFR}). However, the normalization of the outflow velocity from the simulations is too low compared to the data by a factor of $\sim$ 3 -- 5. These simulations predict that only about 10\% of the metals injected by supernovae are incorporated in the warm ionized phase, with the majority being in the hot phase. This is consistent with our results.

All-in-all, there exist both consistencies and differences between our observations and various simulations. To gain more insights into the physics of galactic outflows and their correlations with galaxy properties, incorporation of observations into the recipes of simulations are necessary, which is one of our main goals in future papers. 

\begin{figure*}
\center
	\includegraphics[page = 1, angle=0,trim={0.1cm 0.8cm 0.1cm 0.3cm},clip=true,width=0.5\linewidth,keepaspectratio]{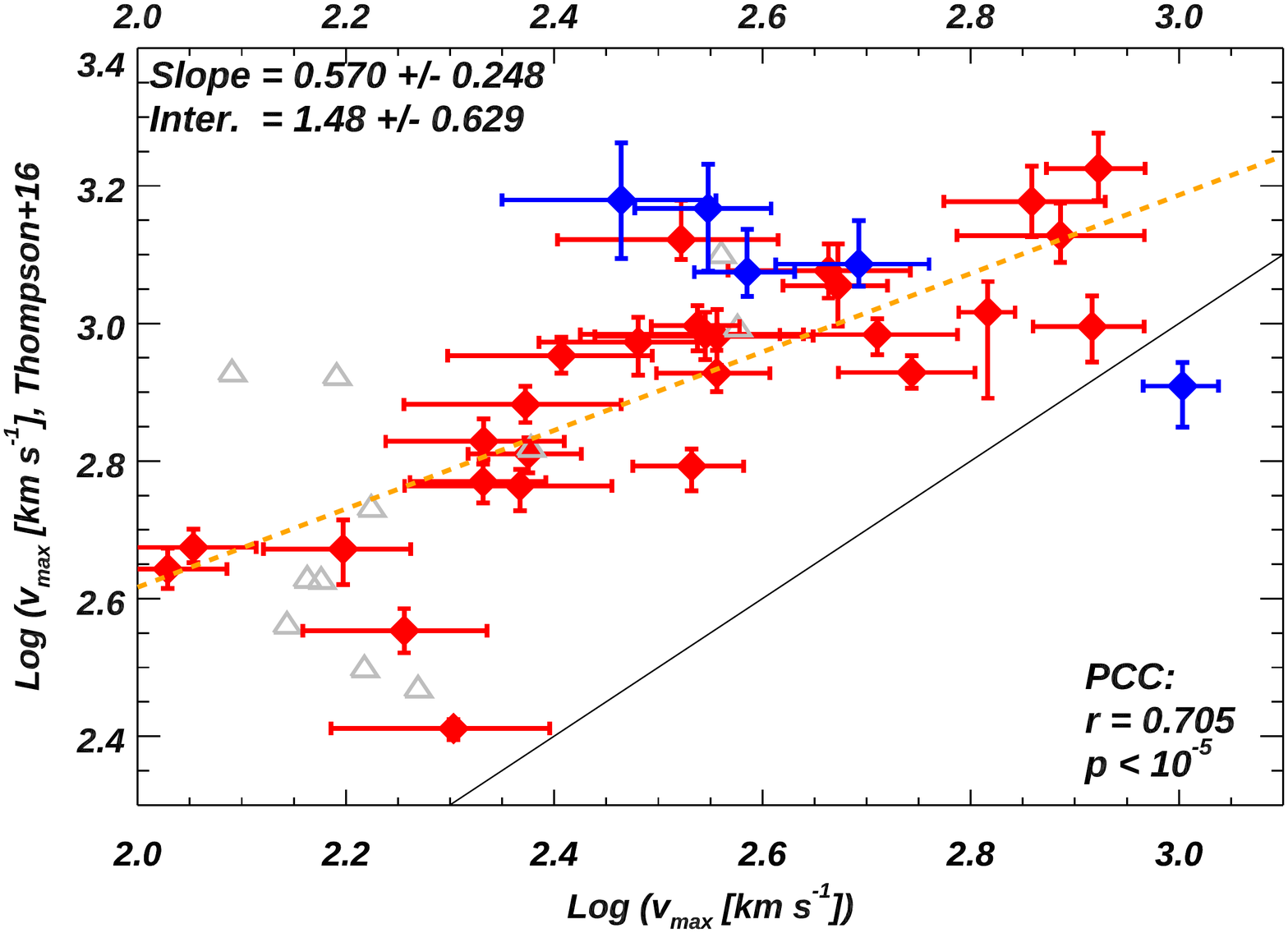}
	\includegraphics[page = 1, angle=0,trim={0.1cm 0.8cm 0.1cm 0.3cm},clip=true,width=0.5\linewidth,keepaspectratio]{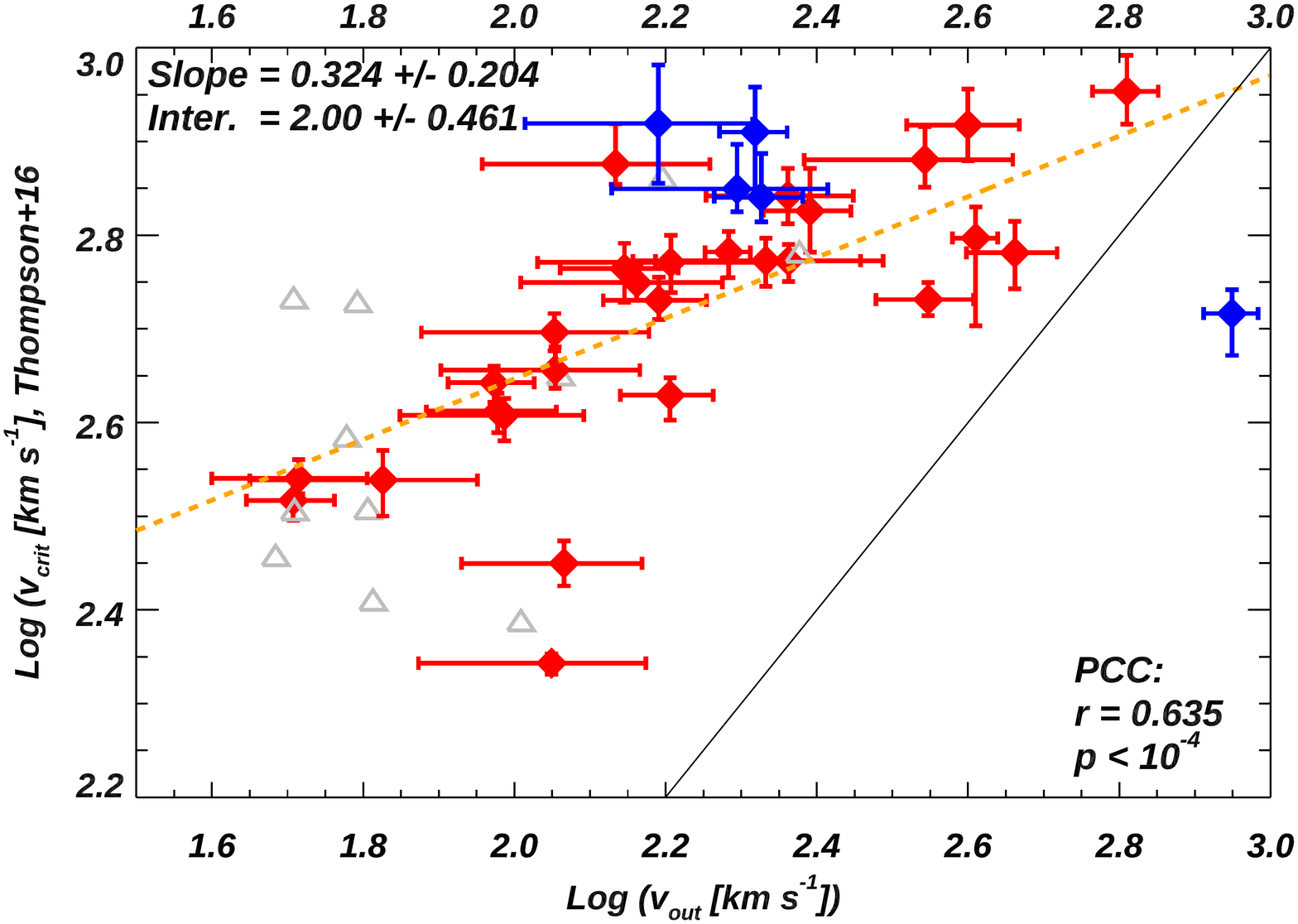}

\caption{\normalfont{Comparisons of predicted outflow velocities from models of cooling winds \citep{Thompson16} to our measured outflow velocities. \textbf{Left:} Comparisons between the predicted maximum outflow velocity [see Equation (\ref{eq:vmax})] to the measured maximum outflow velocity. \textbf{Right:} Comparisons between the predicted critical outflow velocity [see Equation (\ref{eq:vcrit})] to the measured outflow velocity. For both panels, the orange dashed lines represent the best fitting linear correlations, while the black lines show where y = x. The models substantially over-predict the values of the observed outflow velocities. See discussion in Section \ref{sec:Cooling}.} }
\label{fig:Vpred}
\end{figure*}

\subsection{Comparisons to Models of Cooling Winds}
\label{sec:Cooling}
One interesting idea is that the observed outflowing warm gas is produced directly through radiative cooling of the wind fluid. This can happen when there is enough contamination of the wind fluid ($\beta >>1$) that the radiative cooling time of the poisoned wind fluid is shorter than the outflow time of the fluid \citep[e.g.,][]{Wang95,Martin15,Thompson16}. 

\citet{Thompson16} consider two cases. One is a case in which $\beta$ is sufficiently large that radiative cooling becomes important somewhere in the wind (at a radius $>>$ the starburst radius of $R_*$). This leads to a predicted maximum outflow velocity given by:

\begin{equation}
\begin{aligned}\label{eq:vmax}
V_{max} \sim 1250~km/s~ [\alpha~\xi SFR_{10}/((\Omega/4\pi)R_{*,0.3})]^{0.18}
\end{aligned}
\end{equation}

Here $\xi$ is the metallicity of the wind fluid (relative to solar), $SFR_{10}$ is in units of 10 M$_{\odot}$/yr, and $R_{*,0.3}$ is the starburst radius in units of 0.3 kpc.

They also consider a case in which $\beta$ is large enough for cooling to occur at $R_*$.This yield a critical outflow velocity given by:

\begin{equation}
\begin{aligned}\label{eq:vcrit}
V_{crit} \sim 720~km/s~ [\alpha~\xi  SFR_{10}/((\Omega/4\pi)R_{*,0.3})]^{0.135}
\end{aligned}
\end{equation}

In Figure \ref{fig:Vpred}, we compare these two velocities to our observed values of $V_{max} = V_{out} + 0.5 FWHM_{out}$ and $V_{out}$, respectively. We see that in both cases the predicted velocities are significantly larger than the observed values (by typical factors of 3 to 10). The discrepancies are particularly large for the slower outflows. In simple terms, these slow outflows require such large values of $\beta$ that the outflows would be still-born inside the starburst. Also, these models have no natural explanation for the correlation between outflow velocity and galaxy circular velocity (see Figure \ref{fig:VoutCorr}).

The cooling wind models was explored in more detail in high-resolution numerical simulations of a multi-phase starburst-driven wind modeled on the prototype of M 82 \citep{Schneider18}. The simulations produce outflow velocities of the warm phase of $\sim 10^3$ km/s for all cases considered (roughly three to four times higher than our scaling relations in Figure \ref{fig:VoutCorr} for an M82-like starburst). We conclude that these cooling wind models and numerical simulations are not a good match to our data.

\begin{figure}
\center
	\includegraphics[page = 1, angle=0,trim={0.1cm 0.8cm 0.1cm 0.3cm},clip=true,width=1.0\linewidth,keepaspectratio]{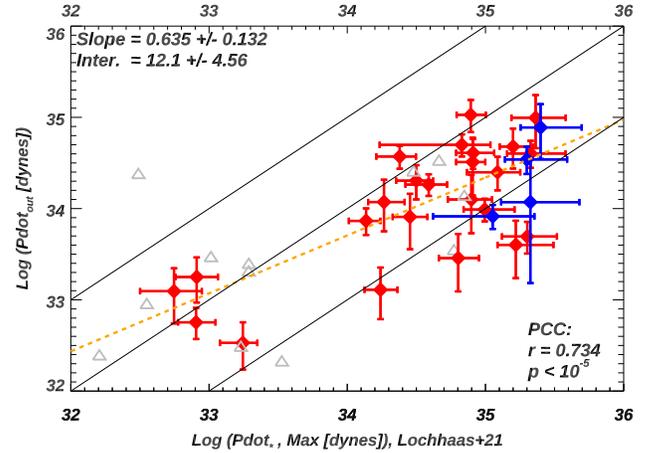}

\caption{\normalfont{Comparisons between the measured momentum flux of outflows (\Pdot) to the predicted maximum wind momentum flux ($\dot{p}_{max}$) from \cite{Lochhaas21}. The three lines show Y = 10, 1, 0.1 X. There is only one galaxy that have \Pdot\ $>$ $\dot{p}_{max}$, which is similar to Figure \ref{fig:XdotCorr}. See Section \ref{sec:Cooling} for discussion. } }
\label{fig:Pdotpred}
\end{figure}

Most recently, \cite{Lochhaas21} examined the amount of momentum that can be carried by a warm outflow in the context of the contamination of the wind fluid ($\beta > 1$) and the resulting radiative losses. Recall that in the simple model described in Section \ref{sec:standardModel}, the amount of momentum flux that can be carried by the wind fluid is $\dot{p}_{wind} = (\alpha \beta)^{1/2} \dot{p}_*$. \cite{Lochhaas21} calculated the maximum allowed value of the product of $\alpha \beta$ under the conditions of increasingly large $\beta$ leading to increasing strong radiative cooling ($\alpha <1$). They find the maximum wind momentum flux to be:

\begin{equation}
\begin{aligned}
\dot{p}_{max} = 7.2 \times 10^{34} (R_{*,0.3})^{0.14} (\alpha SFR_{10})^{0.86} dynes
\end{aligned}    
\end{equation}

We compare our estimates of the outflow momentum flux with $\dot{p}_{max}$ in Figure \ref{fig:Pdotpred}. We find that only one galaxy lies significantly above this relationship. In fact, this plot is very similar to our results in Figure \ref{fig:XdotCorr}, which simply compared the observed momentum flux to that provided by the starburst. The agreement means that $\dot{p}_{max}$ is generally similar to $\dot{p_*}$. That is to say, actual outflows seem to carry the maximum possible momentum allowed for them without being quenched by radiative cooling, and this maximum flux is very similar to the amount input by the starburst ($\alpha\beta \sim 1$). 

\section{Conclusion}
\label{sec:conclusion}

We have reported here the results of our analysis of starburst-driven galactic outflows of warm ionized gas. This was based on data for 45 galaxies taken from the COS Legacy Archive Spectroscopic SurveY (CLASSY), augmented by five additional starbursts with COS data of similarly-high quality and under the same selection criteria. The properties of the outflows were based on fitting the UV resonance absorption-lines, and in particular, using five Si II multiplets and the Si IV 1393, 1402 doublet to derive the column density and covering fraction of these ions as a function of outflow velocity. CLOUDY models were used to derive total Si column densities and these were converted into H column densities using the metallicities derived from the nebular emission-lines. 

The key parameters obtained from this analysis are the mean outflow velocity (\Vout) and the Full Width at Half Maximum (\FWHMout) of the blue-shifted absorption-lines, the total Si and H column densities (\NSi\ and \Nh), and the outflow rates of Si (\MdotSi), mass (\Mdot), momentum (\Pdot), and kinetic energy (\Edot). 

We then examine the scaling relationships between the outflow properties and those of the starburst and its host galaxy. The principle results are as follows:
\begin{itemize}
    \item
    Outflows were detected in roughly 90\% of the sample. This implies that the outflows in starburst galaxies cover most of $4\pi$ steradian (they cannot be well-collimated). 
    \item
    The average value of the covering factor is 0.64, meaning that the effects of partial covering need to be explicitly determined to measure column densities from optically-thick absorption-lines. The values for the covering fractions are very similar for \Siii\ and \siiv.
    \item 
    While the values for \Vout\ are quite consistent among all the transitions we measure, there is significant scatter in the values for \FWHMout\. In particular, we find a systematic trend for \FWHMout\ to decrease as we move from the most to the least optically-thick \Siii\ transitions. This implies that the highest column densities in the outflow are near the characteristic outflow velocity given by \Vout. 
    \item
    We found highly significant correlations of both \Vout\ and \FWHMout\ with star-formation rate (SFR), galaxy stellar mass (\Mstar), and galaxy circular velocity (\Vcir). The best-fit relationship is \Vout\ $\propto$\ \Vcir$^{0.6}$, and the ratio of \Vout/\Vcir\ being $\sim$ 6 and $\sim 2$ for the lowest and highest mass galaxies, respectively.  
    \item
    We found that the outflow rate of Si is (on-average) only about 20\% of the rate at which Si is created and ejected by supernovae in the starburst. We conjecture that most of this ``missing'' Si is in the form of a hotter and more highly-ionized phase of the outflow than what we probe with these data.
    \item 
    Assuming a normal dust-to-gas-phase-metals ratio, the observed Si column densities implied that there is very little dust extinction associated with the observed outflows (mean FUV optical depth of $\sim$ 0.3). Most of the observed extinction is produced by dust in the static ISM.
    \item 
    The average total hydrogen column density (\Nh) is $\sim 10^{20.7}$ cm$^{-2}$, and that neutral hydrogen [N(\hi)] only constitutes 0.1 to 1\% of \Nh. The dominant ion for silicon is \siiii, and 90 to 99\% of the \Siii\ arises in the ionized gas. Based on the derived \Nh(v), we find the column densities peak near \Vout, while the broad wings of outflow profiles have significantly smaller values of \Nh.
    \item
    We found a highly significant correlation between the mass outflow rate (\Mdot) and the SFR, but with a shallow slope (\Mdot $\propto SFR^{0.4}$). Hence the so-called mass-loading factor of the outflow is \Mdot/SFR $\propto SFR^{-0.6}$. We also found that the mass-loading factor is a steep inverse function of both \Vout\ and \Vcir\ (slope $\sim -1.6$), with mass-loading factors of $\sim 10$ in the lowest mass galaxies. Together with the third result above, this supports the idea of a mass-dependent impact of outflows on the evolution of galaxies.
    \item
    We found strong correlations between the rates at which the outflows carry momentum and kinetic energy (\Pdot\ and \Edot) and  the rates at which the starbursts supply momentum and kinetic energy (\PdotStar\ and \EdotStar). The median values are \Pdot\ $\sim$ 30\% \PdotStar\ and \Edot\ $\sim$ 5\% \EdotStar. 
\end{itemize}

We then compared these results to various theoretical models of galactic winds driven by starbursts. We began with a description of the most widely-used model of galactic winds due to \citet{Chevalier85}. In this model the stellar ejecta in the starburst create very hot gas which flows out to create a supersonic wind. This fast-moving and tenuous ``wind-fluid'' interacts with ambient gas which it accelerates to create the warm ionized outflows we observe through the transfer of momentum. We then examined some specific models ranging from very simple analytic ones to state-of-the-art hydrodyamical simulations.

\begin{itemize}
    \item 
    Following \citet{Heckman15}, we evaluated the ratio of the outward force on ambient gas clouds to the inward force of gravity. We found this ratio to have a median value of $\sim$ 30, with only 2 cases having a ratio $<$ 10. Thus, this sample is in the regime of ``strong outflows'' in which gravitational forces are secondary.
    \item
    We considered a simple model of a wind-blown bubble driven by the momentum supplied by the starburst. While this model can fit the relationship between \Vout\ and $SFR$, it fails in other respects, and would produce absorption-lines with \Vout $>>$ \FWHMout (while we find \Vout $\sim$\ 0.5 \FWHMout).
    \item
    We compared our results to a new semi-analytic model of multi-phase galactic winds. We found agreement, but only for models in which the hot wind-fluid that accelerates the warm ionized gas we observe is uncontaminated (made up of pure stellar ejecta) and the clouds it interacts with are massive (10$^5$ to 10$^6$ M$_{\odot}$).
    \item
    Recent high-resolution numerical simulations by \citet[]{Schneider20} produced outflows with some similarities to the data, but whose warm ionized phase was significantly too fast and carried too little mass, momentum, and kinetic energy compared to the data. In contrast, the simulations by \citet[]{Kim20} predict the slope of outflow velocity vs. SFR/Area $\sim$ 0.2, which is similar to our observations. But their normalization of outflow velocity is small compared to the data.
    \item
    Finally, we compared the data to a family of models in which the warm ionized gas is not ambient material, but instead forms directly from the wind fluid (via radiative cooling). These models predicted outflow velocities significantly larger than we observed.
\end{itemize}

In a future paper, we will analyze FUSE spectra of starbursts using the same methodology as in this paper. This will allow us to extend the range of parameter space we can explore to lower values of both SFR/M$*$ and SFR/Area, which will allow us to disentangle the dependence of outflow properties on SFR {\it vs.} M$_*$ {\it vs.} size. We will also analyze the current CLASSY spectra, searching for the possible presence of absorption lines arising from the fine structure levels in the Si II transitions. These could allow us to measure a mean density of the absorbing gas in the outflows for the first time. 

\newpage
\begin{acknowledgements}

The CLASSY team is grateful for the support for this program, HST-GO-15840, that was provided by NASA through a grant from the Space Telescope Science Institute,  which is operated by the Associations of Universities for Research in Astronomy, Incorporated, under NASA contract NAS5- 26555. BLJ thanks support from the European Space Agency (ESA).
The CLASSY collaboration extends special gratitude to the Lorentz Center for useful discussions during the "Characterizing Galaxies with Spectroscopy with a view for JWST" 2017 workshop that led to the formation of the CLASSY collaboration and survey.\\

Funding for SDSS-III has been provided by the Alfred P. Sloan Foundation, the Participating Institutions, the National Science Foundation, and the U.S. Department of Energy Office of Science. The SDSS-III web site is http://www.sdss3.org/.\\

SDSS-III is managed by the Astrophysical Research Consortium for the Participating Institutions of the SDSS-III Collaboration including the University of Arizona, the Brazilian Participation Group, Brookhaven National Laboratory, Carnegie Mellon University, University of Florida, the French Participation Group, the German Participation Group, Harvard University, the Instituto de Astrofisica de Canarias, the Michigan State/Notre Dame/JINA Participation Group, Johns Hopkins University, Lawrence Berkeley National Laboratory, Max Planck Institute for Astrophysics, Max Planck Institute for Extraterrestrial Physics, New Mexico State University, New York University, Ohio State University, Pennsylvania State University, University of Portsmouth, Princeton University, the Spanish Participation Group, University of Tokyo, University of Utah, Vanderbilt University, University of Virginia, University of Washington, and Yale University.\\

This research has made use of the HSLA database, developed and maintained at STScI, Baltimore, USA.

We also thank the anonymous referee for the insightful comments on the paper.

\end{acknowledgements}

\facilities{HST (COS), LBT (MODS), APO (SDSS), KECK (ESI), VLT (MUSE, VIMOS)}
\software{
astropy (The Astropy Collaboration 2013, 2018)
BEAGLE (Chevallard \& Charlot 2006), 
CalCOS (STScI),
dustmaps (Green 2018),
jupyter (Kluyver 2016),
MPFIT (Markwardt 2009),
Photutils (Bradley 2021),
python,
pysynphot (STScI Development Team)}

\clearpage
\begin{turnpage}
\begin{table*}
	\centering
	\caption{Measured Outflow Parameters for Galaxies in the Combined Sample$^{(1)}$}
	\label{tab:mea}
	\begin{tabular}{lllllllllllll} 
		\hline
		\hline
		Object  &	\Vout & \FWHMout & log(N(\Siii)) &  log(N(\siiii)) & log(N(\siiv))  & log(\Nh) & log(N(\hi)) & \barCFv & \Mdot & log(\Pdot) & log(\PdotCrit) & log(\Edot) \\
		\hline
		        & km s$^{-1}$ & km s$^{-1}$ & cm$^{-2}$ & cm$^{-2}$  & cm$^{-2}$  & cm$^{-2}$   & cm$^{-2}$  & & M$\odot$/yr  & dynes & dynes & ergs/s\\ 
		\hline
		(1)&(2)&(3)&(4)&(5)&(6)&(7)&(8)&(9)&(10)&(11)&(12)&(13)\\
		\hline
J0021+0052 & 231$^{+77}_{-77}$ & 566$^{+127}_{-127}$ & 14.39$^{+0.06}_{-0.06}$ & 15.67$^{+0.02}_{-0.02}$ & 15.12$^{+0.06}_{-0.06}$ & 20.78$^{+0.04}_{-0.05}$ & 17.50$^{+0.02}_{-0.02}$ & 0.50$^{+0.10}_{-0.10}$ & 16.5$^{+7.1}_{-6.2}$ & 34.5$^{+0.3}_{-0.4}$ & 33.3$^{+0.1}_{-0.1}$ & 41.8$^{+0.4}_{-0.5}$\\
J0036--3333 & 157$^{+22}_{-22}$ & 413$^{+117}_{-117}$ & 14.39$^{+0.03}_{-0.03}$ & 15.50$^{+0.02}_{-0.02}$ & 14.64$^{+0.02}_{-0.02}$ & 20.49$^{+0.02}_{-0.02}$ & 17.43$^{+0.02}_{-0.02}$ & 0.76$^{+0.04}_{-0.04}$ & 1.9$^{+0.5}_{-0.4}$ & 33.5$^{+0.1}_{-0.2}$ & 32.5$^{+0.1}_{-0.1}$ & 40.8$^{+0.2}_{-0.2}$\\
J0055--0021$^\text{(H15)}$ & 197$^{+63}_{-63}$ & 591$^{+108}_{-108}$ & 15.33$^{+0.04}_{-0.04}$ & 16.03$^{+0.02}_{-0.02}$ & 14.88$^{+0.06}_{-0.06}$ & 21.03$^{+0.03}_{-0.03}$ & 18.40$^{+0.03}_{-0.02}$ & 0.94$^{+0.19}_{-0.19}$ & 36.5$^{+14.9}_{-13.0}$ & 34.9$^{+0.3}_{-0.3}$ & 34.0$^{+0.1}_{-0.1}$ & 42.2$^{+0.4}_{-0.5}$\\
J0127--0619 & \multicolumn{2}{l}{No Outflow$^{*}$} & \dots & \dots & \dots & \dots & \dots & \dots & \dots & \dots & \dots & \dots\\
J0144+0453 & 48$^{+16}_{-16}$ & 182$^{+61}_{-61}$ & 14.59$^{+0.10}_{-0.10}$ & 15.39$^{+0.02}_{-0.02}$ & 14.12$^{+0.10}_{-0.10}$ & 20.97$^{+0.03}_{-0.03}$ & 18.74$^{+0.89}_{-0.24}$ & 0.36$^{+0.07}_{-0.07}$ & 4.1$^{+4.4}_{-1.9}$ & 33.3$^{+0.4}_{-0.4}$ & 32.7$^{+0.2}_{-0.1}$ & 40.0$^{+0.5}_{-0.6}$\\
J0150+1308$^\text{(H15)}$ & 212$^{+28}_{-28}$ & 345$^{+63}_{-63}$ & 14.86$^{+0.04}_{-0.04}$ & 15.78$^{+0.02}_{-0.02}$ & 14.83$^{+0.08}_{-0.08}$ & 20.67$^{+0.04}_{-0.04}$ & 17.80$^{+0.02}_{-0.02}$ & 0.82$^{+0.15}_{-0.15}$ & 17.3$^{+4.8}_{-4.1}$ & 34.5$^{+0.1}_{-0.2}$ & 33.8$^{+0.1}_{-0.1}$ & 41.8$^{+0.2}_{-0.2}$\\
J0337--0502 & \multicolumn{2}{l}{No Outflow$^{*}$} & \dots & \dots & \dots & \dots & \dots & \dots & \dots & \dots & \dots & \dots\\
J0405--3648 & \multicolumn{2}{l}{No Outflow$^{*}$} & \dots & \dots & \dots & \dots & \dots & \dots & \dots & \dots & \dots & \dots\\
J0808+3948 & 646$^{+65}_{-65}$ & 382$^{+127}_{-127}$ & 14.87$^{+0.05}_{-0.05}$ & 15.47$^{+0.02}_{-0.02}$ & 14.40$^{+0.02}_{-0.02}$ & 19.99$^{+0.01}_{-0.02}$ & 17.22$^{+0.02}_{-0.02}$ & 0.70$^{+0.07}_{-0.07}$ & 2.3$^{+0.6}_{-0.5}$ & 34.0$^{+0.1}_{-0.1}$ & 32.0$^{+0.1}_{-0.1}$ & 41.5$^{+0.1}_{-0.2}$\\
J0823+2806 & 136$^{+45}_{-45}$ & 393$^{+131}_{-131}$ & 14.27$^{+0.04}_{-0.04}$ & 15.55$^{+0.02}_{-0.02}$ & 14.92$^{+0.03}_{-0.03}$ & 20.56$^{+0.02}_{-0.02}$ & 17.43$^{+0.02}_{-0.02}$ & 0.85$^{+0.15}_{-0.15}$ & 3.3$^{+1.4}_{-1.2}$ & 33.6$^{+0.3}_{-0.4}$ & 33.0$^{+0.1}_{-0.1}$ & 40.6$^{+0.4}_{-0.5}$\\
J0926+4427 & 353$^{+52}_{-52}$ & 402$^{+130}_{-130}$ & 15.00$^{+0.13}_{-0.13}$ & 15.71$^{+0.02}_{-0.02}$ & 14.67$^{+0.05}_{-0.05}$ & 20.90$^{+0.08}_{-0.09}$ & 18.33$^{+0.05}_{-0.05}$ & 0.50$^{+0.10}_{-0.10}$ & 45.8$^{+15.5}_{-13.1}$ & 35.0$^{+0.2}_{-0.2}$ & 33.4$^{+0.1}_{-0.1}$ & 42.3$^{+0.2}_{-0.2}$\\
J0934+5514 & 112$^{+37}_{-37}$ & 178$^{+59}_{-59}$ & 14.16$^{+0.10}_{-0.10}$ & \dots & 13.91$^{+0.05}_{-0.05}$ & \dots & \dots & \dots & \dots & \dots & \dots & \dots\\
J0938+5428 & 215$^{+72}_{-72}$ & 272$^{+91}_{-91}$ & 15.12$^{+0.14}_{-0.14}$ & 15.74$^{+0.02}_{-0.02}$ & 14.57$^{+0.04}_{-0.04}$ & 20.86$^{+0.06}_{-0.06}$ & 18.50$^{+0.13}_{-0.06}$ & 0.36$^{+0.07}_{-0.07}$ & 12.2$^{+5.4}_{-4.7}$ & 34.1$^{+0.3}_{-0.4}$ & 33.5$^{+0.1}_{-0.1}$ & 41.1$^{+0.4}_{-0.5}$\\
J0940+2935 & 102$^{+34}_{-34}$ & 168$^{+56}_{-56}$ & 14.13$^{+0.21}_{-0.21}$ & 14.68$^{+0.02}_{-0.02}$ & 14.01$^{+0.13}_{-0.13}$ & 19.99$^{+0.08}_{-0.10}$ & 16.89$^{+0.02}_{-0.02}$ & 0.50$^{+0.10}_{-0.10}$ & 0.3$^{+0.1}_{-0.1}$ & 32.4$^{+0.3}_{-0.4}$ & 30.8$^{+0.1}_{-0.1}$ & 39.2$^{+0.4}_{-0.5}$\\
J0942+3547 & 97$^{+26}_{-26}$ & 272$^{+91}_{-91}$ & 13.77$^{+0.05}_{-0.05}$ & 15.03$^{+0.02}_{-0.02}$ & 14.35$^{+0.03}_{-0.03}$ & 20.23$^{+0.02}_{-0.02}$ & 17.19$^{+0.02}_{-0.02}$ & 0.70$^{+0.04}_{-0.04}$ & 0.4$^{+0.1}_{-0.1}$ & 32.5$^{+0.2}_{-0.3}$ & 31.3$^{+0.1}_{-0.1}$ & 39.4$^{+0.3}_{-0.4}$\\
J0944+3424 & \multicolumn{2}{l}{No Outflow$^{*}$} & \dots & \dots & \dots & \dots & \dots & \dots & \dots & \dots & \dots & \dots\\
J0944--0038 & 64$^{+21}_{-21}$ & 163$^{+54}_{-54}$ & 14.57$^{+0.12}_{-0.12}$ & 15.46$^{+0.03}_{-0.03}$ & 14.29$^{+0.11}_{-0.11}$ & 20.95$^{+0.05}_{-0.06}$ & 18.97$^{+0.49}_{-0.55}$ & 0.55$^{+0.11}_{-0.11}$ & 3.3$^{+1.2}_{-1.1}$ & 33.4$^{+0.3}_{-0.4}$ & 32.0$^{+0.1}_{-0.1}$ & 40.3$^{+0.4}_{-0.5}$\\
J1016+3754 & 116$^{+31}_{-31}$ & 128$^{+37}_{-37}$ & 13.64$^{+0.11}_{-0.11}$ & 15.00$^{+0.02}_{-0.03}$ & 14.55$^{+0.03}_{-0.03}$ & 20.74$^{+0.06}_{-0.07}$ & 17.74$^{+0.02}_{-0.02}$ & 0.42$^{+0.08}_{-0.08}$ & 2.2$^{+0.7}_{-0.7}$ & 33.3$^{+0.2}_{-0.3}$ & 31.5$^{+0.1}_{-0.1}$ & 40.1$^{+0.3}_{-0.4}$\\
J1024+0524 & 94$^{+12}_{-12}$ & 286$^{+54}_{-54}$ & 14.58$^{+0.03}_{-0.03}$ & 15.54$^{+0.02}_{-0.02}$ & 14.43$^{+0.05}_{-0.05}$ & 21.00$^{+0.02}_{-0.02}$ & 18.41$^{+0.05}_{-0.03}$ & 0.65$^{+0.05}_{-0.05}$ & 7.2$^{+1.9}_{-1.6}$ & 33.9$^{+0.1}_{-0.2}$ & 32.6$^{+0.1}_{-0.1}$ & 40.9$^{+0.2}_{-0.2}$\\
J1025+3622 & 155$^{+24}_{-24}$ & 409$^{+76}_{-76}$ & 14.79$^{+0.04}_{-0.04}$ & 15.67$^{+0.02}_{-0.02}$ & 14.62$^{+0.03}_{-0.03}$ & 20.84$^{+0.03}_{-0.03}$ & 18.15$^{+0.02}_{-0.02}$ & 0.81$^{+0.12}_{-0.12}$ & 25.2$^{+7.1}_{-6.1}$ & 34.6$^{+0.2}_{-0.2}$ & 33.5$^{+0.1}_{-0.1}$ & 41.8$^{+0.2}_{-0.2}$\\
J1044+0353 & 52$^{+12}_{-12}$ & 123$^{+24}_{-24}$ & 14.84$^{+0.06}_{-0.06}$ & \dots & 13.43$^{+0.17}_{-0.17}$ & \dots & \dots & \dots & \dots & \dots & \dots & \dots\\
J1105+4444 & 115$^{+23}_{-23}$ & 247$^{+60}_{-60}$ & 15.01$^{+0.02}_{-0.02}$ & 15.75$^{+0.02}_{-0.02}$ & 14.46$^{+0.02}_{-0.02}$ & 20.84$^{+0.01}_{-0.01}$ & 18.19$^{+0.03}_{-0.02}$ & 0.84$^{+0.04}_{-0.04}$ & 31.5$^{+6.5}_{-6.4}$ & 34.5$^{+0.2}_{-0.2}$ & 33.9$^{+0.1}_{-0.1}$ & 41.5$^{+0.2}_{-0.3}$\\
J1112+5503 & 349$^{+107}_{-107}$ & 841$^{+230}_{-230}$ & 14.93$^{+0.05}_{-0.05}$ & 15.82$^{+0.02}_{-0.02}$ & 14.83$^{+0.03}_{-0.03}$ & 20.63$^{+0.03}_{-0.03}$ & 17.81$^{+0.01}_{-0.01}$ & 0.80$^{+0.16}_{-0.16}$ & 25.2$^{+10.0}_{-8.8}$ & 35.0$^{+0.2}_{-0.3}$ & 33.5$^{+0.1}_{-0.1}$ & 42.6$^{+0.4}_{-0.5}$\\
J1113+2930$^\text{(H15)}$ & 889$^{+74}_{-74}$ & 237$^{+79}_{-79}$ & 15.39$^{+0.14}_{-0.14}$ & \dots & 12.85$^{+0.23}_{-0.23}$ & \dots & \dots & \dots & \dots & \dots & \dots & \dots\\
J1119+5130 & 65$^{+22}_{-22}$ & 200$^{+54}_{-54}$ & 14.36$^{+0.06}_{-0.06}$ & 15.00$^{+0.03}_{-0.03}$ & 13.68$^{+0.10}_{-0.10}$ & 20.71$^{+0.03}_{-0.04}$ & 18.66$^{+0.20}_{-0.13}$ & 0.32$^{+0.06}_{-0.06}$ & 1.1$^{+0.4}_{-0.4}$ & 32.9$^{+0.3}_{-0.4}$ & 31.4$^{+0.1}_{-0.1}$ & 39.9$^{+0.4}_{-0.5}$\\
J1129+2034 & 51$^{+17}_{-17}$ & 144$^{+15}_{-15}$ & 14.30$^{+0.08}_{-0.08}$ & 15.55$^{+0.03}_{-0.03}$ & 14.39$^{+0.09}_{-0.09}$ & 20.55$^{+0.10}_{-0.12}$ & 18.01$^{+0.04}_{-0.03}$ & 0.46$^{+0.09}_{-0.09}$ & 0.3$^{+0.1}_{-0.1}$ & 32.3$^{+0.3}_{-0.4}$ & 31.5$^{+0.1}_{-0.2}$ & 39.3$^{+0.4}_{-0.6}$\\
J1132+1411 & 60$^{+10}_{-10}$ & 215$^{+69}_{-69}$ & 15.14$^{+0.03}_{-0.03}$ & 15.82$^{+0.02}_{-0.02}$ & 14.56$^{+0.07}_{-0.07}$ & 20.82$^{+0.03}_{-0.03}$ & 18.14$^{+0.05}_{-0.05}$ & 0.76$^{+0.08}_{-0.08}$ & 35.3$^{+6.9}_{-6.6}$ & 34.4$^{+0.1}_{-0.2}$ & 34.0$^{+0.1}_{-0.1}$ & 41.3$^{+0.2}_{-0.2}$\\
J1132+5722 & \multicolumn{2}{l}{No Outflow$^{*}$} & \dots & \dots & \dots & \dots & \dots & \dots & \dots & \dots & \dots & \dots\\
J1144+4012 & 246$^{+33}_{-33}$ & 449$^{+86}_{-86}$ & 15.17$^{+0.04}_{-0.04}$ & 15.79$^{+0.02}_{-0.02}$ & 14.55$^{+0.07}_{-0.07}$ & 20.71$^{+0.03}_{-0.03}$ & 18.12$^{+0.03}_{-0.02}$ & 0.83$^{+0.15}_{-0.15}$ & 24.9$^{+6.7}_{-5.8}$ & 34.6$^{+0.1}_{-0.2}$ & 34.0$^{+0.1}_{-0.1}$ & 41.8$^{+0.2}_{-0.2}$\\
J1148+2546 & 95$^{+19}_{-19}$ & 239$^{+52}_{-52}$ & 15.27$^{+0.20}_{-0.20}$ & 15.62$^{+0.02}_{-0.03}$ & 14.34$^{+0.12}_{-0.12}$ & 21.06$^{+0.06}_{-0.07}$ & 20.01$^{+0.21}_{-0.20}$ & 0.67$^{+0.10}_{-0.10}$ & 25.7$^{+7.0}_{-6.4}$ & 34.3$^{+0.2}_{-0.2}$ & 33.5$^{+0.1}_{-0.1}$ & 41.2$^{+0.2}_{-0.3}$\\
J1150+1501 & 67$^{+22}_{-22}$ & 181$^{+25}_{-25}$ & 14.58$^{+0.03}_{-0.03}$ & 15.62$^{+0.02}_{-0.02}$ & 14.71$^{+0.06}_{-0.06}$ & 20.71$^{+0.03}_{-0.03}$ & 17.90$^{+0.02}_{-0.02}$ & 0.64$^{+0.06}_{-0.06}$ & 1.1$^{+0.4}_{-0.4}$ & 33.1$^{+0.3}_{-0.4}$ & 31.3$^{+0.1}_{-0.1}$ & 40.2$^{+0.4}_{-0.5}$\\
J1157+3220 & 238$^{+49}_{-49}$ & 277$^{+84}_{-84}$ & 14.46$^{+0.03}_{-0.03}$ & 15.49$^{+0.02}_{-0.02}$ & 14.36$^{+0.02}_{-0.02}$ & 20.29$^{+0.01}_{-0.01}$ & 17.46$^{+0.01}_{-0.01}$ & 0.78$^{+0.06}_{-0.06}$ & 7.1$^{+1.5}_{-1.5}$ & 34.1$^{+0.2}_{-0.2}$ & 33.0$^{+0.1}_{-0.1}$ & 41.4$^{+0.2}_{-0.3}$\\
J1200+1343 & 192$^{+13}_{-13}$ & 306$^{+62}_{-62}$ & 14.90$^{+0.08}_{-0.08}$ & 15.86$^{+0.02}_{-0.02}$ & 14.82$^{+0.03}_{-0.03}$ & 20.85$^{+0.05}_{-0.06}$ & 18.07$^{+0.02}_{-0.02}$ & 0.87$^{+0.06}_{-0.06}$ & 10.0$^{+2.6}_{-2.3}$ & 34.3$^{+0.1}_{-0.1}$ & 32.5$^{+0.1}_{-0.2}$ & 41.5$^{+0.1}_{-0.1}$\\
J1225+6109 & 51$^{+17}_{-17}$ & 198$^{+33}_{-33}$ & 14.52$^{+0.15}_{-0.15}$ & 15.41$^{+0.02}_{-0.02}$ & 14.51$^{+0.05}_{-0.05}$ & 20.70$^{+0.09}_{-0.12}$ & 18.15$^{+0.02}_{-0.02}$ & 0.64$^{+0.13}_{-0.13}$ & 2.8$^{+1.2}_{-1.1}$ & 33.5$^{+0.3}_{-0.4}$ & 31.8$^{+0.1}_{-0.2}$ & 40.5$^{+0.4}_{-0.5}$\\
J1253--0312 & 113$^{+38}_{-38}$ & 245$^{+82}_{-82}$ & 14.61$^{+0.10}_{-0.10}$ & 15.69$^{+0.02}_{-0.02}$ & 14.56$^{+0.09}_{-0.09}$ & 21.00$^{+0.03}_{-0.03}$ & 17.90$^{+0.15}_{-0.17}$ & 0.92$^{+0.18}_{-0.18}$ & 7.7$^{+2.7}_{-2.6}$ & 33.9$^{+0.3}_{-0.4}$ & 32.7$^{+0.1}_{-0.2}$ & 41.0$^{+0.4}_{-0.5}$\\
J1314+3452 & 62$^{+21}_{-21}$ & 187$^{+54}_{-54}$ & 15.31$^{+0.17}_{-0.17}$ & 15.73$^{+0.03}_{-0.03}$ & 14.35$^{+0.04}_{-0.04}$ & 20.84$^{+0.03}_{-0.03}$ & 19.87$^{+0.39}_{-0.36}$ & 0.80$^{+0.02}_{-0.02}$ & 0.3$^{+0.1}_{-0.1}$ & 32.5$^{+0.3}_{-0.4}$ & 31.3$^{+0.1}_{-0.1}$ & 39.6$^{+0.4}_{-0.5}$\\
J1323--0132 & \multicolumn{2}{l}{No Outflow$^{*}$} & \dots & \dots & \dots & \dots & \dots & \dots & \dots & \dots & \dots & \dots\\
J1359+5726 & 161$^{+23}_{-23}$ & 359$^{+70}_{-70}$ & 14.99$^{+0.02}_{-0.02}$ & 15.77$^{+0.02}_{-0.02}$ & 14.60$^{+0.03}_{-0.03}$ & 21.05$^{+0.01}_{-0.01}$ & 18.54$^{+0.03}_{-0.03}$ & 0.74$^{+0.04}_{-0.04}$ & 29.5$^{+4.5}_{-4.4}$ & 34.6$^{+0.1}_{-0.1}$ & 33.4$^{+0.1}_{-0.1}$ & 41.7$^{+0.2}_{-0.2}$\\
J1414+0540$^\text{(H15)}$ & 155$^{+52}_{-52}$ & 273$^{+87}_{-87}$ & 14.05$^{+0.06}_{-0.06}$ & 15.48$^{+0.02}_{-0.02}$ & 14.83$^{+0.05}_{-0.05}$ & 20.53$^{+0.01}_{-0.02}$ & 17.95$^{+0.033}_{-0.033}$ & 0.81$^{+0.16}_{-0.16}$ & 6.5$^{+11.1}_{-5.6}$ & 34.1$^{+0.5}_{-0.9}$ & 33.3$^{+0.2}_{-0.4}$ & 41.3$^{+0.6}_{-1.0}$\\
J1416+1223 & 398$^{+68}_{-68}$ & 649$^{+216}_{-216}$ & 14.24$^{+0.06}_{-0.06}$ & 15.47$^{+0.02}_{-0.02}$ & 14.65$^{+0.03}_{-0.03}$ & 20.27$^{+0.02}_{-0.02}$ & 17.24$^{+0.02}_{-0.02}$ & 0.79$^{+0.13}_{-0.13}$ & 2.8$^{+0.8}_{-0.7}$ & 33.7$^{+0.2}_{-0.2}$ & 32.9$^{+0.1}_{-0.1}$ & 40.9$^{+0.2}_{-0.3}$\\

		\hline
		\hline

	\end{tabular}
	\\ [0mm]
	
\end{table*}

\clearpage
\addtocounter{table}{-1}

\begin{table*}
	\centering
	\caption{Continued}
	\begin{tabular}{lllllllllllll} 
		\hline
		\hline
		Object  &	\Vout & \FWHMout & log(N(\Siii)) &  log(N(\siiii)) & log(N(\siiv))  & log(\Nh) & log(N(\hi)) & \barCFv & \Mdot & log(\Pdot) & log(\PdotCrit) & log(\Edot) \\
		\hline
		        & km s$^{-1}$ & km s$^{-1}$ & cm$^{-2}$ & cm$^{-2}$  & cm$^{-2}$  & cm$^{-2}$   & cm$^{-2}$  & & M$\odot$/yr  & dynes & dynes & ergs/s\\ 
		\hline
		(1)&(2)&(3)&(4)&(5)&(6)&(7)&(8)&(9)&(10)&(11)&(12)&(13)\\
		\hline
		
J1418+2102 & 51$^{+7}_{-7}$ & 112$^{+27}_{-27}$ & 15.23$^{+0.20}_{-0.20}$ & 15.34$^{+0.05}_{-0.06}$ & 13.98$^{+0.13}_{-0.13}$ & 21.12$^{+0.08}_{-0.10}$ & 20.54$^{+0.07}_{-0.13}$ & 0.30$^{+0.06}_{-0.06}$ & 1.0$^{+0.3}_{-0.3}$ & 32.8$^{+0.2}_{-0.2}$ & 31.3$^{+0.1}_{-0.2}$ & 39.5$^{+0.2}_{-0.2}$\\
J1428+1653 & 140$^{+25}_{-25}$ & 325$^{+108}_{-108}$ & 14.64$^{+0.07}_{-0.07}$ & 15.72$^{+0.02}_{-0.02}$ & 14.87$^{+0.05}_{-0.05}$ & 20.70$^{+0.04}_{-0.05}$ & 17.69$^{+0.03}_{-0.02}$ & 0.75$^{+0.13}_{-0.13}$ & 20.1$^{+6.2}_{-5.3}$ & 34.4$^{+0.2}_{-0.2}$ & 33.9$^{+0.1}_{-0.1}$ & 41.5$^{+0.2}_{-0.3}$\\
J1429+0643 & 230$^{+51}_{-51}$ & 461$^{+152}_{-152}$ & 14.83$^{+0.06}_{-0.06}$ & 15.83$^{+0.02}_{-0.02}$ & 15.00$^{+0.06}_{-0.06}$ & 20.81$^{+0.04}_{-0.05}$ & 17.80$^{+0.02}_{-0.02}$ & 0.73$^{+0.15}_{-0.15}$ & 22.1$^{+7.6}_{-6.5}$ & 34.7$^{+0.2}_{-0.2}$ & 33.2$^{+0.1}_{-0.1}$ & 42.0$^{+0.3}_{-0.3}$\\
J1444+4237 & 54$^{+18}_{-18}$ & 112$^{+37}_{-37}$ & 14.07$^{+0.08}_{-0.08}$ & 14.99$^{+0.02}_{-0.02}$ & 14.06$^{+0.06}_{-0.06}$ & 20.68$^{+0.03}_{-0.03}$ & 18.25$^{+0.15}_{-0.13}$ & 0.40$^{+0.06}_{-0.06}$ & 44.4$^{+15.3}_{-15.0}$ & 34.4$^{+0.3}_{-0.4}$ & 32.9$^{+0.1}_{-0.1}$ & 41.1$^{+0.4}_{-0.5}$\\
J1448--0110 & 145$^{+43}_{-43}$ & 220$^{+73}_{-73}$ & 14.03$^{+0.06}_{-0.06}$ & 15.27$^{+0.02}_{-0.02}$ & 14.66$^{+0.07}_{-0.07}$ & 20.50$^{+0.03}_{-0.03}$ & 17.57$^{+0.02}_{-0.02}$ & 0.85$^{+0.15}_{-0.15}$ & 1.2$^{+0.5}_{-0.4}$ & 33.1$^{+0.2}_{-0.3}$ & 31.6$^{+0.1}_{-0.1}$ & 40.1$^{+0.4}_{-0.5}$\\
J1521+0759 & 161$^{+54}_{-54}$ & 398$^{+133}_{-133}$ & 14.48$^{+0.05}_{-0.05}$ & 15.49$^{+0.02}_{-0.02}$ & 14.68$^{+0.04}_{-0.04}$ & 20.48$^{+0.03}_{-0.03}$ & 17.49$^{+0.03}_{-0.02}$ & 0.54$^{+0.11}_{-0.11}$ & 3.6$^{+1.5}_{-1.3}$ & 33.5$^{+0.3}_{-0.4}$ & 33.0$^{+0.1}_{-0.1}$ & 40.3$^{+0.4}_{-0.5}$\\
J1525+0757 & 408$^{+28}_{-28}$ & 495$^{+59}_{-59}$ & 15.21$^{+0.14}_{-0.14}$ & 15.93$^{+0.02}_{-0.02}$ & 14.82$^{+0.04}_{-0.04}$ & 20.92$^{+0.06}_{-0.06}$ & 18.21$^{+0.04}_{-0.02}$ & 0.87$^{+0.15}_{-0.15}$ & 20.6$^{+5.6}_{-4.8}$ & 34.7$^{+0.1}_{-0.1}$ & 33.9$^{+0.1}_{-0.1}$ & 42.1$^{+0.1}_{-0.1}$\\
J1545+0858 & 113$^{+33}_{-33}$ & 203$^{+51}_{-51}$ & 15.16$^{+0.15}_{-0.15}$ & 15.77$^{+0.03}_{-0.03}$ & 14.62$^{+0.06}_{-0.06}$ & 21.38$^{+0.05}_{-0.06}$ & 20.19$^{+0.27}_{-0.36}$ & 0.61$^{+0.06}_{-0.06}$ & 14.4$^{+5.8}_{-5.0}$ & 34.1$^{+0.2}_{-0.3}$ & 32.8$^{+0.1}_{-0.1}$ & 40.9$^{+0.3}_{-0.5}$\\
J1612+0817 & 459$^{+63}_{-63}$ & 732$^{+158}_{-158}$ & 15.43$^{+0.11}_{-0.11}$ & \dots & \dots & \dots & \dots & \dots & \dots & \dots & \dots & \dots\\
J2103--0728$^\text{(H15)}$ & 208$^{+21}_{-21}$ & 290$^{+97}_{-97}$ & 15.18$^{+0.12}_{-0.12}$ & 15.83$^{+0.02}_{-0.02}$ & 14.64$^{+0.07}_{-0.07}$ & 20.47$^{+0.04}_{-0.04}$ & 17.82$^{+0.01}_{-0.02}$ & 0.79$^{+0.10}_{-0.10}$ & 4.2$^{+1.1}_{-0.9}$ & 33.9$^{+0.1}_{-0.1}$ & 33.4$^{+0.1}_{-0.1}$ & 41.2$^{+0.2}_{-0.2}$\\

		\hline
	\multicolumn{13}{l}{%
  	\begin{minipage}{22cm}%
		Notes: (1)\ \ Measured outflow information for 45 galaxies from the CLASSY sample and 5 galaxies from \cite{Heckman15}. The latter is marked as (H15). Descriptions for each column: (2) and (3) Measured outflow velocity (the absolute value) and FWHM, respectively (Section \ref{sec:2GFits}); (4) and (6) Measured column density of outflows from \Siii\ and \siiv\ lines (Section \ref{sec:ColumnDensity}); (5), (7), (8) CLOUDY models predicted column density for \siiii, total hydrogen, and \hi, respectively (Section \ref{sec:PI}); (9) The average covering fraction of outflows (Section \ref{sec:ColumnDensity}); (10), (11), and (13): The mass, momentum, energy rate of outflows, respectively (Section \ref{sec:TotalMassRate} and \ref{sec:TotalPERate}); (12) the critical momentum flux to drive the outflows discussed in Section \ref{sec:compModel}.\\
    	* \ \ These objects are marked as ``No Outflow'' because less than half of their observed absorption troughs show blue-shifted outflow component. Note that the rest ($<$ half) of their absorption troughs may still show outflow features. See discussion in Section \ref{sec:2GFits}.\\
  	\end{minipage}%
	}
	\end{tabular}
	\\ [0mm]
	
\end{table*}

\end{turnpage}
\newpage

\begin{table*}
	\centering
	\caption{Ancillary Parameters for Galaxies in the Combined Sample$^{(1)}$}
	\label{tab:anc}
	\begin{tabular}{lllllll} 
		\hline
		\hline
		Object  &	$z$ & r$_{50}$ & r$_{50}$  & \Vcir\ & log SFR/\Mstar\ & log SFR/A\\
		\hline
		        &  & (\arcsec) & (kpc)    &(km s$^{-1}$)  & (yr$^{-1}$)  & (M$\odot$/yr/kpc$^{2}$)\\ 
		\hline
		&(2)&(3)&(4)&(5)&(6)&(7)\\
		\hline
J0021+0052 & 0.09839 & 0.25 & 0.45 & 72.1$^{+20.8}_{-8.0}$ & -8.02$^{+0.21}_{-0.40}$ & 0.98$^{+0.14}_{-0.16}$\\
J0036--3333 & 0.02060 & 0.28 & 0.11 & 74.1$^{+12.1}_{-11.7}$ & -8.12$^{+0.33}_{-0.30}$ & 2.09$^{+0.23}_{-0.21}$\\
J0055--0021$^\text{(H15)}$ & 0.16744 & 0.20 & 0.56 & 98.3$^{+22.3}_{-9.6}$ & -7.93$^{+0.38}_{-0.35}$ & 1.34$^{+0.36}_{-0.19}$\\
J0127--0619 & 0.00550 & 0.15 & 0.02 & 57.0$^{+6.0}_{-6.3}$ & -9.49$^{+0.22}_{-0.21}$ & 1.88$^{+0.16}_{-0.17}$\\
J0144+0453 & 0.00532 & 3.54 & 0.43 & 27.5$^{+5.8}_{-4.1}$ & -8.46$^{+0.52}_{-0.41}$ & -0.88$^{+0.51}_{-0.41}$\\
J0150+1308$^\text{(H15)}$ & 0.14668 & 0.25 & 0.63 & 115.3$^{+21.6}_{-13.6}$ & -8.29$^{+0.39}_{-0.32}$ & 1.11$^{+0.35}_{-0.20}$\\
J0337--0502 & 0.01346 & 1.62 & 0.46 & 18.5$^{+2.7}_{-2.8}$ & -7.37$^{+0.26}_{-0.22}$ & -0.44$^{+0.43}_{-0.84}$\\
J0405--3648 & 0.00280 & 6.43 & 0.50 & 13.7$^{+2.9}_{-2.3}$ & -8.42$^{+0.39}_{-0.42}$ & -2.01$^{+0.30}_{-0.34}$\\
J0808+3948 & 0.09123 & 0.08 & 0.13 & 73.1$^{+8.8}_{-13.2}$ & -7.85$^{+0.39}_{-0.25}$ & 2.27$^{+0.27}_{-0.20}$\\
J0823+2806 & 0.04730 & 0.28 & 0.25 & 87.2$^{+11.6}_{-17.4}$ & -7.90$^{+0.46}_{-0.24}$ & 1.88$^{+0.33}_{-0.17}$\\
J0926+4427 & 0.18030 & 0.23 & 0.66 & 57.8$^{+10.9}_{-10.5}$ & -7.73$^{+0.33}_{-0.29}$ & 0.59$^{+0.16}_{-0.14}$\\
J0934+5514 & 0.00264 & 1.53 & 0.11 & 10.9$^{+1.6}_{-0.9}$ & -7.79$^{+0.14}_{-0.22}$ & -0.42$^{+0.07}_{-0.09}$\\
J0938+5428 & 0.10210 & 0.28 & 0.51 & 75.0$^{+15.8}_{-8.3}$ & -8.10$^{+0.25}_{-0.35}$ & 0.84$^{+0.20}_{-0.21}$\\
J0940+2935 & 0.00171 & 3.06 & 0.18 & 14.6$^{+4.5}_{-2.1}$ & -8.71$^{+0.44}_{-0.58}$ & -1.34$^{+0.37}_{-0.42}$\\
J0942+3547 & 0.01483 & 0.33 & 0.10 & 25.8$^{+5.6}_{-3.3}$ & -8.31$^{+0.24}_{-0.35}$ & 0.45$^{+0.15}_{-0.21}$\\
J0944+3424 & 0.02005 & 3.74 & 1.52 & 39.5$^{+6.4}_{-9.3}$ & -8.20$^{+0.77}_{-0.36}$ & -1.17$^{+0.65}_{-0.28}$\\
J0944--0038 & 0.00487 & 2.34 & 0.28 & 15.9$^{+2.9}_{-3.9}$ & -7.60$^{+0.45}_{-0.31}$ & -0.46$^{+0.16}_{-0.19}$\\
J1016+3754 & 0.00391 & 1.52 & 0.15 & 14.8$^{+2.3}_{-2.4}$ & -7.89$^{+0.32}_{-0.28}$ & -0.34$^{+0.18}_{-0.18}$\\
J1024+0524 & 0.03326 & 0.40 & 0.26 & 32.1$^{+5.5}_{-7.0}$ & -7.67$^{+0.39}_{-0.28}$ & 0.57$^{+0.15}_{-0.17}$\\
J1025+3622 & 0.12720 & 0.35 & 0.77 & 62.1$^{+12.2}_{-9.5}$ & -7.83$^{+0.31}_{-0.30}$ & 0.46$^{+0.20}_{-0.17}$\\
J1044+0353 & 0.01286 & 0.38 & 0.10 & 15.6$^{+3.0}_{-3.7}$ & -7.39$^{+0.43}_{-0.28}$ & 0.60$^{+0.17}_{-0.14}$\\
J1105+4444 & 0.02148 & 4.11 & 1.79 & 66.9$^{+11.7}_{-11.8}$ & -8.29$^{+0.37}_{-0.37}$ & -0.61$^{+0.22}_{-0.28}$\\
J1112+5503 & 0.13153 & 0.20 & 0.46 & 100.5$^{+13.9}_{-19.7}$ & -7.99$^{+0.41}_{-0.28}$ & 1.48$^{+0.27}_{-0.22}$\\
J1113+2930$^\text{(H15)}$ & 0.17514 & 0.37 & 1.07 & 92.3$^{+16.5}_{-12.1}$ & -8.60$^{+0.28}_{-0.41}$ & 0.0$^{+0.20}_{-0.34}$\\
J1119+5130 & 0.00444 & 2.18 & 0.13 & 15.3$^{+3.0}_{-1.4}$ & -8.35$^{+0.18}_{-0.34}$ & -0.62$^{+0.11}_{-0.21}$\\
J1129+2034 & 0.00466 & 0.38 & 0.04 & 36.8$^{+7.4}_{-8.1}$ & -8.46$^{+0.67}_{-0.47}$ & 1.56$^{+0.56}_{-0.39}$\\
J1132+1411 & 0.01763 & 8.86 & 3.19 & 54.8$^{+7.5}_{-9.3}$ & -8.25$^{+0.39}_{-0.31}$ & -1.37$^{+0.27}_{-0.24}$\\
J1132+5722 & 0.00510 & 0.84 & 0.10 & 21.9$^{+4.2}_{-3.1}$ & -8.38$^{+0.42}_{-0.38}$ & 0.11$^{+0.35}_{-0.28}$\\
J1144+4012 & 0.12695 & 0.40 & 0.89 & 122.3$^{+26.2}_{-13.9}$ & -8.37$^{+0.34}_{-0.35}$ & 0.82$^{+0.31}_{-0.21}$\\
J1148+2546 & 0.04524 & 1.31 & 1.17 & 38.2$^{+6.8}_{-7.8}$ & -7.61$^{+0.37}_{-0.30}$ & -0.40$^{+0.15}_{-0.17}$\\
J1150+1501 & 0.00250 & 1.29 & 0.09 & 16.0$^{+3.5}_{-2.7}$ & -8.17$^{+0.36}_{-0.41}$ & -0.08$^{+0.23}_{-0.29}$\\
J1157+3220 & 0.01120 & 2.89 & 0.68 & 69.5$^{+9.0}_{-13.3}$ & -8.07$^{+0.53}_{-0.28}$ & 0.51$^{+0.42}_{-0.21}$\\
J1200+1343 & 0.06690 & 0.18 & 0.22 & 37.6$^{+12.0}_{-10.1}$ & -7.37$^{+0.49}_{-0.46}$ & 1.27$^{+0.18}_{-0.21}$\\
J1225+6109 & 0.00233 & 2.91 & 0.21 & 19.3$^{+3.4}_{-3.9}$ & -8.19$^{+0.43}_{-0.36}$ & -0.50$^{+0.26}_{-0.26}$\\
J1253--0312 & 0.02267 & 0.85 & 0.39 & 27.4$^{+4.5}_{-7.8}$ & -7.09$^{+0.53}_{-0.27}$ & 0.58$^{+0.15}_{-0.15}$\\
J1314+3452 & 0.00285 & 0.30 & 0.03 & 25.9$^{+4.0}_{-4.6}$ & -8.24$^{+0.63}_{-0.31}$ & 1.72$^{+0.56}_{-0.24}$\\
J1323--0132 & 0.02246 & 0.23 & 0.10 & 11.2$^{+0.8}_{-1.8}$ & -7.03$^{+0.28}_{-0.12}$ & 0.47$^{+0.13}_{-0.11}$\\
J1359+5726 & 0.03390 & 1.10 & 0.74 & 45.5$^{+8.6}_{-8.4}$ & -7.98$^{+0.33}_{-0.32}$ & -0.11$^{+0.14}_{-0.20}$\\
J1414+0540$^\text{(H15)}$ & 0.08190 & 0.23 & 0.36 & 107.6$^{+17.0}_{-19.7}$ & -8.12$^{+0.51}_{-0.32}$ & 1.67$^{+0.91}_{-0.48}$\\
J1416+1223 & 0.12316 & 0.13 & 0.27 & 100.3$^{+19.2}_{-19.5}$ & -8.02$^{+0.41}_{-0.34}$ & 1.91$^{+0.27}_{-0.23}$\\
J1418+2102 & 0.00857 & 0.40 & 0.08 & 10.6$^{+2.8}_{-2.9}$ & -7.35$^{+0.51}_{-0.38}$ & 0.31$^{+0.19}_{-0.17}$\\
J1428+1653 & 0.18170 & 0.35 & 1.04 & 98.3$^{+16.6}_{-9.5}$ & -8.34$^{+0.24}_{-0.35}$ & 0.39$^{+0.21}_{-0.27}$\\
J1429+0643 & 0.17350 & 0.15 & 0.43 & 59.0$^{+9.0}_{-12.3}$ & -7.38$^{+0.39}_{-0.24}$ & 1.35$^{+0.19}_{-0.14}$\\
J1444+4237 & 0.00219 & 8.20 & 7.33 & 12.6$^{+1.5}_{-1.3}$ & -8.42$^{+0.18}_{-0.20}$ & -4.47$^{+0.08}_{-0.11}$\\
J1448--0110 & 0.02738 & 0.23 & 0.12 & 26.7$^{+4.6}_{-6.4}$ & -7.22$^{+0.44}_{-0.27}$ & 1.41$^{+0.17}_{-0.16}$\\
J1521+0759 & 0.09426 & 0.28 & 0.47 & 67.7$^{+15.0}_{-11.8}$ & -8.05$^{+0.34}_{-0.34}$ & 0.80$^{+0.20}_{-0.18}$\\
J1525+0757 & 0.07579 & 0.25 & 0.35 & 137.3$^{+44.6}_{-23.5}$ & -9.06$^{+0.37}_{-0.81}$ & 1.10$^{+0.26}_{-0.70}$\\
J1545+0858 & 0.03772 & 0.33 & 0.24 & 25.1$^{+4.8}_{-6.3}$ & -7.15$^{+0.47}_{-0.29}$ & 0.80$^{+0.20}_{-0.16}$\\
J1612+0817 & 0.14914 & 0.20 & 0.51 & 113.7$^{+21.3}_{-19.3}$ & -8.20$^{+0.37}_{-0.38}$ & 1.37$^{+0.26}_{-0.29}$\\
J2103--0728$^\text{(H15)}$ & 0.13689 & 0.10 & 0.24 & 159.5$^{+48.4}_{-25.4}$ & -9.00$^{+0.44}_{-0.64}$ & 1.74$^{+0.36}_{-0.51}$\\

		\hline
	\multicolumn{7}{l}{%
  	\begin{minipage}{11cm}%
		Notes: (1)\ \ Ancillary parameters for 45 galaxies from the CLASSY sample and 5 galaxies from \cite{Heckman15}. The latter is marked as (H15). The derivation of these parameters are discussed in Section \ref{sec:anc}. Descriptions for each column: (2) Redshift of the galaxy derived from UV emission lines; (3) and (4) Adopted half-light radius for each galaxy in units of \arcsec\ and kpc, respectively, which are derived from either HST/COS or optical images; (5) Measured galaxy circular velocity; (6) and (7) The log of specific star-formation rate (SFR) and SFR per unit area, respectively, which are derived from SED fittings.\\
  	\end{minipage}%
	}\\
	\end{tabular}
	\\ [0mm]
	
\end{table*}
\clearpage


\typeout{} 
\bibliography{main}{}
\bibliographystyle{aasjournal}

\clearpage

\end{document}